 \documentclass[journal,comsoc]{IEEEtran}
\IEEEoverridecommandlockouts

\usepackage[noadjust]{cite}
\usepackage{amsmath,amssymb,amsfonts}
\usepackage{graphicx}
\usepackage{textcomp}
\usepackage{xcolor}
\usepackage{amsmath}
\newtheorem{theorem}{\bf Theorem}

\newtheorem{lemma}{\bf Lemma}
\newtheorem{definition}{\bf Definition}
\newtheorem{remark}{\bf Remark}
\newtheorem{corollary}{\bf Corollary}

 \usepackage{algorithm}
 \usepackage{algpseudocode}

\usepackage{changes}
\usepackage{verbatim}
\def\BibTeX{{\rm B\kern-.05em{\sc i\kern-.025em b}\kern-.08em
 T\kern-.1667em\lower.7ex\hbox{E}\kern-.125emX}}

\newcommand{\vect}[1]{{\mathbf{#1}}}
\newcommand{\mat}[1]{{\mathbf{#1}}}

\newcommand{\cR}{\mathcal{R}}

\newcommand{\cS}{\mathcal{S}}
\newcommand{\cT}{\mathcal{T}}

\newcommand{\cG}{\mathcal{G}}

\begin{document}

\title{Coded Computing for Half-Duplex Wireless Distributed Computing Systems via Interference Alignment%Coded Computing for Wireless Distributed Computing Systems via Interference Alignment%How to Achieve Zero Communication Latency for Wireless Distributed Computing

\author{
	\IEEEauthorblockN{Youlong Wu, Zhenhao Huang, Kai Yuan, Shuai Ma, and Yue Bi \\}
	\thanks{This work was in part presented at the \textit{IEEE International Symposium on Information Theory (ISIT)}, Espoo, Finland, July 2022 \cite{yuan2022coded}. %\cite{Kai'22}.
			
	Youlong Wu, Zhenhao Huang and Kai Yuan are with the School of Information Science and Technology, ShanghaiTech University, Shanghai 201210, China. (e-mail: \{wuyl1, huangzhh, yuank\}@shanghaitech.edu.cn).
   
   	Shuai Ma is with the Peng Cheng Laboratory, Shenzhen 518055, China. (e-mail: mashuai001@cumt.edu.cn).
   			
	Yue Bi is with the School of Electronic Information and Electrical Engineering, Shanghai Jiao Tong University, Shanghai 200240, China, and also with LTCI, Telecom Paris, IP Paris, Palaiseau 91120, France; (bi@telecom-paris.fr)
			
}
}
}

\maketitle

{\begin{abstract} 
		Distributed computing frameworks such as MapReduce and Spark are often used to process large-scale data computing jobs. In wireless scenarios, exchanging data among distributed nodes would seriously suffer from the communication bottleneck due to limited communication resources such as bandwidth and power. To address this problem, we propose a {coded parallel computing} (CPC) scheme for distributed computing systems where distributed nodes exchange information over a half-duplex wireless interference network. The CPC scheme achieves the multicast gain by utilizing coded computing to multicast coded symbols {intended to} multiple receiver nodes and the cooperative transmission gain by allowing multiple {transmitter} nodes to jointly deliver messages via interference alignment. To measure communication performance, we apply the widely used latency-oriented metric: \emph{normalized delivery time (NDT)}. It is shown that CPC can significantly reduce the NDT by jointly exploiting the parallel transmission and coded multicasting opportunities.  Surprisingly, when $K$ tends to infinity and the computation load is fixed, CPC approaches zero NDT while all state-of-the-art schemes achieve positive values of NDT. 
        {Finally, we establish an information-theoretic lower bound for the NDT-computation load
        trade-off over \emph{half-duplex} network, and prove our scheme achieves the minimum NDT  within a multiplicative gap of $3$, i.e., our scheme is order optimal.
        %The numerical results demonstrate that our scheme achieves better performance compared with other schemes.
        }

	\end{abstract}}


\section{Introduction}
Distributed computing frameworks can jointly exploit the storage and computing capabilities of distributed computing nodes and thus are often used to process large-scale data computing jobs. For instance, distributed computing has been widely used in real-life applications \cite{2010Large}, such as telephone networks or wireless networks \cite{nikoletseas2011theoretical}, network applications \cite{10.1145/2491245} and various parallel computing models such as cluster computing \cite{valentini2013overview}, grid computing \cite{sadashiv2011cluster} and cloud computing \cite{hussain2013survey}. 
However, large-scale computing tasks inevitably suffer from communication bottlenecks due to limited communication resources, i.e., communication latency for exchanging data severely deteriorates the speed of distributed computing. As pointed out in \cite{zhang2013performance}, when running TeraSort and Self-Join applications, respectively, on heterogeneous Amazon EC2 clusters, 65\% and 70\% of the overall job execution time is spent in the data exchange phase. 

Recent results have shown that introducing computational redundancy and coding techniques can significantly reduce communication latency. In particular, Li \emph{et al}. in \cite{CDC} proposed a \emph{coded distributed computing} (CDC) scheme that achieves the optimal computation-communication trade-off in MapReduce systems. It is shown that the CDC scheme with {computation load} $ r\leq 5$ (the total number of computed Map functions at all nodes normalized by the total number of input files) can achieve 1.97$\sim$3.39 times speedup for TeraSort on Amazon EC2 compared with the uncoded scheme \cite{CDC}. Due to its optimality and practical feasibility, the CDC approach has attracted wide interest in the literature
\cite{howtoallocate,DGA,pda,cmdc,cdcnc,storagecomtradeoff}. For instance, the resource allocation in the coding distributed computing scheme was considered and the optimal number of servers and the optimal data shuffling scheme were given to minimize the total execution time \cite{howtoallocate}. Saurav \emph{et al.} proposed a coded computing scheme in a distributed graph processing to reduce the communication load \cite{DGA}. The placement delivery array (PDA) was applied to reduce the number of sub-files and output functions of the CDC scheme \cite{pda}. Chen \emph{et al.} studied a central server-aided MapReduce system and proposed an optimal scheme that achieves the minimum communication load \cite{cmdc}. %Attia \emph{et al.} \cite{attia2019near} proposed a near-optimal encoding data shuffling scheme for distributed machine learning, thereby obtaining the optimal lower bound on the communication cost of computing nodes under different storage modes. 
Heterogeneous distributed computing systems with different storage sizes and computing capabilities were studied in \cite{nccd,ccoh,bflexible}. Distributed computing systems with stragglers problem (i.e., when the computation or communication capabilities of some nodes are much smaller than other nodes, the task duration would be determined by the worst computing nodes) have been investigated in \cite{narra2019slack,bitar2020stochastic,maity2019robust,ferdinand2018hierarchical}.%\cite{narra2019slack,yang2019timely,bitar2020stochastic,maity2019robust,charles2017approximate,ferdinand2018hierarchical,das2018c}.

The communication bottleneck will become more challenging for wireless distributed computing systems such as 5G beyond or 6G systems, due to the increasing number of connective devices and limited communication resources such as transmit bandwidth and power. To reduce the communication overhead, 
the authors in  \cite{CWDC} considered a wireless MapReduce system where all edge nodes communicate with a central base station via wireless \emph{orthogonal} links, and an optimal coded wireless distribution coding was proposed to achieve the minimum uplink and downlink communication loads. 
The \cite{oneshot'IT,biDoF,bi2022bounds}   exploited the multiplex gain based on zero-forcing and interference alignment techniques for wireless MapReduce systems. In particular, Li \emph{et al}. proposed a \emph{one-shot linear} (OSL) scheme by exploiting side information cancellation and zero-forcing \cite{oneshot'IT}, and prove that the OSL scheme achieves the optimal computation-communication trade-off under one-shot delivery. Bi and Wigger \cite{biDoF,bi2022bounds} further reduced the communication latency compared to \cite{CWDC,oneshot'IT} by allowing multiple nodes to jointly deliver coded signals to all other nodes via interference alignment. 
It is worth mentioning that all the aforementioned schemes are proposed for \emph{full-duplex} systems where every node can simultaneously transmit and receive signals, while most of the current wireless communication systems are \emph{half-duplex}, i.e., every node cannot transmit and receive information at the same time in the same bandwidth.
In fact, since every node cannot transmit and receive information simultaneously, two fundamental questions arise in the half-duplex wireless MapReduce system: 
\begin{itemize}
\item First, how many users should serve as the transmitter groups and receiver groups respectively in each transmission slot, to enlarge the cooperative transmission gain?    
\item Second,   how to design the coded message such that each message could be decoded by more intended receivers and achieve a larger multicast gain? 
\end{itemize}

In view of these facts, this paper considers a $K$-user half-duplex wireless MapReduce-type system, and investigates how to further reduce the communication latency (measured by normalized delivery time (NDT) \cite{NDT'IT,taoWireless}). The main contributions are summarized as follows. 

\begin{itemize} %\{\mathcal{T}_p,\mathcal{R}_p\}
	\item We propose a coding strategy, namely \emph{coded parallel computing} (CPC), which achieves a multicast gain by utilizing coded computing to multicast coded symbols where each symbol contains multiple information intended to different receiver nodes. Besides, the CPC achieves a  \emph{cooperative transmission gain} by allowing multiple transmitter nodes to jointly deliver messages, followed by using the {interference neutralization and alignment} approach to cancel interference caused by parallel transmission.   We show that there exists a trade-off between the \emph{coded multicast gain} and \emph{cooperative transmission gain}, and obtain the optimal trade-off by solving an optimization problem on minimizing the NDT of our scheme.    Our theoretical analysis shows that focusing on only one question while ignoring the other would degrade communication efficiency.
  
	\item {The CPC scheme always significantly outperforms the state-of-the-art schemes, including the original CDC scheme, the half-duplex OSL scheme \cite{oneshot'IT}, and the half-duplex BW scheme \cite{bi2022bounds}\footnote{ The original BW scheme is delicately proposed for full-duplex systems where $K$ users simultaneously transmit signals, and it is still unknown how to extend it to the half-duplex case while keeping the low NDT. For fair comparisons,   in this paper we compute the NDT of the BW scheme by multiplying the NDT of the full-duplex BW by a factor $2$, see Remark \ref{ReHalfDuplex}.}. Unlike the NDTs of {CDC} and OSL, which both monotonically increase with $\textit{K} $, the NDT of CPC will decrease with $K$, demonstrating that our CPC scheme exhibits more robust behavior when increasing the total number of nodes $K$. In fact, our scheme can even improve the \emph{full-duplex} OSL under a mild condition $K \geq 2(r+1+\sqrt{r^2+1})$.   When $ r \ll K $ and $K \to \infty$, the NDT of our CPC scheme approaches 0, while the NDTs of CDC  and half-duplex OSL both tend to be $1/r$. }
  
    \item {We also establish an information-theoretic lower bound for the NDT-computation load trade-off over \emph{half-duplex} network, and prove that the obtained NDT is within a multiplicative gap of $3$ to the optimum, i.e., our scheme is order optimal.  }

\end{itemize}

{Due to the properties of low communication latency and robustness to the increasing number of distributed nodes, our scheme could be potentially useful in wireless applications, such as autonomous control and navigation for unmanned vehicles and aerial vehicles, augmented reality enabled by mobile smartphones and nearby fog nodes, and smart city management with wireless sensors. Please see more discussions about wireless  CDC applications in  \cite{CWDC,Datla2012WirelessDC,surveyCDC}.} %

\subsection{Related Works on Coded Caching Problems}
{Coded distributed computing problem is quite close to the coded caching problem \cite{MN}, as both introduce redundant data placement and coded multicasting, except that the latter involves storage, computation, and communication while the former ignores the computation aspect.}
Coded caching was originally proposed to reduce the heavy traffic load and improve the user's latency experience. The key idea is to shift the network traffic to the low congestion periods. Users store partial contents in cache memory during the low congestion periods leading to some overlapping caching between users. Then one or multiple servers send coded signals to multiple users such that each user can recover its demanded file by exploiting the local cache during the traffic peak time.  The coded caching problem is similar to the coded distributed computing problem as both introduce redundant data placement and coded multicasting, except that the latter involves storage, computation, and communication while the former ignores the computation aspect. The coded caching technique has been widely studied in wireless scenarios \cite{oneshot'IT,taoWireless, degrees, Nader2017, Roig2017}. In \cite{taoWireless, degrees, Nader2017, Roig2017}, the authors studied a multi-server coded caching setup where receivers connect with multiple transmitters, both equipped with cache memories, through an interference wireless network. By combining coded caching, zero-forcing, interference alignment, and interference cancellation, different content delivery schemes are proposed to reduce communication latency. 

Compared with the above works on wireless coded caching, our work differs from them in three aspects. First, in \cite{taoWireless, degrees, Nader2017, Roig2017}, the nodes are either transmitters or receivers, while in our work each node could play the roles of both the transmitter and receiver. As we consider the half-duplex mode, the transmission groups and receiver groups would be dynamic in different transmission slots.  Second, the transmitters in \cite{taoWireless, degrees, Nader2017, Roig2017} store all input files, while in our model each transmitter group is not necessary to store all input files, leading to the different data placement and delivery strategy. Third,   there exists a trade-off between the multicast gain and cooperative gain (see discussion in Remark \ref{gainchange}), which does not exist in previous works. How to determine the optimum sizes and index sets of transmission groups and receiver groups is a totally new problem not seen in previous works. We obtain the optimal trade-off by solving an optimization problem on minimizing the communication latency of our scheme.

\subsection{Organization and Notations}
The remainder of this work is organized as follows. The problem formulation is given in Section \ref{problem formulation}.   Section \ref{CPCwithIA} describes the coded parallel transmission strategy and analyzes its performance. Section \ref{sec:converse} presents a lower bound for NDT-computation load trade-off and proves the order optimality of our scheme. Section \ref{numerical results} gives some numerical results. The work is concluded in Section \ref{conclusion}. 

\emph{Notations:} We use italic, boldface, and calligraphic letters to denote scalar values, vectors (matrices), and sets, respectively. Let $\mathbb{N}^+$ denote the set of positive integers. The operation $(\cdot)^{T}$ denotes the transpose. The sets of real numbers and complex numbers are denoted by $\mathbb{R}$ and $\mathbb{C}$, respectively. Let $\mathcal{CN}(m,\delta^2)$ denote the complex Gaussian distribution with a mean of $m$ and variance of $\delta^2$. The operator $|\cdot|$ is the cardinality of a set or the absolute value of a scalar number. For any $k\in\mathbb{N}^{+}$, define $[k]\triangleq \{1,2,\cdots,k\}$.  We use $\mathbb{F}_{2^{F}}$ to denote the set of binary sequences with length $F$. For a finite set $\mathcal{A}$ and any $n\in\mathbb{Z}^+$, define $[\mathcal{A}]^n$ as the collection of all the subset of $\mathcal{A}$ with cardinality $n$, i.e. $[\mathcal{A}]^t \triangleq \{\cT\colon \cT \subset \mathcal{A}, |\cT|=t\}$. In particular, $[[n]]^t$ denotes the set of all size-$t$ subsets of $[n]$.

\section{Problem Formulation} \label{problem formulation}

Consider a half-duplex wireless MapReduce-type system where $K$ nodes, each equipped with a single antenna, exchange information via a wireless network. Consider the same MapReduce-type task as in \cite{CDC}. It contains $Q\in\mathbb{N}^+ $ output functions and $ N\in\mathbb{N}^+ $ input files $ \omega_{1},...,\omega_{N} \in \mathbb{F}_{2^{\textit{F}}}$, for some $F\in\mathbb{N}^+ $. Let  $ {\mathcal{M}}_{k} \subseteq \{ \omega_{1},...,\omega_{N}\}$  be the set of files stored at node $k\in [K] $ and $ {\mathcal{W}}_{k} \subseteq [Q]$ be the index set of output functions computed by node $ k $. Similar to \cite{CDC}, we define
\begin{IEEEeqnarray}{rCl}
	 \eta_1 \triangleq \frac{N}{\binom{K}{r}} \text{ and } \eta_2 \triangleq \frac{Q}{K}.
\end{IEEEeqnarray}
We also assume $\eta_1$ and $\eta_2$ are integers in Section \ref{CPCwithIA} such that all files and outputs can be symmetrically assigned to $K$ nodes, while we only assume $\eta_2$ is an integer in Section \ref{sec:converse}.  The wireless MapReduce task contains the following three phases:

\subsubsection{Map phase} Node $k\in[K]$ maps each file $w_n\in\mathcal{M}_k$ into length-$B$ intermediate values (IVs) as follows:
%\begin{IEEEeqnarray}{rCl}
$v_{q,n}= g_{q,n}(w_n)\in\mathbb{F}_{2^B},\forall q\in[Q], $ %\nonumber
%\end{IEEEeqnarray}
where $g_{q,n}$ is a Map function.

We use the same definition of \emph{computation load} as in \cite{CDC}.
\begin{definition}
	Define the \emph{computation load} as {the total number of mapped files at all nodes normalized by $N$}, i.e.,
	%\begin{IEEEeqnarray}{rCl}
	$r = \sum_{k=1}^{K} \frac{|\mathcal{M}_{k}|}{N}.$ % \nonumber
	%\end{IEEEeqnarray}
\end{definition}

After the map phase, node $k$ already knows IVs $(v_{q,n}:w_n\in\mathcal{M}_k,q\in[Q])$ and requires $(v_{q,n}:w_n\notin\mathcal{M}_k,q\in\mathcal{W}_k)$.

\subsubsection{Shuffle phase}
 Let $T$ be the total communication block length in the Shuffle phase and $h_{j,m}(t) \in \mathbb{C}$ represent the channel gain from transmitter $m$ to receiver $j$ at time $t\in[T]$. {In each block, all the $K$ nodes are divided into transmitters and receivers. We assume the channel state information (CSI) is perfectly known by all nodes as in \cite{oneshot'IT,biDoF,bi2022bounds}, where the CSI could be estimated by using pilot sequences  \cite{Liu2017MassiveCW} or via the deep learning technique \cite{kulkarnideepchannel}.} In the Shuffle phase, transmitter $m$ generates input signals $X_m(t)$ based on CSI and the local IVs $(v_{q,n}:w_n\in\mathcal{M}_m,q\in[Q])$. The communication over the network is modeled as
\begin{IEEEeqnarray}{rCl}
	Y_{j}(t) = \sum\nolimits_{m=1}^Kh_{j,m}(t)X_m(t)+Z_{j}(t),
\end{IEEEeqnarray}
where $Y_{j}(t)\in \mathbb{C}$ is the received signal at receiver $j\in[K]$, $X_m(t)\in \mathbb{C}$ is the transmitted signal sent by transmitter $m\in[K]$ subject to a power constraint $\mathbb{E}[|X_m(t)|^2]\leq P$, and $Z_{j}(t)\sim \mathcal{CN}(0,1)$ denotes the additive white Gaussian noise. We consider the \emph{half-duplex} communication, where each node can not receive or transmit signal at the same time, i.e., at time $t\in[T]$, $ Y_{k}(t) \cdot X_k(t) = 0 $ for all $ k \in [K] $.

{At the end of the  Shuffle phase, node $ k $ decodes the desired IVs based on received signal $ \textbf{Y}_k \triangleq (Y_k(1), ... ,Y_k(T) )$ and the locally mapped IVs $(v_{q,n}:q\in[Q],w_n\in\mathcal{M}_k )$}, i.e., %More specifically, each node $k$ uses decoding functions $ \chi_k$ to decode the desired IVs as
 \begin{IEEEeqnarray*}{rCl}
	(\hat{v}_{q,n} : q \in \mathcal{W}_k, w_n\notin\mathcal{M}_k) = \chi_k\big( \textbf{Y}_k ,(v_{q,n}:q\in[Q], w_n\in\mathcal{M}_k )\big),\nonumber
\end{IEEEeqnarray*}  
where $ \chi_k$ is a decoding function.

\subsubsection{Reduce Phase}

For any $\textit{q} \in \mathcal{W}_k$, node $ k $ computes the desired output functions through the Reduce operation {$ \textit{u}_q = \phi_q(v_{\textit{q},1},...,v_{\textit{q},N})$}, for all $ q \in \mathcal{W}_k $.

To characterize the transmission latency of the network, we introduce the definition NDT similar to that in \cite{taoWireless}.
\begin{definition}[$\textit{Normalized Delivery Time}$] The normalized delivery time is defined as
	\begin{IEEEeqnarray}{rCl}\label{eqNDT}
		\delta(r) = \lim_{P\to\infty} \lim_{B\to\infty} \frac{T}{NQB/ \log P}.
	\end{IEEEeqnarray}
	%\end{IEEEeqnarray}
\end{definition}

Given the computation load $r$, the NDT $\delta(r) $ is said to be \emph{feasible} if there exists a wireless MapReduce scheme consisting of the Map and Shuffle operations such that all desired IVs can be decoded at the intended nodes with  error probability 
\begin{IEEEeqnarray}{rCl} \label{eq:error_computing}
	\textnormal{Pr} \left( \exists k \in [K], q \in \mathcal{W}_k, w_n \notin \mathcal{M}_k \colon \hat{v}_{q, n} \neq v_{q,n} \right) \rightarrow 0
\end{IEEEeqnarray}	
 as $B$ increases. 
Define the \emph{minimum} NDT as
%\begin{IEEEeqnarray}{rCl}
$\delta^*(r) = \text{inf}\{\delta(r):\delta(r) ~\text{is feasible}\}$, {where $\text{inf}\{\mathcal{S}\}$ denotes the infimum of a set $\mathcal{S}$}.
%\end{IEEEeqnarray}

 Similar to \cite[Remark 1]{taoWireless}, we can interpret NDT as follows. 
\begin{remark}\label{Rem1}
 Let 
 	\begin{IEEEeqnarray}{rCl}
 		R \triangleq \frac{\sum_{k\in[K]}|\left\{ (q, n) \colon q \in \mathcal{W}_k, w_n \notin \mathcal{M}_k \right\}| \cdot B}{KNQB}
 	\end{IEEEeqnarray}
 denote the average \emph{communication load} per user normalized by $NQB$, and $d$ be the per-user degree of freedom (DoF) of the communication channel, i.e., {the per-user capacity in high signal-to-noise regime can be approximately written as $(d\cdot\log P + o(\log P))$.} Then, the communication time is $T=\frac{RNQB}{d\cdot \log P + o(\log P)}$, and we have 
		\begin{IEEEeqnarray}{rCl}\label{eqNDTLoad_per}
			\delta(r) = {R}/{d}.
		\end{IEEEeqnarray}  
\end{remark}
For example, {the uncoded time-division multiple access (TDMA) broadcast scheme, where each user  broadcasts uncoded messages to other users in different time slots}, achieves the NDT
	\begin{IEEEeqnarray}{rCl}\label{uncoded}
		\delta_\text{uncoded}(r,K) =\frac{R}{d}=\frac{(1-\frac{r}{K}) \frac{1}{K} }{ \frac{1}{K} } = 1-\frac{r}{K}.
	\end{IEEEeqnarray}  
	The CDC scheme \cite{CDC} achieves the NDT
	\begin{IEEEeqnarray}{rCl}\label{TDCDC}
		\delta_\text{CDC}(r,K)=\frac{1}{r}\left(1-\frac{r}{K}\right),
	\end{IEEEeqnarray} because the communication load is $\frac{1}{K}(1-\frac{r}{K})$ and per-receiver DoF $d=\frac{r}{K}$. 
	The full-duplex OSL scheme \cite{oneshot'IT} achieves the NDT 
	\begin{IEEEeqnarray}{rCl}\label{TDoneshot}
		\delta^{\text{FD}}_\text{OSL}(r,K)=\frac{1}{\min\{K,2r\}}\left(1-\frac{r}{K}\right).
	\end{IEEEeqnarray}
	The full-duplex  BW scheme in  \cite{bi2022bounds} achieves the NDT 
	\begin{IEEEeqnarray}{rCl}\label{BiFullDuplex}
		\delta^{\text{FD}}_\text{BW}(r,K)=\left\{
		\begin{array}{ll}
			\!\!(1-\frac{r}{K}) \frac{1}{K}, & \text{if}~r\geq \frac{K}{2}, \\
			\!\!(1-\frac{r}{K}) \frac{r(K-1)+K-r-1}{r(K-1)^2+r(K-2)}, &  \text{if}~ r<\frac{K}{2}. 
		\end{array}
		\right.
	\end{IEEEeqnarray}
	
	{
\begin{remark} \label{ReHalfDuplex}Both the schemes of \cite{oneshot'IT} and \cite{bi2022bounds}  are proposed for full-duplex systems where $K$  nodes will transmit and receive signals simultaneously. To adapt these schemes into a half-duplex manner, the resulting NDT will increase by at least a factor of 2, since the full duplex can at most double the channel capacity compared to half-duplex communications \cite{FullDuplexMagazine}\footnote{We acknowledge the reviewer for pointing out that the OSL scheme can be modified to work in half-duplex mode with exactly a double NDT  of full-duplex OSL. 
When $r\geq K/2$, the $K$ nodes can be divided into the transmitter group and the receiver group in different time blocks, while all the desired intermediate values for the nodes in the receiver group can be mapped by the transmitter group. With the similar beamforming design in the original scheme, the intermediate values will be zero-forcing or canceled with side information at unintended receivers. Similar methods can be applied when $r<K/2$ by alternately letting $K-2r$ nodes be inactive in different transmission rounds.}. 
In this paper, our focus is to delicately design a communication-efficient half-duplex scheme, instead of extending their schemes to the half-duplex system. For a fair comparison, we denote the NDTs of the half-duplex OSL and  half-duplex BW scheme as 
\begin{IEEEeqnarray}{rCl}\label{OSLHalfDuplex}
\delta^{\text{HD}}_\text{OSL}(r,K)=\frac{2}{\min\{K,2r\}}\left(1-\frac{r}{K}\right),
\end{IEEEeqnarray}
and 
\begin{IEEEeqnarray}{rCl}\label{BiHalfDuplex}
		\delta^{\text{HD}}_\text{BW}(r,K)=\left\{
		\begin{array}{ll}
			\!\!(1-\frac{r}{K}) \frac{2}{K}, & \text{if}~r\geq \frac{K}{2}, \\
			\!\!(1-\frac{r}{K}) \frac{2\left(r(K-1)+K-r-1\right)}{r(K-1)^2+r(K-2)}, &  \text{if}~ r<\frac{K}{2}, 
		\end{array}
		\right.
	\end{IEEEeqnarray}
	respectively. As we will see in Section \ref{CompareExisting},  our proposed schemes  greatly improves the $\delta^{\text{HD}}_\text{BW}(r,K)$ and $\delta^{\text{HD}}_\text{OSL}(r,K)$. 
\end{remark}
}

\section{Coded Parallel Transmission with Interference Alignment} \label{CPCwithIA}
In this section, we first present an illustrative example and then present our general CPC scheme.
\subsection{An Illustrative Example} \label{exam}
%\begin{sloppypar}
Consider a MapReduce-type system with $K=6$, $Q=6$, $N=20$,  $ r=3 $, file placement 
    \begin{IEEEeqnarray}{rcl}
        \mathcal{M}_1 &=& \{1,2,3,4,5,6,7,8,9,10\},\nonumber\\
        \mathcal{M}_2 &=& \{1,2,3,4,11,12,13,14,15,16\},\nonumber\\
        \mathcal{M}_3 &=& \{1,5,6,7,11,12,13,17,18,19\},\nonumber\\
        \mathcal{M}_4 &=& \{2,5,8,9,11,14,15,17,18,20\},\nonumber\\
        \mathcal{M}_5 &=& \{3,6,8,10,12,14,16,17,19,20\},\nonumber\\
        \mathcal{M}_6 &=& \{4,7,9,10,13,15,16,18,19,20\},\nonumber
    \end{IEEEeqnarray}
and output assignment $ \mathcal{W}_k = k, k \in [Q] $. %As shown in Fig.~\ref{fig:KQ6N10mapdiv}
%\end{sloppypar}
Let $v_{d_j,\mathcal{U} } \triangleq \{v_{q,n}:q\in\mathcal{W}_j, w_n \in \cap_{i\in\mathcal{U}} \mathcal{M}_i\}$, $|\mathcal{U}|=3$ denote the IVs stored at all nodes in $ \mathcal{U} $ and required by node $ j $, e.g., $ v_{d_4,\{1,2,5\}} = v_{4,3}$. In the Shuffle phase, partition all 6 nodes into {two groups: }a 3-node transmitter group $\mathcal{T}_p$ and a 3-node receiver group $\mathcal{R}_p$ with {$p\in \{1,\dots,20\}$}, as there are $ \tbinom{6}{3} = 20 $ partitions $ \pi_p = \{\mathcal{T}_p, \mathcal{R}_p \}$, {as shown in Table \ref{tab_par}}. 
\begin{comment}
\begin{table}[htbp]
	\centering
	\caption{$20$ partitions for $6$ nodes with $|\mathcal{T}_p|=|\mathcal{R}_p|=3$.}
	\begin{tabular}{|c|c|c||c|c|c||c|c|c||c|c|c|}
		\hline
		$\pi_{p}$ & $\mathcal{T}_p$  & $\mathcal{R}_p$ & $\pi_{p}$ & $\mathcal{T}_p$  & $\mathcal{R}_p$ & $\pi_{p}$ & $\mathcal{T}_p$  & $\mathcal{R}_p$ & $\pi_{p}$ & $\mathcal{T}_p$  & $\mathcal{R}_p$ \\ \hline
		$\pi_{1}$ & $\{1,2,3\}$ & $\{4,5,6\}$ & $\pi_{2}$ & $\{1,2,6\}$ & $\{4,5,3\}$ & $\pi_{3}$ & $\{1,3,5\}$ & $\{2,4,6\}$ & $\pi_{4}$ & $\{1,5,6\}$ & $\{2,3,4\}$ \\ \hline
		$\pi_{5}$ & $\{2,3,5\}$ & $\{1,4,6\}$ & $\pi_{6}$ & $\{2,5,6\}$ & $\{1,3,4\}$ & $\pi_{7}$ & $\{2,3,4\}$ & $\{1,5,6\}$ & $\pi_{8}$ & $\{2,4,6\}$ & $\{1,3,5\}$ \\ \hline
		$\pi_{9}$ & $\{1,2,5\}$ & $\{3,4,6\}$ & $\pi_{10}$ & $\{1,2,6\}$ & $\{3,4,5\}$ & $\pi_{11}$ & $\{1,2,4\}$ & $\{3,5,6\}$ & $\pi_{12}$ & $\{2,3,6\}$ & $\{1,4,5\}$ \\ \hline
		$\pi_{13}$ & $\{3,4,5\}$ & $\{1,2,6\}$ & $\pi_{14}$ & $\{3,4,6\}$ & $\{1,2,5\}$ & $\pi_{15}$ & $\{3,5,6\}$ & $\{1,2,4\}$ & $\pi_{16}$ & $\{4,5,6\}$ & $\{1,2,3\}$ \\ \hline
		$\pi_{17}$ & $\{1,2,4\}$ & $\{3,5,6\}$ & $\pi_{18}$ & $\{2,4,5\}$ & $\{1,3,6\}$ & $\pi_{19}$ & $\{1,3,4\}$ & $\{2,5,6\}$ & $\pi_{20}$ & $\{1,3,6\}$ & $\{2,4,5\}$ \\ \hline
	\end{tabular}
	\label{tab_par}
\end{table}
\end{comment}
\begin{table}[htbp]
	\centering
	\caption{$20$ partitions for $6$ nodes with $|\mathcal{T}_p|=|\mathcal{R}_p|=3$.}
	\begin{tabular}{|c|c|c||c|c|c|}
		\hline
		$\pi_{p}$ & $\mathcal{T}_p$  & $\mathcal{R}_p$ & $\pi_{p}$ & $\mathcal{T}_p$  & $\mathcal{R}_p$ \\\hline
		$\pi_{1}$ & $\{1,2,3\}$ & $\{4,5,6\}$ & $\pi_{2}$ & $\{1,2,6\}$ & $\{4,5,3\}$ \\\hline
        $\pi_{3}$ & $\{1,3,5\}$ & $\{2,4,6\}$ & $\pi_{4}$ & $\{1,5,6\}$ & $\{2,3,4\}$ \\ \hline
		$\pi_{5}$ & $\{2,3,5\}$ & $\{1,4,6\}$ & $\pi_{6}$ & $\{2,5,6\}$ & $\{1,3,4\}$ \\\hline
        $\pi_{7}$ & $\{2,3,4\}$ & $\{1,5,6\}$ & $\pi_{8}$ & $\{2,4,6\}$ & $\{1,3,5\}$ \\ \hline
		$\pi_{9}$ & $\{1,2,5\}$ & $\{3,4,6\}$ & $\pi_{10}$ & $\{1,2,6\}$ & $\{3,4,5\}$ \\\hline
        $\pi_{11}$ & $\{1,2,4\}$ & $\{3,5,6\}$ & $\pi_{12}$ & $\{2,3,6\}$ & $\{1,4,5\}$ \\\hline
		$\pi_{13}$ & $\{3,4,5\}$ & $\{1,2,6\}$ & $\pi_{14}$ & $\{3,4,6\}$ & $\{1,2,5\}$ \\\hline
        $\pi_{15}$ & $\{3,5,6\}$ & $\{1,2,4\}$ & $\pi_{16}$ & $\{4,5,6\}$ & $\{1,2,3\}$ \\ \hline
		$\pi_{17}$ & $\{1,2,4\}$ & $\{3,5,6\}$ & $\pi_{18}$ & $\{2,4,5\}$ & $\{1,3,6\}$ \\\hline
        $\pi_{19}$ & $\{1,3,4\}$ & $\{2,5,6\}$ & $\pi_{20}$ & $\{1,3,6\}$ & $\{2,4,5\}$ \\\hline
	\end{tabular}
	\label{tab_par}
\end{table}
The size of transmitter cooperation group $\mathcal{B}$ is $|\mathcal{B}|=2$ and the size of receiver multicast group $\mathcal{D}$ is $|\mathcal{D}|=2$. The IV $ v_{d_j,\mathcal{U}} $ will be sent by transmitter cooperation groups $\mathcal{B} \subseteq \mathcal{U}$ in each partition $\pi_p$ where $p$ satisfies $ \mathcal{B} \subseteq \mathcal{T}_p$ and $ \{j\} \cup \mathcal{U} \backslash \mathcal{B} \subseteq \mathcal{R}_p$. Thus, evenly split each $v_{d_j,\mathcal{U}} $ into $ \tbinom{3}{2}\tbinom{6-3-1}{3-2}=6$ segments $ v^{(p,\mathcal{B})}_{d_j,\mathcal{U}\backslash \mathcal{B}}$ where $\mathcal{B}\subseteq\mathcal{U}$ and $ p$ representing the index of partition satisfying $\{j\} \cup \mathcal{U} \backslash \mathcal{B} \subseteq \mathcal{R}_p$, see definition in \eqref{definegen}, e.g., $v_{d_4,\{1,2,5\}} = \big( v_{d_4,\{5\}}^{(1,\{1,2\})}, v_{d_4,\{5\}}^{(2,\{1,2\})}, v_{d_4,\{2\}}^{(3,\{1,5\})},  v_{d_4,\{2\}}^{(4,\{1,5\})}, v_{d_4,\{1\}}^{(5,\{2,5\})}, v_{d_4,\{1\}}^{(6,\{2,5\})} \big) $. 

%according to the number of times $ k \in \mathcal{U} $ appears at $ \mathcal{R}$ and $ \{j\} \cup \mathcal{U} \backslash \{k\} $ appears at $ \mathcal{T} $ simultaneously , divide equally into disjoint segments $ v^{(p,k)}_{d_j,\mathcal{U}}$, $ p $ represents the partition index, $ k $ represents the specific transmitter (specific definition sees \eqref{definegen}), e.g., $ v_{1,\{2,4\}} = (v_{1,\{2,4\}}^{(1,4)}, v_{1,\{2,4\}}^{(1,4)}, v_{1,\{2,4\}}^{(2,4)}, v_{1,\{3,2\}}^{(3,2)})$. 

Let $ V^{(p)}_{\mathcal{D},\mathcal{B}} $ denote the encoded message sent by node group $ \mathcal{B} $ to receiver node $ j \in \mathcal{D} $ in partition $ \pi_p $. Generate $ \{V^{(p)}_{\mathcal{D},\mathcal{B}}\} $ according to \eqref{Eqencoding}, e.g., 
%\begin{align*}
	%V_{\{4,5\},\{1,2\}}^{(1)} &\!\!=\! v_{d_4,\{5\}}^{(1,\{1,2\})} \!\!\oplus\! v_{d_5,\{4\}}^{(1,\{1,2\})} \!,\!\!\!\!\! & V_{\{4,5\},\{1,2\}}^{(2)} &\!\!=\! v_{d_4,\{5\}}^{(2,\{1,2\})} \!\!\oplus\! v_{d_5,\{4\}}^{(2,\{1,2\})} , \\
	%V_{\{2,4\},\{1,5\}}^{(3)} &= v_{d_4,\{2\}}^{(3,\{1,5\})} \oplus v_{d_2,\{4\}}^{(3,\{1,5\})} , & V_{\{2,4\},\{1,5\}}^{(4)} &= v_{d_4,\{2\}}^{(4,\{1,5\})} \oplus v_{d_2,\{4\}}^{(4,\{1,5\})} , \\
	%V_{\{1,4\},\{2,5\}}^{(5)} &= v_{d_4,\{1\}}^{(5,\{2,5\})} \oplus v_{d_1,\{4\}}^{(5,\{2,5\})} , & V_{\{1,4\},\{2,5\}}^{(6)} &= v_{d_4,\{1\}}^{(6,\{2,5\})} \oplus v_{d_1,\{4\}}^{(6,\{2,5\})} .
%\end{align*} 
\begin{align*}
	V_{\{4,5\},\{1,2\}}^{(1)} &= v_{d_4,\{5\}}^{(1,\{1,2\})} \oplus v_{d_5,\{4\}}^{(1,\{1,2\})} , \\
        V_{\{4,5\},\{1,2\}}^{(2)} &= v_{d_4,\{5\}}^{(2,\{1,2\})} \oplus v_{d_5,\{4\}}^{(2,\{1,2\})} , \\
	V_{\{2,4\},\{1,5\}}^{(3)} &= v_{d_4,\{2\}}^{(3,\{1,5\})} \oplus v_{d_2,\{4\}}^{(3,\{1,5\})} , \\
        V_{\{2,4\},\{1,5\}}^{(4)} &= v_{d_4,\{2\}}^{(4,\{1,5\})} \oplus v_{d_2,\{4\}}^{(4,\{1,5\})} , \\
	V_{\{1,4\},\{2,5\}}^{(5)} &= v_{d_4,\{1\}}^{(5,\{2,5\})} \oplus v_{d_1,\{4\}}^{(5,\{2,5\})} , \\
        V_{\{1,4\},\{2,5\}}^{(6)} &= v_{d_4,\{1\}}^{(6,\{2,5\})} \oplus v_{d_1,\{4\}}^{(6,\{2,5\})} .
\end{align*} This message generation ensures that each node $j\in\mathcal{R}_p$ can decode the desired segments $v^{(p,\mathcal{B})}_{d_j, \mathcal{D}\backslash \{j\}}$ if $ V^{(p)}_{\mathcal{D},\mathcal{B}} $ are successfully obtained. For example, node 4 can decode 
\begin{align*}
	v_{d_4,\{5\}}^{(1,\{1,2\})}, v_{d_4,\{5\}}^{(2,\{1,2\})}, v_{d_4,\{2\}}^{(3,\{1,5\})}, 
	v_{d_4,\{2\}}^{(4,\{1,5\})}, v_{d_4,\{1\}}^{(5,\{2,5\})}, v_{d_4,\{1\}}^{(6,\{2,5\})} 
\end{align*}
by XORing $V_{\{4,5\},\{1,2\}}^{(1)}$, $V_{\{4,5\},\{1,2\}}^{(2)}$, $V_{\{2,4\},\{1,5\}}^{(3)}$, $V_{\{2,4\},\{1,5\}}^{(4)}$, $V_{\{1,4\},\{2,5\}}^{(5)}$, and $V_{\{1,4\},\{2,5\}}^{(6)}$ with the local segments $v_{d_5,\{4\}}^{(1,\{1,2\})}$, $v_{d_5,\{4\}}^{(2,\{1,2\})}$, $v_{d_2,\{4\}}^{(3,\{1,5\})}$, $v_{d_2,\{4\}}^{(4,\{1,5\})}$, $v_{d_1,\{4\}}^{(5,\{2,5\})}$, and $v_{d_1,\{4\}}^{(6,\{2,5\})}$, respectively. {For a receiver $j$ in one partition $\pi_{p}$, $p\in[20]$, the total size of messages desired by receiver $j$, i.e., $\{V^{(p)}_{\mathcal{D},\mathcal{B}}: d\in\mathcal{D},\mathcal{B}\in\mathcal{T}_p,\mathcal{D}\subseteq\mathcal{R}_p\} $, is $\frac{B}{6}\binom{|\mathcal{R}_p-1|}{|\mathcal{D}-1|}\cdot \binom{|\mathcal{T}_p|}{|\mathcal{B}|}=B$ bits, and {the communication load in this partition for one receiver is $R_p=B/NQB=1/120$.}

\begin{figure*}
	\centering
	\includegraphics[width=0.9\linewidth]{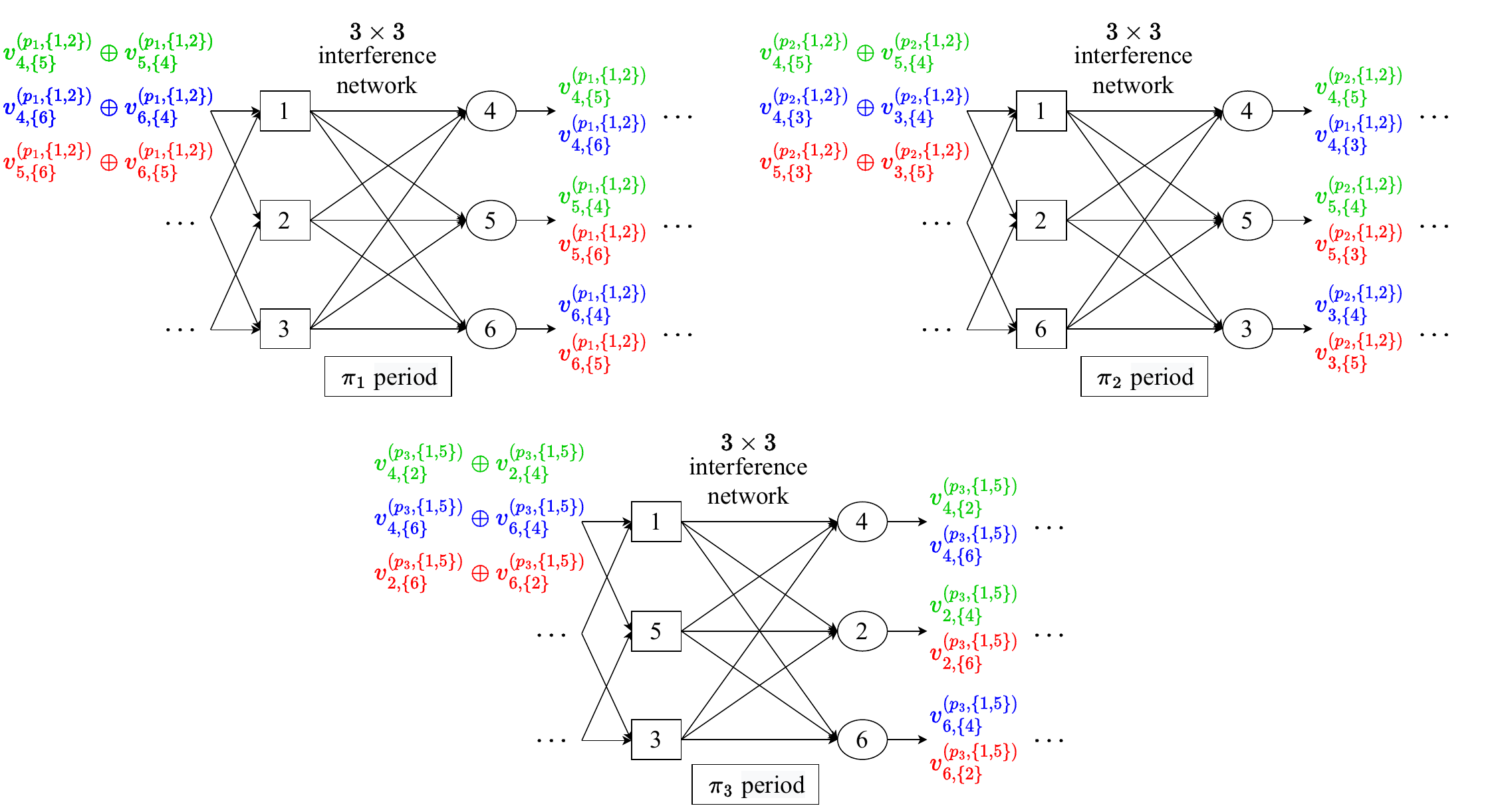}
	\caption{An example of the communication over the wireless network in partition $\pi_1$, $\pi_2$, $\pi_3$ with $K_\textnormal{t}=K_\textnormal{r}=3$, $Q=6$, $t=2$, $s=2$. In $\pi_{1}$ period, the transmitter cooperation group $\{1,2\}$, $\{1,3\}$ and $\{2,3\}$ sequentially send their message. The message sent by node group $\{1,2\}$ in period $\pi_{1}$ can be seen in the figure. The receiver nodes can decode their messages according to the interference neutralization and alignment approach. A similar transmission strategy is adopted for different cooperation groups and different periods in this example. Note that the interference management strategy is different according to the relation between $s$, $t$ and $K_{\textnormal{r}}$.}
	\label{KQ6N10}
\end{figure*}

The communication in all partitions has the same structure, see Fig.~\ref{KQ6N10}. 
In this example, a total of $\binom{K_\textnormal{t}}{|\mathcal{B}|}\binom{K_\textnormal{r}}{|\mathcal{D}|}=\binom{3}{2}\binom{3}{2}$ messages need to be sent in $\pi_{1}$. These messages will be divided into $3$ groups which are sent sequentially in a time division manner according to the different transmitter groups. We use the symbol $2$ extension to transmit the $3$ messages for each group. Each receiver node desires $2$ messages out of $3$ messages for each group. For each symbol, it can be neutralized at the undesired receiver so that each receiver can successfully decode $2$ desired messages in $2$ time slots leading to the per-receiver DoF of 1.  
{
Take the transmitter cooperation group $\{1,2\}$ in period $\pi_1$ for example. Denote $\mathbf{V}_{\{4,5\},\{1,2\}}$, $\mathbf{V}_{\{4,6\},\{1,2\}}$ and $\mathbf{V}_{\{5,6\},\{1,2\}}$ as the transmitted symbols encoded from messages $V^{(1)}_{\{4,5\},\{1,2\}}$, $V^{(1)}_{\{4,6\},\{1,2\}}$ and $V^{(1)}_{\{5,6\},\{1,2\}}$, respectively. For simplicity, we omit the time index of the precoder.  In each time slot $u\in [2]$,  receiver $j\in\{4,5,6\}$ receives the signal as follows (neglecting the noise), e.g.,
\begin{IEEEeqnarray}{rcl}\label{eq:eg_y4_i}
    %&y_4(u)& \nonumber\\
    %&&\hspace{-8pt}= h_{4,1}(u)( w_{\{4,5\},\{1,2\},1}\mathbf{V}_{\{4,5\},\{1,2\}} + w_{\{4,6\},\{1,2\},1}\mathbf{V}_{\{4,6\},\{1,2\}} ) \nonumber\\
    %&&\hspace{-5pt}+ h_{4,2}(u)( w_{\{4,5\},\{1,2\},2}\mathbf{V}_{\{4,5\},\{1,2\}} + w_{\{4,6\},\{1,2\},2}\mathbf{V}_{\{4,6\},\{1,2\}} ) \nonumber\\
    %&&\hspace{-5pt}+ h_{4,1}(u)w_{\{5,6\},\{1,2\},1}\mathbf{V}_{\{5,6\},\{1,2\}}   \nonumber\\
    %&&\hspace{-5pt}+ h_{4,2}(u)w_{\{5,6\},\{1,2\},2}\mathbf{V}_{\{5,6\},\{1,2\}} 
    &y_4(u)& \nonumber\\
    &&\hspace{-13pt}= ( h_{4,1}(u) w_{\{4,5\},\{1,2\},1} + h_{4,2}(u) w_{\{4,5\},\{1,2\},2} ) \mathbf{V}_{\{4,5\},\{1,2\}} \nonumber\\
    &&\hspace{-10pt}+ ( h_{4,1}(u)w_{\{4,6\},\{1,2\},1} + h_{4,2}(u)w_{\{4,6\},\{1,2\},2} ) \mathbf{V}_{\{4,6\},\{1,2\}} \nonumber \\
    &&\hspace{-10pt}+ ( h_{4,1}(u)w_{\{5,6\},\{1,2\},1} + h_{4,2}(u)w_{\{5,6\},\{1,2\},2} ) \mathbf{V}_{\{5,6\},\{1,2\}}\nonumber\\
\end{IEEEeqnarray}
%\begin{IEEEeqnarray}{rcl}\label{eq:eg_y4_i} 
%	y_4(u)
%	&&= \sum_{m=1}^{2}\left(h_{4,m}(u)w_{\{4,5\},\{1,2\},m}(u) \right) \mathbf{V}_{\{4,5\},\{1,2\}} \nonumber\\
%	&&\hspace{+10pt}+ \sum_{m=1}^{2}\left(h_{4,m}(u)w_{\{4,6\},\{1,2\},m}(u) \right) \mathbf{V}_{\{4,6\},\{1,2\}} \nonumber \\
%	&&\hspace{+10pt}+ \sum_{m=1}^{2}\left(h_{4,m}(u)w_{\{5,6\},\{1,2\},m}(u) \right) \mathbf{V}_{\{5,6\},\{1,2\}}\nonumber\\
%\end{IEEEeqnarray}
where $h_{j,m}(u)$ is the channel realization, and $w_{\mathcal{D},\mathcal{B},m}$ is the precoder of symbol $\mathbf{V}_{\mathcal{D},\mathcal{B}}$ at transmitter $m$. Since $\mathbf{V}_{\{5,6\},\{1,2\}}$ is desired by receiver $5,6$ and unwanted by receiver $4$,  the precoder should satisfy 
\begin{IEEEeqnarray}{rcl}\label{eq:eg_neu}
    \begin{cases}
    & h_{4,1}(u)w_{\{5,6\},\{1,2\},1} + h_{4,2}(u)w_{\{5,6\},\{1,2\},2} = 0,\\
    & h_{5,1}(u)w_{\{5,6\},\{1,2\},1} + h_{5,2}(u)w_{\{5,6\},\{1,2\},2} \neq 0, \\
    & h_{6,1}(u)w_{\{5,6\},\{1,2\},1} + h_{6,2}(u)w_{\{5,6\},\{1,2\},2} \neq 0. 
    \end{cases}
\end{IEEEeqnarray}
To neutralize the interference in \eqref{eq:eg_y4_i}, One solution of \eqref{eq:eg_neu} is $w_{\{5,6\},\{1,2\},1} = -h_{4,2}(u)$, $w_{\{5,6\},\{1,2\},2} = h_{4,1}(u)$. Similarly, we can design the precoder of other symbols, i.e., $w_{\{4,5\},\{1,2\},1} = -h_{6,2}(u)$, $w_{\{4,5\},\{1,2\},2} = h_{6,1}(u)$, $w_{\{4,6\},\{1,2\},1} = -h_{5,2}(u)$ and $w_{\{4,6\},\{1,2\},2} = h_{5,1}(u)$.
After interference neutralization, the received signal in \eqref{eq:eg_y4_i} can be rewritten as 
\begin{IEEEeqnarray}{rcl}
    &\tilde{y}_4(u)& \nonumber\\
    &&\hspace{-13pt}= ( h_{4,1}(u) w_{\{4,5\},\{1,2\},1} + h_{4,2}(u) w_{\{4,5\},\{1,2\},2} ) \mathbf{V}_{\{4,5\},\{1,2\}} \nonumber\\
    &&\hspace{-10pt}+ ( h_{4,1}(u)w_{\{4,6\},\{1,2\},1} + h_{4,2}(u)w_{\{4,6\},\{1,2\},2} ) \mathbf{V}_{\{4,6\},\{1,2\}}. \nonumber
\end{IEEEeqnarray}
From $\tilde{y}_4(u)$, $\mathbf{V}_{\{4,5\},\{1,2\}}$ and $\mathbf{V}_{\{4,6\},\{1,2\}}$ can be decoded by receiver $4$ in $2$-symbol extension with probability $1$ \cite[Lemma 3]{Annapureddy2012}.
Meanwhile, the receiver $5$ and receiver $6$ can decode their desired symbols. Thus, a per-receiver DoF of $1$ is achieved.
}
According to \eqref{eqNDTLoad_per}, the NDT of each user in one partition is $1/120$. Thus, this scheme achieves the total NDT of $1/120\cdot \binom{K}{|\mathcal{R}_p|}=1/6$.

\subsection{The General Coded Parallel Scheme with Interference Alignment}\label{SecScheme}
%Based on the definitions above, 
We first present the CPC scheme with $r\in \{1,\cdots K\}$ and then extend the result to the general case $1\leq r\leq K$.

\emph{Map phase}:
Apply the same data placement as in \cite{CDC}, where the entire data is placed symmetrically. Divide the $N$ files evenly into $\binom{K}{r} $ disjoint groups, each group contains $\eta_{1}$ files. In the Map phase, each node $k \in [K]$ generates local IVs $\{v_{q,n} : q \in [Q], w_n \in \mathcal{M}_k\}$ with
$\left| \mathcal{M}_k \right| = \tbinom{K-1}{r-1} \eta_{1} = \frac{r N}{K}. $
%\end{IEEEeqnarray*} 

%Make each file equal in size and be placed $ r$ times repeatedly. In order to make the files symmetrically placed, $\mathcal{D}$ is equal to the set of all the subsets of ${[K]}$ containing $ r$ elements, i.e, any $ {\mathcal{R} \in \mathcal{D}}$, $|\mathcal{R}| = r$. Therefore, in the placement phase, the number of files $N$ is a multiple of $\tbinom{K}{r}$. Here, for simplicity, set $N = \tbinom{K}{r}$, which reduces multiset $\mathcal{D}$ into a set. 

%Before the encoding phase, each node k is placed $\tbinom{K-1}{r-1}$ files, and each file $ \omega_n \in \mathcal{M}_k$ can be mapped to $Q$ IVs. After the decoding phase, node $k$ first obtains all $\{ v_{k, n}: k \in \mathcal{R}_n \in \mathcal{D} \}$ and then obtains the objective output function calculation result $ \textit{u}_k $ through reduction.

\emph{Shuffle Phase}:
After the Map phase, each node is responsible for computing $ \eta_2$ output functions. Each node $k$ knows IVs $\{v_{q,n}:q\in [Q], n\in\mathcal{M}_k\}$ and requires $ \binom{K-1}{r}\eta_2$ IVs $\{v_{q,n}:q\in\mathcal{W}_k, n\notin\mathcal{M}_k\}$.
Given some integer $K_\textnormal{r} \in[K]$ which will be determined in \eqref{optKr}, divide $K$ nodes into $ \tbinom{K}{K_\textnormal{t} } $ partitions $\{\pi_p:p\in[ \tbinom{K}{K_\textnormal{t} }]\}$. Each partition contains a group of transmitters $\mathcal{T}_p$ and a group of receivers $\mathcal{R}_p$, i.e., $\pi_p=\{\mathcal{T}_p,\mathcal{R}_p\}, ~\forall p=1,\ldots, \tbinom{K}{K_\textnormal{t} }$,
where $\mathcal{T}_p\subseteq[K]$ with size $|\mathcal{T}_p|=K_\textnormal{t}$ and $\mathcal{R}_p=[K]\backslash \mathcal{T}_p$ with size $|\mathcal{R}_p|=K_\textnormal{r}=K-K_\textnormal{t}$.

%\begin{IEEEeqnarray}{rCl} \label{definevju}	
%\end{IEEEeqnarray}
%{\color{green}For all $K$ users, since all users are responsible for different reduce functions, indicating that each user requires different IVs, there are in total $K\binom{K-1}{r}\eta_1\eta_2$ IVs needed to be exchanged during the Shuffle phase.} For simplicity, we describe our scheme for the case $\eta_1=1$ and $\eta_2=1$. The scheme can be easily extended to the general case. 

%\text{LCP}(\tbinom{K}{K_\textnormal{r} },r) , where $\text{LCP}(\tbinom{K}{K_\textnormal{r} },r) $ denotes the least common multiple of $\tbinom{K}{K_\textnormal{r} }$ and $r$
 %In the example in Section \ref{}, the 6 users are divided into 15 partitions.

%For example, we have In the example in Section \ref{}, the 6 users are divided into 15 partitions.

\subsubsection{Message Generation}
Recall the definition of the IVs required by node $j$ and available at $r$ nodes in $\mathcal{U} \subseteq[K]\backslash\{j\}:|\mathcal{U}|=r$ as $v_{d_j,\mathcal{U} } \triangleq \{v_{q,n}:q\in\mathcal{W}_j, w_n \in \cap_{i\in\mathcal{U}} \mathcal{M}_i\}$.  Given some integer $s,t$ satisfying $s+t=r+1$, divide each IV $v_{d_j,\mathcal{U}}$ into $\tbinom{r}{t}\tbinom{K-r-1}{K_\textnormal{r}-s } $ disjoint segments $v^{(p,\mathcal{B})}_{d_j,\mathcal{U}\backslash\mathcal{B} }$ of equal size with $\mathcal{B}\subseteq \mathcal{T}_p\cap \mathcal{U}$ and all $p\in[\tbinom{K}{K_\textnormal{r}}]$ satisfying $|\mathcal{B}|=t$ and   $\{j\}\cup\mathcal{U}\backslash \mathcal{B} \subseteq \mathcal{R}_p$, i.e.,
\begin{IEEEeqnarray}{rCl} \label{definegen}
 \hspace{-15pt}v_{d_j,\mathcal{U}}\!=\! \big(v^{(p,\mathcal{B})}_{d_j,\mathcal{U}\backslash\mathcal{B}}\!:\!  \mathcal{B}\!\!\subseteq\!\! \mathcal{U}\!\cap\! \mathcal{T}_p, p \!\!\in\!\! [\tbinom{K}{K_\textnormal{r}}],  \{j\} \!\cup\! \mathcal{U} \backslash \mathcal{B} \!\subseteq\!  \mathcal{R}_p \big),
	%v_{q,\mathcal{R}_n}=\Big(v^{(p,k)}_{q,n}:p\in[N_\text{g} ],k\in[g-1]\Big),\\%v_{q,n}=\Big(v^{(p,k)}_{q,n}:p\in[(g-1)N_\text{g} ],k\in[r]\Big),
\end{IEEEeqnarray}
where {$v^{(p,\mathcal{B})}_{d_j,\mathcal{U}\backslash\mathcal{B}} $ is assigned to transmitter cooperation group $\mathcal{B}$ in partition $p$ and the size of each segment is $ {\eta_1\eta_2B}/\big({\tbinom{r}{t}\tbinom{K-r-1}{K_\textnormal{r}-s } } \big)$ bits.}
The reason for such division is as follows: In the Shuffle phase, each \emph{multicast group} consisting of a transmitter subset $\mathcal{B}\subseteq\mathcal{T}_p$ and receiver group $ \mathcal{D}\subseteq \mathcal{R}_p$ with size $|D|=s$ will appear $\tbinom{K-r-1}{K_\textnormal{r}-s } $ times among all partitions $\{\pi_p\}$. Besides, segments of each IV are known by $r$ users. These segments will be sent by $\tbinom{r}{t}$ transmitter cooperation groups. Thus, we need to split each IV into segments to avoid repetitive transmission of any parts of any IV.

For each partition $\pi_p=\{\mathcal{T}_p,\mathcal{R}_p\}$, each transmitter in $\mathcal{B}$ generates an encoded message intended to the destination nodes in $\mathcal{D}\subseteq\mathcal{R}_p$ with size $|\mathcal{D}|=s$ as follows:
\begin{IEEEeqnarray}{rCl}\label{Eqencoding}
	V^{(p)}_{\mathcal{D},\mathcal{B}} = \oplus_{j \in \mathcal{D}} v^{(p,\mathcal{B})}_{d_j, \mathcal{D}\backslash\{j\} },
\end{IEEEeqnarray}
where the size of $V^{(p)}_{\mathcal{D},\mathcal{B}} $ is $ {\eta_1\eta_2B}/\big({\tbinom{r}{t}\tbinom{K-r-1}{K_\textnormal{r}-s } } \big) $ bits. Compared with CDC, the sub-packaging level of CPC is $\tbinom{K-r-1}{K_\textnormal{r}-s }\tbinom{r}{t}/r$ times larger.
% which is reasonable when $K_\textnormal{r}$ is close to $r$ or $K$. 

%{\color{green}
%In Fig \ref{} of the example in Section \ref{}, node 5 and 6 generate messages $(V_{123,5},V_{124,5},V_{134,5},V_{234,5})$ and $(V_{123,5},V_{124,5},V_{134,5},V_{234,5})$, respectively. 
%}

If receiver $j\in\mathcal{D}$ successfully obtains $V^{(p)}_{\mathcal{D},\mathcal{B}} $, it can decode its desired segments $v^{(p,\mathcal{B})}_{d_j, \mathcal{D}\backslash\{j\} }$ by XORing $ V^{(p)}_{\mathcal{D},\mathcal{B}}$ with the local $s-1$ segments
$\{v^{(p,\mathcal{B})}_{d_m, \mathcal{D}\backslash\{m\} }, m\in \mathcal{D}\backslash\{j\}\}.$ {Thus, the communication load for each receiver in every partition $\pi_p$, $p=1,\ldots,\binom{K}{K_\textnormal{r}}$ is }
	\begin{align}\label{TotalBits_per}
		R_{p} = \frac{\eta_1\eta_2B}{\tbinom{K-r-1}{K_\textnormal{r}-s }\tbinom{r}{t}} \binom{K_\textnormal{r}-1}{s-1}\binom{K_\textnormal{t}}{t}\frac{1}{NQB} = \frac{1}{K_\textnormal{r}}\left({1-\frac{r}{K}}\right)\frac{1}{\tbinom{K}{K_\textnormal{r}}}.
	\end{align} %bits. 

\begin{remark}\label{MessageGroup}
	In each partition $\pi_p$, there are $\binom{K_\textnormal{t}}{t}$ transmitter cooperation groups, each of these groups $\mathcal{B}\in \mathcal{T}_p$ {needs to deliver} $\binom{K_\textnormal{r}}{s}$ independent encoded messages $\{V^{(p)}_{\mathcal{D},\mathcal{B}}: \mathcal{D}\subseteq\mathcal{R}_p\}$, each intended to a $s$-receiver group $\mathcal{D}$. Each receiver node $j\in\mathcal{R}_p$ wishes to decode $\binom{K_\textnormal{t}}{t}\binom{K_\textnormal{r}-1}{s-1}$ messages $\{V^{(p)}_{\mathcal{D},\mathcal{B}}: \mathcal{B}\subseteq \mathcal{T}_p, j \in \mathcal{D}\subseteq\mathcal{R}_p\}$ with the presence of $\binom{K_\textnormal{t}}{t}\binom{K_\textnormal{r}-1}{s}$ interference messages $\{V^{(p)}_{\mathcal{D},\mathcal{B}}: \mathcal{B}\subseteq\mathcal{T}_p, \mathcal{D}\subseteq\mathcal{R}_p\backslash\{j\}\}$.
\end{remark}

In the following part, we show that given partition $\pi_p$, how receiver nodes in $\mathcal{R}_p$ decode their desired messages $\{V^{(p)}_{\mathcal{D},\mathcal{B}}: \mathcal{B}\subseteq\mathcal{T}_p, \mathcal{D}\subseteq\mathcal{R}_p\}$ over the wireless interference network by interference alignment approach \cite{taoWireless,degrees}.

\subsubsection{Message transmission over wireless network}
First recall the definition of $\binom{K_\textnormal{t}}{t}\times \binom{K_\textnormal{r}}{s}$ cooperative X-multicast channel introduced in \cite[Definition 2]{taoWireless}.
\begin{definition}\label{DefXchannel}
	The channel satisfying the following features is called $\binom{K_\textnormal{t}}{t}\times \binom{K_\textnormal{r}}{s}$ cooperative X-multicast channel:
	\begin{itemize}
		\item there are $K_\textnormal{t}$ transmitters and $K_\textnormal{r}$ receivers;
		\item each set of $s$ ($s\leq K_\textnormal{r}$) receivers forms a receiver multicast group;
		\item each set of $t$ ($t\leq K_\textnormal{t}$) transmitters forms a transmitter cooperation group;
		\item each transmitter cooperation group has an independent message to send to each receiver multicast group.
	\end{itemize}
\end{definition}

From Remark \ref{MessageGroup} we know, the channel is turned into the $\binom{K_\textnormal{t}}{t}\times \binom{K_\textnormal{r}}{s}$ cooperative X-multicast channel in each partition $\pi_p$, as depicted in Fig. \ref{ex}. While each $t$ nodes group $\mathcal{B}\in \mathcal{T}_p$ has messages $\{V^{(p)}_{\mathcal{D},\mathcal{B}}: \mathcal{D}\subseteq\mathcal{R}_p\}$ intended for the $s$ nodes group $\mathcal{D} \in \mathcal{R}_{P}$, the nodes in $\mathcal{T}_p$ and $\mathcal{R}_p$ change in the different partitions. The same nodes group pair $(\mathcal{B},\mathcal{D})$ will appear multiple times among all the partitions. The aforementioned method of message generation ensures the transmitted messages are independent. An achievable per-receiver DoF of this channel is given below. {Note that in \cite{taoWireless} each file is cached by $r$ users and $t$ transmitters, and the size of the receiver multicast group is $r+1$.  While in our scheme,  each file is cached by $r$ nodes (playing roles of transmitters and receivers in different partitions) and the size of the receiver multicast group is $s=r+1-t$ since $t$ out of $r$ users operate as transmitters.}

\begin{lemma} \label{corollary1}
	(\cite[Lemma 1]{taoWireless}) An achievable per-receiver DoF of the $\binom{K_\textnormal{t}}{t}\times \binom{K_\textnormal{r}}{s}$ cooperative X-multicast channel is given by
	\begin{align}
		d_{s,t}=\left\{
		\begin{array}{ll}
			1, & s+t\ge K_\textnormal{r}+1, \\
			\frac{\binom{K_\textnormal{r}-1}{s-1}\binom{K_\textnormal{t}}{t}t}{\binom{K_\textnormal{r}-1}{s-1}\binom{K_\textnormal{t}}{t}t+1}, & s+t= K_\textnormal{r}, \\
			\max\left\{d'_{s,t},\frac{s+t}{K_\textnormal{r}}\right\}, & s+t\le K_\textnormal{r}-1,
		\end{array}
		\right.\label{lemma d}
	\end{align}
	where
	%\begin{align}
		%d'_{s,t}\triangleq\max\limits_{1\le t'\le t}\left\{\frac{\binom{K_\textnormal{r}-1}{s-1}\binom{K_\textnormal{t}}{t'}\binom{K_\textnormal{r}-s}{t'-1}t'}
		%{\binom{K_\textnormal{r}-1}{s-1}\binom{K_\textnormal{t}}{t'}\binom{K_\textnormal{r}-s}{t'-1}t'+\binom{K_\textnormal{r}-1}{s}\binom{K_\textnormal{r}-s-1}{t'-1}\binom{K_\textnormal{t}}{t'-1}}\right\}.
	%\end{align}
        \begin{align}
            d'_{s,t}\triangleq\max\limits_{1\le t'\le t}\left\{ \frac{1}{ 1 + \frac{ K_\textnormal{r}-s-t'+1 }{ s(K_\textnormal{t}-t'+1) } } \right\}.
        \end{align}
\end{lemma}

\begin{figure}[!t]
	\centering
	\includegraphics[width=0.8\linewidth]{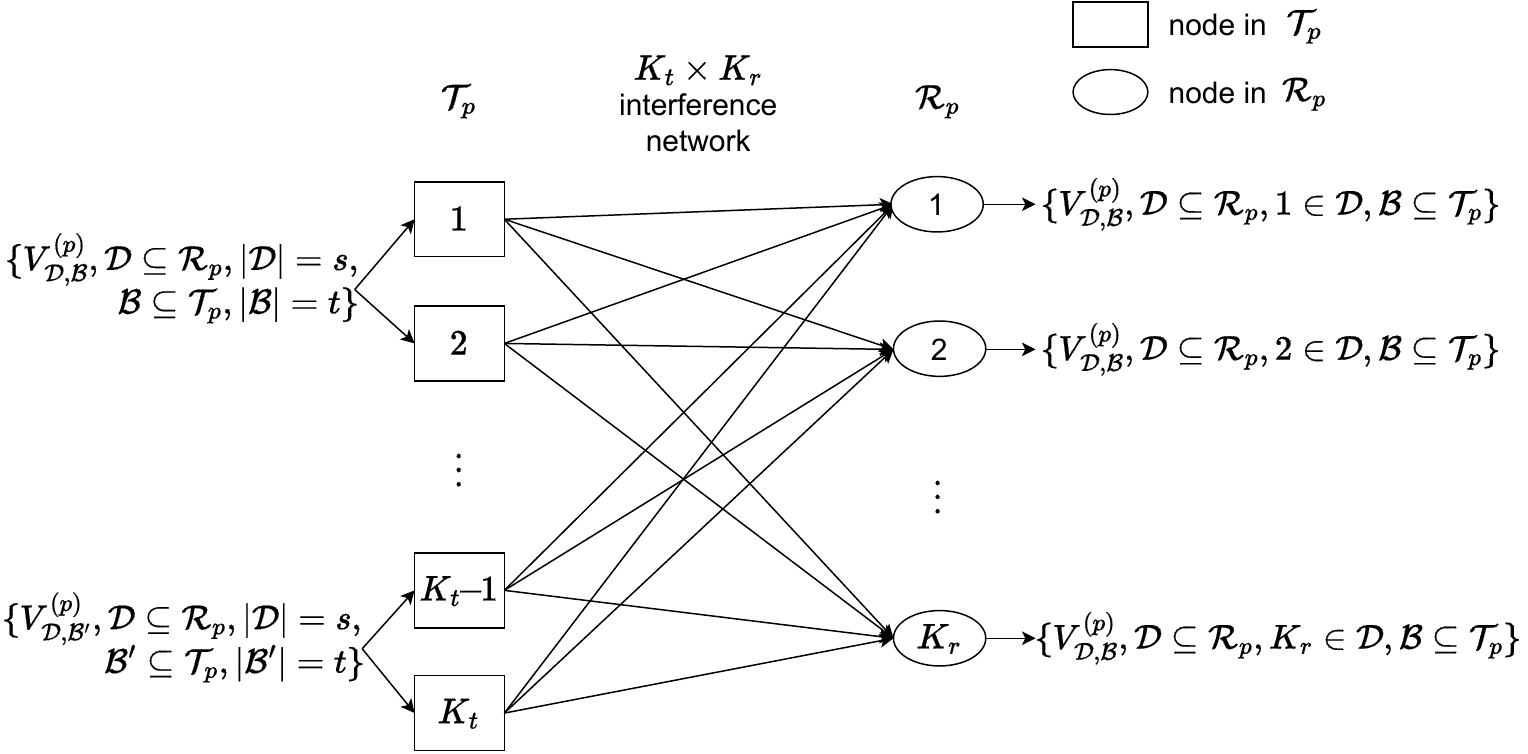}
	\caption{The cooperative X-multicast channel in a partition $\pi_p= \{\mathcal{T}_p,\mathcal{R}_{p}\} $. Each transmitter group $ \mathcal{B} \in \mathcal{T}_p $ sends message $V_{\mathcal{D},\mathcal{B}}^{(p)}$ to the receiver group $\mathcal{D}\in \mathcal{R}_p$.}%The left half of the figure shows that at certain slot, each broadcast group has only one node sending messages to other nodes in the group, which can be compared to $ \lfloor \frac{K}{r+1} \rfloor $-User interference channel in the right half of the figure.}
	\label{ex}
\end{figure}

References \cite{taoWireless,degrees} considered the $\binom{K_\textnormal{t}}{t}\times \binom{K_\textnormal{r}}{s}$ cooperative X-multicast channel and proposed interference neutralization and alignment approaches which enable all receivers to successfully decode their desired messages. The neutralization and alignment strategy used in our scheme is similar to that in \cite{taoWireless} leading to the per-receiver DoF of \eqref{lemma d}. Please refer to the detailed explanation in Appendix \ref{IAproof}.%our extended version \cite[Appendix A]{arXiv}.

From \eqref{TotalBits_per} and $\delta=R/d$ in \eqref{eqNDTLoad_per}, we obtain the NDT in a partition $\delta_p = \frac{R_p}{d_{s,t}}$, and total NDT is
\begin{align}\label{eqNDTCPC}
	\delta_{\text{CPC}}(r,K) &= \delta_p \cdot \binom{K}{K_\textnormal{r}} = \min_{s, K_\textnormal{r}}\! \left(1-\frac{r}{K}\right)\frac{1}{K_\textnormal{r}d_{s,t}}, \\
	\textrm{s.t.} \qquad & s \leq K_\textnormal{r}, t \leq K_\textnormal{t}, s+t=r+1, K_\textnormal{t} \leq K-s, \nonumber
\end{align} 
where $K_\textnormal{r}=K-K_\textnormal{t}$, which is equivalent to \eqref{d(r,k)}. %The algorithm of the general achievable scheme is described in Algorithm \ref{alg:achievable1}.

Now we follow the similar approach in \cite{CDC} to extend our results to general case $1 \!\!\leq \!\!r \!\!\leq\!\! K$. First, expand $ r \!\!=\!\! \alpha \lfloor r \rfloor \!+\! (1\!-\!\alpha) \lceil r \rceil$ for some $ 0 \!\!\leq\!\! \alpha \!\!\leq\!\! 1 $. Partition the $ K $ nodes into two different types of partitions: $(\mathcal{T}_p^{(1)},\mathcal{R}_p^{(1)})$: $|\mathcal{T}_p^{(1)}| \!=\! K_\textnormal{t}^{(1)}$, $|\mathcal{R}_p^{(1)}|\! =\! K_\textnormal{r}^{(1)} \!=\! K \!-\! K_\textnormal{t}^{(1)}$ and $(\mathcal{T}_p^{(2)}$, $\mathcal{R}_p^{(2)})$: $|\mathcal{T}_p^{(2)}| \!\!=\!\! K_\textnormal{t}^{(2)}, |\mathcal{R}_p^{(2)}| \!=\! K_\textnormal{r}^{(2)} \!=\! K \!-\! K_\textnormal{t}^{(2)}$, where $ K_\textnormal{t}^{(1)} \!=\! \mathop{\arg\min}_{K_\textnormal{t}^{(1)} \in [K- \lfloor r \rfloor]} \delta_{\text{CPC}}(\lfloor r \rfloor, K) $ and $ K_\textnormal{t}^{(2)} \!=\! \mathop{\arg\min}_{K_\textnormal{t}^{(2)} \in [K- \lceil r \rceil]} \delta_{\text{CPC}}(\lceil r \rceil, K) $. Next partition the set of $ N $ input files into two disjoint subsets $\mathcal{N}_1 $ and $ \mathcal{N}_2 $ of sizes $ |\mathcal{N}_1| = \alpha N $ and $ |\mathcal{N}_2| = (1-\alpha) N $.
Then we apply the CPC scheme described above to the files in $ \mathcal{N}_1 $ with computation load $ \lfloor r \rfloor $ and the {first type} of partition $ (\mathcal{T}_p^{(1)},\mathcal{R}_p^{(1)}) $, and to the files in $ \mathcal{N}_2 $ with computation load $ \lceil r \rceil $ and the second type of partition $(\mathcal{T}_p^{(2)},\mathcal{R}_p^{(2)})$, respectively.
This results in NDT of
\begin{IEEEeqnarray}{rCl} \label{anyr}
	\alpha {\delta}(\lfloor r \rfloor, K)+(1-\alpha){\delta}(\lceil r \rceil, K),
\end{IEEEeqnarray}
where $ {\delta}(r,K) $ is the NDT achieved by CPC in \eqref{eqNDTCPC} for integer-valued $ r \in [K] $.
From \eqref{anyr}, we see that for any $r$ $(1 \leq r \leq K) $, the NDT $ \alpha {\delta}(\lfloor r \rfloor, K)+(1-\alpha){\delta}(\lceil r \rceil, K) $ is achievable. Thus, for general $1\leq r\leq K$, the lower convex envelop of points $ \left\{(r,\delta_{\text{CPC}(r,K)}): \text{for} \ r \in [K] \right\}$ is achievable.

\subsubsection{Coding Complexity}
We now discuss the coding complexity of the proposed scheme. The IVs are encoded and decoded with XOR operations.  The number of XOR operations required for one user to generate one message $V^{(p)}_{\mathcal{D},\mathcal{B}} = \oplus_{j \in \mathcal{D}} v^{(p,\mathcal{B})}_{d_j, \mathcal{D}\backslash\{j\} }$ is $s-1$. Because each user has $\tbinom{K-K_r-1}{t-1}\tbinom{K_r}{s}$ messages to be sent in each partition and there is a total of $\tbinom{K}{K_r}$ partitions, the total number of XOR operations for each user for encoding is $\tbinom{K}{K_r}\cdot\tbinom{K-K_r-1}{t-1}\tbinom{K_r}{s}\cdot(s-1)$. Similarly, the number of XOR operations required for each user to decode the desired message $v^{(p,\mathcal{B})}_{d_j, \mathcal{D}\backslash\{j\} }$ from $V^{(p)}_{\mathcal{D},\mathcal{B}}$ is $s-1$. Because each user has $\tbinom{K-K_r}{t}\tbinom{K_r-1}{s-1}$ desired messages in each partition and there is a total of $\tbinom{K}{K_r}$ partitions, the total number of XOR operations for each user to decode is $\tbinom{K}{K_r}\cdot\tbinom{K-K_r}{t}\tbinom{K_r-1}{s-1}\cdot(s-1)$. Thus, the coding complexity for each user is: $\tbinom{K}{K_r}(s-1)\big(\tbinom{K-K_r-1}{t-1}\tbinom{K_r}{s} + \tbinom{K-K_r}{t}\tbinom{K_r-1}{s-1}\big)|V^{(p)}_{\mathcal{D},\mathcal{B}}| = \tbinom{K}{K_r}(s-1)\big(\tbinom{K-K_r-1}{t-1}\tbinom{K_r}{s} + \tbinom{K-K_r}{t}\tbinom{K_r-1}{s-1}\big) {\eta_1\eta_2B}/\big({\tbinom{r}{t}\tbinom{K-r-1}{K_\textnormal{r}-s } } \big)$.

\subsubsection{Stragglers Effect}%we can exploit the redundant file placement (i.e., each file is stored by $r$ nodes).Let $\mathcal{S}\subseteq\mathcal{T}_p$ denote the set of stragglers in the transmitter group with $|\mathcal{S}|\leq t-1$.
	The CPC scheme can be modified to combat up to $t-1$ stragglers (i.e., some nodes that complete the Map Phase slower than normal nodes). For each partition, if the stragglers are on the receiver side, they have no effect on the transmission. The stragglers can still receive the signals and decode the desired messages after completing their Map computations. Now we consider the stragglers that appear in transmitter groups. Let $\mathcal{S}\subseteq\mathcal{T}_p$ denote the set of stragglers in the transmitter group with $|\mathcal{S}|\leq t-1$.  Since each file is stored by $r$ nodes, any messages owned by the transmitter cooperation group $\mathcal{B}$ can be sent by the smaller transmitter cooperation groups $\mathcal{B} \backslash \mathcal{S}$. Thus, we divide the coding messages $\{V^{(p)}_{\mathcal{D},\mathcal{B}}\}_{\mathcal{D}\subseteq \mathcal{R}_{p},\mathcal{B}\subseteq \mathcal{T}_{p}}$ as follows and send them in   a time-division manner   	
    \begin{align*}
		\{V^{(p)}_{\mathcal{D},\mathcal{B}}\}_{\mathcal{D}\subseteq \mathcal{R}_{p},\mathcal{B}\subseteq \mathcal{T}_{p}}
		= \{ V^{(p)}_{\mathcal{D},\mathcal{B}_0},...,V^{(p)}_{\mathcal{D},\mathcal{B}_{|S|}} \}_{\mathcal{D}\subseteq \mathcal{R}_{p}, \mathcal{B}_0, ..., \mathcal{B}_{|\mathcal{S}|}\subseteq \mathcal{T}_{p}},
    \end{align*}
	where $\mathcal{B}_0\cap\mathcal{S}=\emptyset$, $\mathcal{B}_1\cap\mathcal{S} = 1,\cdots, \mathcal{B}_{|\mathcal{S}|}\cap\mathcal{S} = |\mathcal{S}|$, $|\mathcal{B}_0|=|\mathcal{B}_1|=\cdots=|\mathcal{B}_{|\mathcal{S}|}| = t$. Here $\mathcal{B}_i\subseteq\mathcal{T}_p$ represents the transmitter cooperation groups with exactly $i\in\{0,1,...,t-1\}$ stragglers.
	For messages $\{V^{(p)}_{\mathcal{D},\mathcal{B}_{i}}\}_{\mathcal{D}\subseteq \mathcal{R}_{p}, \mathcal{B}_i\subseteq \mathcal{T}_{p}}$, we let the nodes in $\mathcal{B}_{i} \backslash \mathcal{S}$ apply the same transmission strategy, but over the $\tbinom{K_t-|S|}{t-i}\times\tbinom{K_r}{s}$ cooperative X-multicast channel. After receiving all messages sent by the transmitter groups $\mathcal{B} \backslash \mathcal{S}$, each receiver in the receiver group can recover $\{V^{(p)}_{\mathcal{D},\mathcal{B}}\}_{\mathcal{D}\subseteq \mathcal{R}_{p},\mathcal{B}\subseteq \mathcal{T}_{p}} $ and decode the desired information in a way the same as the original CPC. %The corresponding NDT can be derived with the similar analysis and is omitted. %will be increased due to the decrease in multiplexing gain. 

{From \eqref{TotalBits_per}, \eqref{lemma d}, \eqref{eqNDTCPC}, and \eqref{anyr}, we achieve the NDT in the following theorem. }

\begin{theorem} \label{theorem1}
	For the $K$-user wireless MapReduce system with computation load $r\in[K]$, the following NDT is achievable:
	\begin{align}\label{d(r,k)}
		\delta(r,K) = \min_{K_\textnormal{r} \in [1:K], t \in [1:r]} \delta_{\textnormal{CPC}}(r,t,K,K_\textnormal{r}),
	\end{align}
	where
	\begin{align}\label{eqDelay}
		{\delta}_\textnormal{CPC} & (r,t,K,K_\textnormal{r}) =
		\begin{cases}
			\frac{1}{K_\textnormal{r}} (1 - \frac{r}{K}) , & r \geq K_r, \\
			\frac{1}{K_\textnormal{r}} (1 - \frac{r}{K}) ( 1+ \frac{1}{\binom{r}{t} \binom{K - K_\textnormal{r}}{t} t} ), & r = K_r - 1, \\
			\frac{1}{K_\textnormal{r}} (1 - \frac{r}{K}) \min\{ \frac{1}{\tau_{r,t}} , \frac{K_\textnormal{r}}{r} \} , & r \leq K_r - 2,
		\end{cases}
		%	\\\text{s.t.} &
		%	\begin{cases}
		%		s \leq K_r \\
		%		t \leq K_t \\
		%		s + t = r + 1 \\
		%		K_t \leq K - s 
		%	\end{cases}
		%\min_{K_\textnormal{t}\in [K\!-\!r]} \frac{1}{r}\cdot \big({1}- \frac{r}{K}\big)\cdot \frac{ K_\textnormal{t} r + K_\textnormal{r}-r }{K_\textnormal{t}K_\textnormal{r}},~
	\end{align}
	with $t \leq K-K_\textnormal{r}$, and $\tau_{r,t}$ is given by
            \begin{align}\label{tau_rt}
                & \tau_{r,t} \triangleq \max\limits_{1\le j\le t}\left\{ \frac{1}{ 1 + \frac{ K_\textnormal{r}+t-r-j }{ (r+1-t)(K-K_\textnormal{r}-j+1) } } \right\}.
            \end{align}
	For general $1\leq r\leq K$, the lower convex envelope of the above points is achievable.
\end{theorem}
\begin{IEEEproof}
	{By using the scheme present above, and  substituting \eqref{TotalBits_per}, \eqref{lemma d} into  \eqref{eqNDTCPC}, then the NDT of  \eqref{eqNDTCPC} can be written as \eqref{d(r,k)}.}
\end{IEEEproof}

Recall \eqref{eqNDTCPC}, the $\delta_{\text{CPC}}(r,t,K,K_\textnormal{r})$ can be written as
\begin{align}\label{eqDelay2}
	\delta_{\text{CPC}}(r,t,K,K_\textnormal{r}) = \frac{1}{s} (1 - \frac{r}{K}) \frac{s}{K_\textnormal{r} d_{s,t}}, 
\end{align}
where   $d_{s,t}$ is the per-receiver DoF given in Lemma \ref{corollary1}. Like the factor $\frac{1}{r}$ appears in $\delta_\text{CDC}(r,K)$ in \eqref{TDCDC}, the first factor here $\frac{1}{s}$ is referred to as the \emph{coded multicasting gain}, because each node will generate and multicast coded symbols, each of which contains useful information to $s$ receivers.

The second factor in ${\delta}_\textnormal{CPC}(r,t,K)$ is $\big({1}- \frac{r}{K}\big)$, appearing also in $\delta_\text{uncoded}(r,K)$ in \eqref{uncoded}. We call it the \emph{local computation gain} as each node will map $r/K$ fraction of total $N$ files. Observe that CDC and OSL also embrace this local computation gain.

Finally, the third factor in ${\delta}_\textnormal{CPC}(r,K)$ is $\frac{s}{K_\textnormal{r} d_{s,t}}$ which we call the \emph{cooperative transmission gain}. Actually the reciprocal of this gain is equal to the sum DoF of the $\binom{K_\textnormal{t}}{t} \times \binom{K_\textnormal{r} }{ s }$, coming from the fact that $K_\textnormal{t}$ nodes will cooperatively and simultaneously transmit coded symbols to $ K_\textnormal{r} $ receivers, with each coded symbols intended by $s$ receivers.

{The problem in Theorem \ref{theorem1} is an integer programming problem and we can derive the minimum NDT as presented in the following theorem.}
\begin{theorem} \label{minmul_throrem2}
	For the $K$-user wireless MapReduce system with computation load $r\in [K]$, by solving integer programming problem \eqref{d(r,k)}, the minimum NDT is given by
	\begin{align}\label{eqMin}
		\delta_{min} = \min\left\{ \textnormal{NDT}_1(r,K), \textnormal{NDT}_2(r,K) \right\},
	\end{align}
	where $\textnormal{NDT}_1(r,K) , \textnormal{NDT}_2(r,K)$ are given by
	\begin{align}
		\textnormal{NDT}_1(r,K) = & \frac{1}{r+1} (1 - \frac{r}{K}) ( 1+ \frac{1}{\binom{r}{t^*} \binom{K-r-1}{t^{*}} t^{*}} ), \label{ndt1} \\
		\textnormal{NDT}_2(r,K) = & \frac{1}{r} (1 - \frac{r}{K})\frac{(K-K_\textnormal{r}^*) r +K_\textnormal{r}^*- r}{(K-K_\textnormal{r}^*) K_\textnormal{r}^*}, \label{ndt2}
	\end{align}
	with $t^{*} = \lfloor 1+\frac{-r^2+rK-r}{K} \rfloor$ and $K_\textnormal{r}^{*}$ is given by
	\begin{align}\label{optKr}
		K_\textnormal{r}^{*} = \begin{cases}
			\lfloor \frac{K+1}{2} \rfloor , & r=1, \\
			\lfloor \frac{rK-r+\frac{r-1}{2} - \sqrt{r(K-1)(K-r) +\frac{(r-1)^2}{4}}}{r-1} \rfloor , & r > 1.
		\end{cases}
	\end{align}
\end{theorem}
\begin{IEEEproof}
 Please see the proof in Appendix \ref{proof_th2}.
\end{IEEEproof}

\begin{remark} \label{gainchange} {From \eqref{eqMin}, \eqref{ndt1} and \eqref{ndt2}, we observe that the minimum NDT decreases with the $r$. However, the gains mentioned above exhibit different behaviors with respect to $K$. Specifically, if $t$ is fixed to $1$ which is proved in the next theorem to be the optimal value in most cases, the \emph{coded multicasting gain}, \emph{local computation gain} and \emph{cooperative transmission gain} equal to $\frac{1}{r}$, $1-\frac{r}{K}$ and $\frac{(K-K_\textnormal{r}) r +K_\textnormal{r}- r}{(K-K_\textnormal{r}) K_\textnormal{r}}$, respectively.} When $K$ increases, the value of local computation gain will monotonically increase (bigger value means larger latency and worse performance), but the value of  $\frac{(K-K_\textnormal{r}^*) r +K_\textnormal{r}^*- r}{(K-K_\textnormal{r}^*) K_\textnormal{r}^*}$ will decrease when $ K \gg r $, e.g., $ r =2$, $ K>2+\sqrt{4+2\sqrt{2}} $. Thus, there exists a trade-off between local computation gain and cooperative transmission gain.
\end{remark}

{The following theorem shows that the optimal value of $t$ equals 1 in most cases.}
\begin{theorem} \label{teq1_theorem3}
	For the wireless MapReduce system with computation load $1 < r \leq K$, if $K \leq 5$ or $K \geq \max\{r+4+\frac{4}{r-1}, \frac{r+4+\sqrt{r^2+16r}}{2}\}$, $r\in\mathbb{N}^+$, the minimum NDT for proposed scheme $\delta_{min}$ is attained when $t = 1$ and
	\begin{align}\label{eqMin_rk}
		\delta_{min} = \frac{1}{r} (1 - \frac{r}{K}) \frac{(K-K_\textnormal{r}^{*}) r +K_\textnormal{r}^{*}- r}{(K-K_\textnormal{r}^{*}) K_\textnormal{r}^{*}},
	\end{align}
	with the $K_\textnormal{r}^{*}$ is given by \eqref{optKr}.
\end{theorem}

\begin{IEEEproof}
	Please see the proof in Appendix \ref{proof_th3}.
\end{IEEEproof}

\begin{remark}
	%t>1
{When $r$ is close to $K$, $\textnormal{NDT}_1$ will be smaller than $\textnormal{NDT}_2$, i.e.,  the choice $t>1$ could lead to smaller NDT than the choice $t=1$. For example, when $K=8$, $r=5$, $\textnormal{NDT}_1(5,8) = 0.065625$ with $t^{*}=2$ while $\textnormal{NDT}_2(5,8) = 0.06875$ with   $K_\textnormal{r}^{*} = 6$ and $t=1$,  satisfying $\textnormal{NDT}_2  \geq \textnormal{NDT}_1$.}% than $\textnormal{NDT}_1(5,8)$.
\end{remark}

\subsection{Comparison with Existing Works} \label{CompareExisting}% NDT in Theorem \ref{theorem1} with that of 
{The following theorem shows that our scheme outperforms the state-of-the-art schemes including the CDC scheme, half-duplex OSL scheme, and half-duplex BW scheme, and tends to zero NDT when $K$ is sufficiently large. 
\begin{theorem} \label{RemCompa}For the wireless MapReduce system with arbitrary computation load $r$ and $K\in\mathbb{N}^+$, let $\bar{\delta}_{\text{CPC}}(r,K) = \delta_{\text{CPC}}(r,1,K,K_\textnormal{r})$,
	we have
%	\begin{IEEEeqnarray}{rCl}\label{CPCCDC}
%		&\bar{\delta}_{\text{CPC}}(r,K) \leq {\delta}_\text{CDC}(r,K)\leq \delta^{\text{HD}}_\text{OSL}(r,K), %{\color{blue}\leq \delta^\text{half-duplex}_\text{BW}(r,K).}
%        %& \bar{\delta}_{\text{CPC}}(r,K) \leq \delta^{\text{HD}}_\text{BW}(r,K).
%	\end{IEEEeqnarray}
%        \begin{IEEEeqnarray}{rCl}\label{CPCBW}
%            & \bar{\delta}_{\text{CPC}}(r,K) \leq \delta^{\text{HD}}_\text{BW}(r,K),
%        \end{IEEEeqnarray}
        \begin{IEEEeqnarray}{rCl}
        	&\bar{\delta}_{\text{CPC}}(r,K) \leq {\delta}_\text{CDC}(r,K)\leq \delta^{\text{HD}}_\text{OSL}(r,K),&\label{CPCCDC}\\ %{\color{blue}\leq \delta^\text{half-duplex}_\text{BW}(r,K).}
        	%& \bar{\delta}_{\text{CPC}}(r,K) \leq \delta^{\text{HD}}_\text{BW}(r,K).
        	& \bar{\delta}_{\text{CPC}}(r,K) \leq \delta^{\text{HD}}_\text{BW}(r,K).&\label{CPCBW}
        \end{IEEEeqnarray}
	Moreover, if $r\ll K$, then
	\begin{IEEEeqnarray}{rCL}\label{eqCPCinft}%= \frac{1}{r},\quad &
	&&\lim_{K\to\infty}\delta^{\text{HD}}_\text{BW}(r,K) =\lim_{K\to\infty} \bar{{\delta}}_{\text{CPC}}(r,K) = 0, \nonumber\\
		&&\lim_{K\to\infty} \delta_\text{CDC}(r,K) =  \lim_{K\to\infty}\delta^{\text{HD}}_\text{OSL}(r,K) = \frac{1}{r}. %\nonumber\\
       % &&\lim_{K\to\infty}\delta^{\text{FD}}_\text{OSL}(r,K) = \frac{1}{2r},        
		%&&  {\color{blue}\lim_{K\to\infty} \delta^\text{half-duplex}_\text{BW}(r,K)=1,} \ \ \lim_{K\to\infty} \bar{{\delta}}_{\text{CPC}}(r,K) = 0.
	\end{IEEEeqnarray}
\end{theorem}
\begin{IEEEproof}
	Please see the proof in  Appendix \ref{prooftheorem3}.
\end{IEEEproof}
}

The result $ \lim_{K\to\infty} \bar{\delta}_\text{CPC} = 0$ comes from the fact that if $r\ll K $ and $ K \to \infty $, then $ K_\textnormal{t},K_\textnormal{r}$ can be infinite, {the cooperative transmission gain $\frac{(K-K_\textnormal{r}) r +K_\textnormal{r}- r}{(K-K_\textnormal{r}) K_\textnormal{r}}\to 0$, the multiplication of coded multicasting gain and local computation gain $\frac{1}{r}(1-\frac{r}{K})\to \frac{1}{r}$. Thus, the NDT tends to $0$ as $K\to \infty$. Moreover, in Appendix \ref{prooftheorem3}, we showed our half-duplex scheme can even outperform the full-duplex OSL,  i.e., 
	%\begin{IEEEeqnarray}{rCl}\label{CPCOneshot}
	$	\bar{\delta}_{\text{CPC}}(r,K) \leq {\delta}^{\text{FD}}_\text{OSL}(r,K)$,  if $K \geq 2(r+1+\sqrt{r^2+1})$.
		%~\text{if $1\leq r\leq \frac{K^*_tK^*_r\!-\!2K^*_r}{2K^*_t-2}$ and $K\geq 4$}.
%	\end{IEEEeqnarray} } 

The CDC scheme only benefits from \emph{coded multicasting gain} since it allows one node to send one coded message through a shared link. 
The CPC scheme and the full-duplex OSL scheme allow multiple nodes to transmit messages simultaneously through a wireless channel such that they benefit from DoF enhancement. With the side information cancellation and zero-forcing, the full-duplex OSL scheme achieves the sum DoF of $\min\{K,2r\}$, however the full-duplex OSL scheme only obtains a gain $\frac{1}{2r}$ coming from DoF enhancement when $r < \frac{K}{2}$. 
The NDT of the half-duplex OSL scheme and the half-duplex BW scheme is larger than the NDT of our CPC scheme, furthermore, the half-duplex OSL scheme is no better than the original CDC scheme, indicating that our scheme can better exploit the distributed computing nodes in the half-duplex system.
\begin{remark}
	Since  $\delta_{\text{CPC}}(r, K)$ is decreasing with the computation load $r$ (increasing $r$ means smaller communication load and larger cooperative transmission gain), our  CPC scheme achieves zero NDT if $K\to\infty $, not necessary $r\ll K$.   We focus on $r\ll K$ since the computation load $r$ should be relatively small to reduce code generation complexity in practical implementations \cite{CDC}. 
\end{remark}

\section{A Lower Bound of NDT for Half-duplex Wireless MapReduce-type Systems}\label{sec:converse}
{
In this section, we present a lower bound of NDT for the half-duplex system. Let
\begin{IEEEeqnarray}{rCl} \label{eq:ndt_lower_bound1}
	\delta_{\textnormal{Lb1}}( r)  \!\triangleq\! \left\{
	\begin{aligned}
		&\frac{1}{ K}  \left(2 - \frac{2}{ K} \right)  &&\text{if }  r \!=\!1,\\
		&  \frac{1}{ K}  \Big( 1 - \frac{ r}{ K}  + \!\!\!\!\max_{t \in[\lfloor \frac{K}{2} \rfloor] }\!\!\!\textnormal{lowc}( C_t(  r) )\Big)&& \text{if }  r \!\in\! (1,\lceil \frac{ K}{2} \rceil) ,\\
		&\frac{1}{ K}  \left( 1 - \frac{ r}{ K} \right)  &&\text{if }   r \!\in\! [\lceil\frac{K}{2}\rceil, K],
	\end{aligned}
	\right.\nonumber\\
\end{IEEEeqnarray}
where  
\begin{IEEEeqnarray}{rCl}\label{eq:func_C}
	C_t(i) \triangleq \left\{ \begin{aligned}
 &\frac{\binom{ K-i}{t-i}\cdot ( K-t)}{\binom{ K}{t}\cdot t}, &\text{if } i &\in [t] \\
 &0, &\text{if } i &\in [K]\backslash[t] 
 \end{aligned}
 \right.
\end{IEEEeqnarray}
and $\text{lowc}(f(t))$ denotes the lower convex envelope of points $(t, f(t))$. Also, let
\begin{IEEEeqnarray}{rCl} \label{eq:ndt_lower_bound2}
	\delta_{\textnormal{Lb2}}( r)  \triangleq \frac{1}{K-1}\left( 1 - \frac{r}{K}\right).
\end{IEEEeqnarray}
%To find a lower bound for $1/d$, we first proof the following lemma. 
\begin{theorem}\label{thm:converse}
    For a given computation load $r$, the $K$-user wireless MapReduce system has the following lower bound for the minimum NDT:
    \begin{IEEEeqnarray}{rCl}\label{eq:thm_converse}
        \delta^*(r) \geq \max\left\{ \delta_{\textnormal{Lb1}}( r), \delta_{\textnormal{Lb2}}( r) \right\}.
    \end{IEEEeqnarray}
\end{theorem}
\begin{IEEEproof}
We first prove the   bound  $ \delta^*(r) \geq \delta_{\textnormal{Lb2}}( r)$. %As the average communication load $R = \left(1 - \frac{r}{K}\right)$ and the maximal number of receiver is $K-1$, which lead to the maximal DoF $d_{max} = K-1$. We can deduce the bound by $R / d_{max}$.
 From \cite{CDC}, we know that the minimum per-user communication load is $R = \frac{1}{K}\left(1 - \frac{r}{K}\right)$. As the maximal number of receiver is $K-1$, which leads to the maximal per-user DoF $d_{max} = \frac{K-1}{K}$, we can deduce the bound by $R / d_{max}$.

Then, we prove the bound   $ \delta^*(r) \geq \delta_{\textnormal{Lb1}}( r)$. Given any file assignment (map phase), we have the following lemma. 
%Define We start by proving the following lemma.  As there exist only a finite number of file assignment scheme for a given number of input file $N$, we can thus first consider a fixed
\begin{lemma}\label{lma:dof_upper_bound}
	%In a  wireless distributed computing system with K nodes, 
	Consider two disjoint sets $\cT$ and $\cR$ of same size 
	\begin{equation}
		|\cT| = |\cR|.
	\end{equation}
Let $\mathcal{M}_t \subseteq [  N]$ be the set of files known only to nodes $\cT$ but not to any other node, $\mathcal{M}_r \subseteq [  N]$ be the set of files known only to nodes $\cR$ but not to any other node,  $\mathcal{M}_j$ for $j \in [ K]$ be the set of files known to the node $j$, $\mathcal{M}$ be the set of files known only to maximal $|\cT|$ nodes in the set $[ K] \backslash (\cT \cup \cR)$. Then, partition the set of all IVs $\mathcal{A}$ into the following disjoint subsets:
\begin{IEEEeqnarray}{rCl}
	\mathcal{W}_{r} & \triangleq & \{a_{j,m}\}_{\substack{j \in \mathcal{R}\\m \in [  N]\backslash \mathcal{M}_j }}, \\
	\mathcal{W}_{t1} & \triangleq  & \{a_{j,m}\}_{\substack{j \in [ K]\backslash (\cR \cup \cT) \\ m \in \mathcal{M}_{t}\backslash\mathcal{M}_j}},\\
	\mathcal{W}_{t2} & \triangleq  & \{a_{j,m}\}_{\substack{j \in \cT \\ m \in \mathcal{M}_r\backslash\mathcal{M}_j}}.%\\
	%	\mathcal{W}_c & \triangleq & \mathcal{A} \backslash (	\mathcal{W}_r  \cup 	\mathcal{W}_t).
\end{IEEEeqnarray}
For any sequence of distributed computing systems:
\begin{equation}\label{eqn:thm_1}
	\bar{d} \triangleq \varlimsup_{P \to \infty}  \varlimsup_{  T \to \infty}\frac{ {\eta_2}  B  }{   T \log   P} \leq\frac{ |\mathcal{T}|}{ |\mathcal{W}_{r}| +  |\mathcal{W}_{t1}| + |\mathcal{W}_{t2}|}.
\end{equation}
\end{lemma}
Here $\mathcal{W}_{r}$ denotes the set of all IVs deduced from all files and intended to nodes in $\cR$. $\mathcal{W}_{t1}$ denotes the set of IVs deduced from files in $\cT$ and intended for nodes not in $\cT$ or in $\cR$. And $\mathcal{W}_{t2}$ denotes the set of all IVs deduced from files assigned to $\cR$ and intended to nodes in  $\cT$.
\begin{IEEEproof}
Lemma \ref{lma:dof_upper_bound} can be proved by a multiple-access channel argument and the half-duplex property of the system. The detailed proof is given in Appendix \ref{sec:dof_upper_bound_proof}. 
\end{IEEEproof}

To implement Lemma \ref{lma:dof_upper_bound}, we denote $\mathcal{B}_{\mathcal{S}}^j$ the set of IVs that are computed exclusively at nodes in set $\mathcal{S} \subseteq [K]$ and intended for reduce function $j$.  Define $b_{\mathcal{S}} =|\mathcal{B}_{\mathcal{S}}^j|$, which does not depend on the index of the reduce function  $j \in [ K]\backslash\mathcal{S}$. 

Choose two disjoint subsets $\cT$ and $\cR$ of same size $|\cT|=|\cR|$. By  Lemma \ref{lma:dof_upper_bound}, and rewriting the sets $\mathcal{W}_{t1}$, $\mathcal{W}_{t2}$ and $\mathcal{W}_r$ in the lemma in terms of the sets $\{\mathcal{B}_{\mathcal{S}}^j\}$,  we obtain: 
\begin{IEEEeqnarray}{rCl}
	\frac{|\cT|}{\bar{d}} &\geq&\!\! \sum_{\mathcal{S}  \subseteq [K]} \; \sum_{j\in \cR \backslash \mathcal{S}} |\mathcal{B}_{\mathcal{S}}^j| 
    +\!\! \sum _{\mathcal{G} \subseteq \cT} \; \sum_{j \in  [K]\backslash (\cR \cup \cT)} \!\!\!\!\!\!|\mathcal{B}_{\mathcal{G}}^j| +\!\! \sum _{\mathcal{G} \subseteq \cR} \; \sum_{j \in  \cT } \!\! |\mathcal{B}_{\mathcal{G}}^j|\IEEEeqnarraynumspace \nonumber\\
	&=& \sum_{\mathcal{S}  \subseteq [K]} |\cR\backslash \mathcal{S}| \cdot b_{\mathcal{S}} +\! \sum _{\mathcal{G} \subseteq \cT} (K - 2|\cT|)   b_{\mathcal{G}} +\! \sum _{\mathcal{G} \subseteq\cR} |\cT|   b_{\mathcal{G}}.    \nonumber%\eqref{eq:converse_constrain} \label{eq:converse_constrain}
\end{IEEEeqnarray}
% \eqref{eq:converse_constrain} 
Summing up the above equalityover all  sets  $\cT$ and $\cR$  of constant size $|\cT|=|\cR| = t\leq  K/2$, we obtain:
\begin{IEEEeqnarray}{rCl}
	\lefteqn{\binom{ K}{t} \binom{K-t}{t}\cdot \frac{t}{\bar{d}} } \quad \nonumber \\ 
	&\geq& \sum_{\cT \in[[ K]]^t}  \;\; \sum_{\cR \in [ [ K]\backslash \cT]^t}  \;\; \sum_{\cS \subseteq [K]} |\cR \backslash \cS| \cdot b_{\cS} \nonumber \\
	&&+ \sum_{\cT \in[[ K]]^t} \;\;  \sum_{\cR \in [ [ K]\backslash \cT]^t}  \;\; \sum_{\cG \subseteq \cT} (K - 2t) \cdot b_{\mathcal{G}}  \nonumber \\
	&&+	\sum_{\cT \in[[ K]]^t} \;\;  \sum_{\cR \in [ [ K]\backslash \cT]^t}  \;\; \sum_{\cG \subseteq \cR } t \cdot b_{\mathcal{G}} 
	\IEEEeqnarraynumspace \nonumber \\
	& \geq& \sum_{\cS \subseteq [K]} \binom{ K}{t} \binom{K-t}{t}\cdot \frac{t}{K}(K-|\cS|) \cdot b_{\cS} \nonumber \\
	&&+ \sum_{\substack{\cG \subseteq [K],|\cG|\leq t}} \binom{K-|\mathcal{G}|}{t - |\mathcal{G}|}\binom{K-t}{t}\cdot (K - t) \cdot b_{\mathcal{G}},  \label{eq:converse_sum_constrain}
\end{IEEEeqnarray}
where we define $\binom{a}{0}=1$ for any positive integer $a$. Then applying the following relations
\begin{IEEEeqnarray}{rCl}
	\sum_{\mathcal{S} \subseteq [ K]} b_{\mathcal{S}} =   N, \quad \sum_{\mathcal{S} \subseteq [ K]} |\mathcal{S}| \cdot b_{\mathcal{S}} \leq  r \cdot   N, 
\end{IEEEeqnarray}
 dividing both sides of \eqref{eq:converse_sum_constrain} by $\binom{ K}{t}  \binom{ K -t}{t} \cdot t$, and defining $b_i \triangleq \sum_{\substack{\mathcal{S}\in[ [ K]]^i}} b_{\mathcal{S}}$,
%\begin{IEEEeqnarray}{rCl}
%	b_i \triangleq \sum_{\substack{\mathcal{S}\in[ [ K]]^i}} b_{\mathcal{S}},
%\end{IEEEeqnarray}
for any $t\in [\lfloor K/2\rfloor]$, we obtain:
\begin{IEEEeqnarray}{rCl}\label{eq:min}
	\frac{1}{\bar{d}}& \geq &   N - \frac{r \cdot   N}{ K} +  \min_{\substack{b_1,\ldots, b_{ K} \in \mathbb{Z}^+\colon \\ \sum_{i=1}^{ K} b_i=   N \\ \sum_{i=1}^{ K} i b_i \leq  r   N}}
	\sum_{i=1}^{ K} C_t(i) \cdot b_i, \quad  t \in [ \lfloor K/2 \rfloor], \IEEEeqnarraynumspace
\end{IEEEeqnarray}
where the function $C_t(i)$ is defined in \eqref{eq:func_C}

For any $t\in  [ \lfloor K/2 \rfloor]$, the sequence of coefficients $C_{t}(1), C_{t}(2), \ldots, C_{t}( K) $ is convex and non-increasing, which can be proved by the recurrence relation of binomial coefficient. Based on this convexity, it can be shown (see \cite[Apendix E]{bi2022bounds}) that for any $ r < t+1$ there exists a solution to the minimization problem in  \eqref{eq:min} putting only positive masses on $b_{\lfloor  r\rfloor}^*$ and $b_{\lceil  r\rceil}^*$ in the unique  way satisfying 
\begin{IEEEeqnarray}{rCl}
	b_{\lfloor  r\rfloor}^* + b_{\lceil  r\rceil}^*  =    N, \quad \lfloor  r\rfloor b_{\lfloor  r\rfloor}^* +  \lceil  r\rceil  b_{\lceil  r\rceil}^* =   r   N .
\end{IEEEeqnarray}
For $r \geq t+1$, the optimal solution is obtained by choosing $b^*_{\lfloor r \rfloor}$.

For $r = 1$, we choose $t = 1$, and for $r \in (1, \lceil \frac{K}{2} \rceil)$, we choose the $t$ that maximize the bound. Then, we obtain the lower bound for NDT in the theorem.

\end{IEEEproof}
}
% To simplify the bound in \eqref{eq:min}, we fix the value of $t$ by choosing its optimum values. For $ r \geq 2$, we choose $t= \lfloor  K/2 \rfloor$. For $ r \geq 2$, for $ r=1$ we choose  $t=1$, and for $ r \in (1,2)$ we  maximize over the value of $t$. Finally, we obtain the following lower bound for NDT.
{
\begin{corollary}
    For the considered $K$-user wireless MapReduce system with computation load $r\in[K]$, the multiplicative gap between the achievable NDT and the converse bound is less than $3$, i.e., $\frac{\delta(r, K)}{\delta^*(r,K)}<3$.
\end{corollary}
\begin{IEEEproof}
    The proof is based on the following observation. 
    When $r=1$, $\delta(1, K)\leq NDT_{2}(1,K) = \frac{1}{K}(1-\frac{1}{K})(4-\frac{1}{K})< 2\cdot \frac{1}{K}(2-\frac{2}{K}) = 2\cdot\delta^*(1,K)$.
    
    When $r\geq 2$ and ignoring the integer constraints, we have the following relations
    \begin{IEEEeqnarray}{rCl}
        \delta(r, K) &\leq& NDT_{2}(r,K) \nonumber\\
        &\leq& \frac{1}{r}(1-\frac{r}{K})\frac{ (K-\frac{K}{2})r+\frac{K}{2}-r }{ (K-\frac{K}{2})\frac{K}{2} } \nonumber\\
        &=& \frac{1}{K}(1-\frac{r}{K})\big[2+2(\frac{1}{r}-\frac{2}{K})\big],
    \end{IEEEeqnarray}
    and
    %\begin{IEEEeqnarray}{rCl}
        $\delta^*(r,K) \geq \frac{1}{K}(1-\frac{r}{K})$.
    %\end{IEEEeqnarray}
    Then \begin{IEEEeqnarray}{rCl}
        \frac{\delta(r, K)}{\delta^*(r,K)} \leq 2+2(\frac{1}{r}-\frac{2}{K}) \leq 3-\frac{4}{K} < 3.
    \end{IEEEeqnarray}
    Therefore, $ {\delta(r, K)}/{\delta^*(r,K)} < 3$.
\end{IEEEproof}

}

\section{Numerical Results} \label{numerical results}
{In this section, we numerically compare the NDTs of the Uncoded TDMA in \eqref{uncoded}, the CDC scheme \eqref{TDCDC}, the half-duplex OSL scheme in \eqref{OSLHalfDuplex}, the half-duplex BW scheme in \eqref{BiHalfDuplex}, our CPC scheme in \eqref{eqMin_rk}, and our lower bound in \eqref{eq:thm_converse}}.  {Recall that the original BW scheme is delicately proposed for full-duplex systems, and it is still unknown how to extend it to the half-duplex case while preserving low NDT. Here,  for fair comparisons,  we compute the NDT of the BW scheme by multiplying the NDT of its full-duplex BW by a factor $2$, see Remark \ref{ReHalfDuplex}.}

In Fig.~\ref{conlusion}, we compare the NDT of each scheme when $ r $ changes given $ K = Q = 50 $. It is obvious that when $ r $ is small, our CPC scheme achieves the best performances, but still higher than the lower bound, and all schemes except the uncoded scheme, exhibit similar performance when $r$ is sufficiently large. When $ r =2 $, $ \delta_\text{CDC}(2,50) = {0.48}$, $ \delta_\text{CPC}(2,50) = {0.0544}$, the CPC scheme reduces the NDT by nearly $ 8.82 $ times than the CDC scheme. As $ r $ increases, the complexity of the system will increase \cite{CDC} and $ \delta_\text{CPC}(2,50) =0.0544 < \delta_\text{CDC}(13,50) =0.0569$, which means that the CPC scheme is more robust {to the increase of communication load and coding complexity} than the CDC scheme. Our CPC scheme achieves a much smaller NDT than the half-duplex OSL scheme and a slightly smaller NDT than the half-duplex BW scheme, identical to the analysis in Theorem \ref{RemCompa}. Although the NDT of the half-duplex BW scheme is close to our CPC scheme, it is in fact unclear how to extend the full-duplex BW scheme to the half-duplex case with the double NDT. %only increasing by a factor of $2$.

Fig.~\ref{r2differentK} compares the NDT of various schemes with $ r =2 $ when changing $ K $.  It can be seen that the CPC scheme outperforms the state-of-the-art schemes, including the CDC scheme, the half-duplex OSL scheme, and the half-duplex BW scheme. We can see that the half-duplex OSL scheme is no better than the original CDC scheme. Also, we observe that the NDT of the CPC scheme first increases with $ K $ and then decreases with $ K $. This is due to the trade-off between the local computation gain (increasing with $ K $) and the cooperative transmission gain (decreasing with $ K $), see Remark \ref{gainchange}. When $K$ is large, the cooperative transmission gain will be dominant. Moreover, the NDT of our CPC scheme decreases with $ K$ when $ K \geq 6 $, indicating it can better exploit the computing and communication resources.
%unlike all other schemes whose NDTs are increasing with the $K$
%When $ K $ is small and close to $ r $ since the number of receivers $ K_\textnormal{r} $ needs to be at least $ r $ to make full use of the coded multicast gain, which makes the number of transmitters $ K_\textnormal{t} $ less and the parallel transmission gain is poor. When $ K $ increases, the number of transmitters and receivers in the network increases, $ d_{\text{sum}} $ becomes greater, parallel transmission gain effect obtained is significantly improved, which ultimately enables our scheme to obtain better NDT. This is consistent with Theorem \ref{RemCompa}. Meanwhile, with the increase of $ K $, the NDT of the Uncoded, CDC, and OSL schemes all increase accordingly, becoming a monotonous increasing trend. But when $ K \geq 6 $, the CPC scheme decreases monotonically with $ K $. Among the three gains included in $ \delta_{\text{CPC}} $, coded multicast gain has nothing to do with $ K $, local computation gain value increases with the increase of $ K $, and the value of parallel transmission gain decreases with the increase of $ K $. {\color{blue}The magnitude of parallel transmission gain changed with $ K $ is greater than the magnitude of local computation gain, and finally, the NDT decreases with the increase of $ K $.} This indicates that the proposed CPC scheme can better exploit the computing resource. 

%Compared with the OSL scheme \cite{oneshot'IT} we observe that when $ r $ is small ($r\leq k/4 $), our CPC scheme achieves a much smaller NDT than the OSL
\begin{figure}[!t]
	\centering
	%\begin{minipage}[t]{0.45\linewidth}
	\centering
	\includegraphics[scale=0.4]{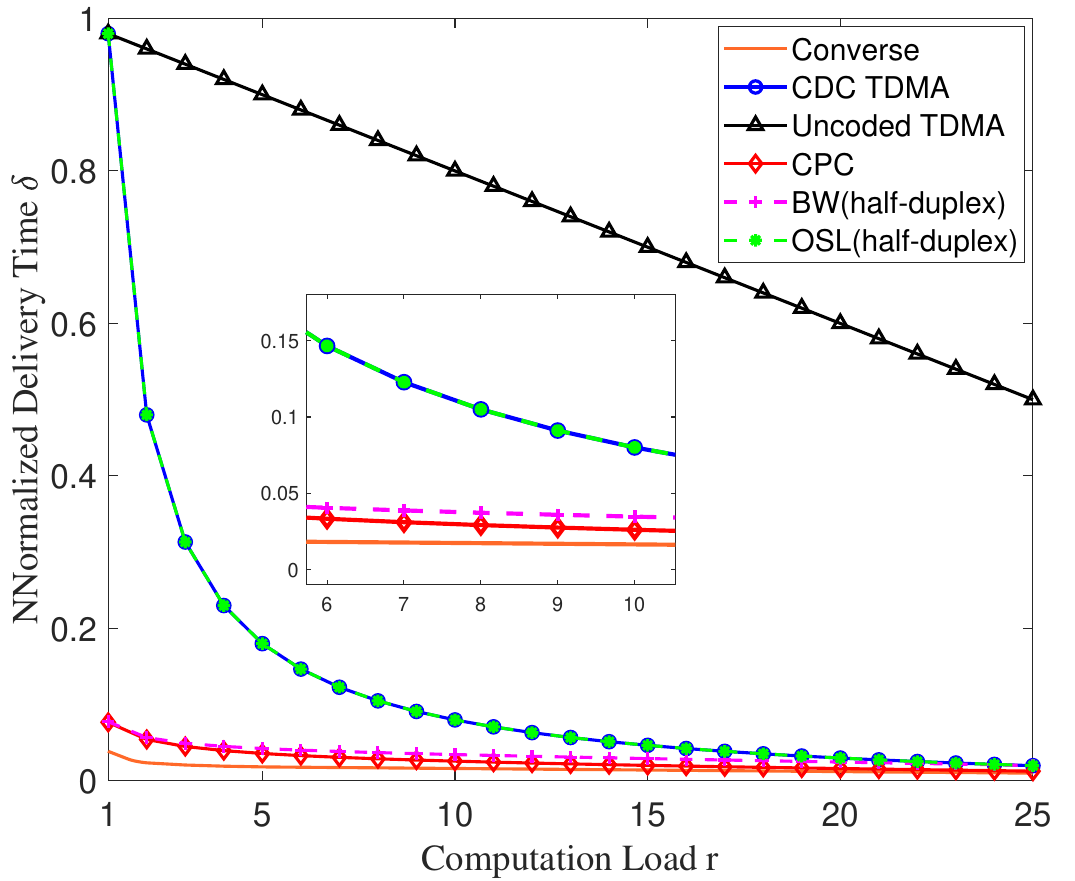}
	\caption{{The NDT of various schemes and lower bound versus $ r $ when $K=Q=50$.}}
	\label{conlusion}
\end{figure}
\begin{figure}[!t]
	%\end{minipage}%
        %\hspace{0.08\linewidth}
	%\begin{minipage}[t]{0.45\linewidth}
	\centering
        \includegraphics[scale=0.4]{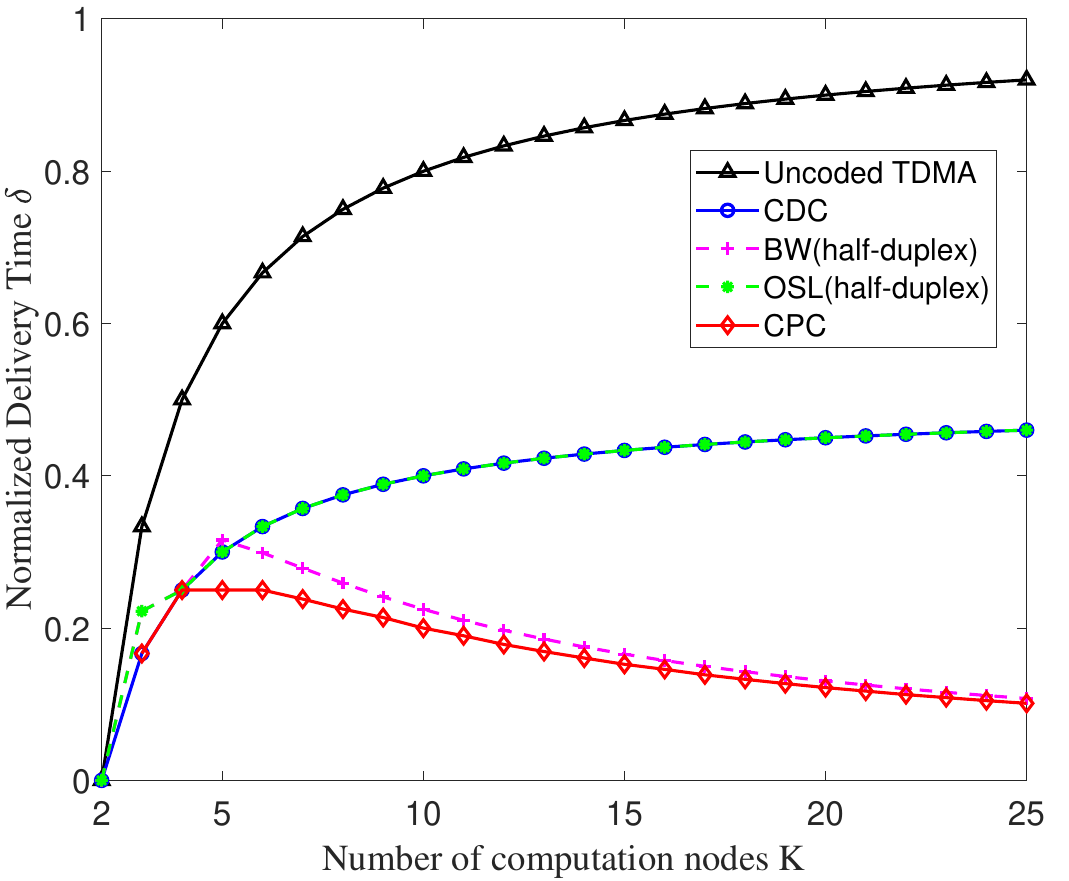}
	\caption{The NDT of various schemes versus $ K $ when $r =2 $ and $ Q=K $.}
	\label{r2differentK}
	%\end{minipage}
\end{figure}

Fig.~\ref{differentK} compares the NDT of the CPC scheme under different $ r $ as the number of computation nodes $ K $ increases. Similar to Fig.~\ref{r2differentK}, the trade-off between local computation gain and cooperative transmission gain results in the NDT of the CPC scheme first increasing and then decreasing with $ K $. Moreover, the CPC scheme approaches zero NDT as $K\to\infty$, identical to the result in \eqref{eqCPCinft}.
Fig.~\ref{differentKr} compares the NDT of the CPC scheme with different sizes of transmitter cooperation group $t = 1, 2, 3$ when $4\leq r \leq 10$ and $25\leq K \leq 50$. The NDT with different $t$ gets small from increasing $r$ and $K$, and the performance with $t = 1$ is better. This is due to the trade-off between the coded multicast gain and the DoF enhancement. When $r$ is relatively small, the coded multicast gain becomes dominant. More receivers share the same file (i.e., larger $r \!\!-\!\! t$), and more improvement can be obtained from coded multicast.
\begin{figure}[!t]
    %\begin{minipage}[t]{0.5\linewidth}
    \centering
    \includegraphics[scale=0.5]{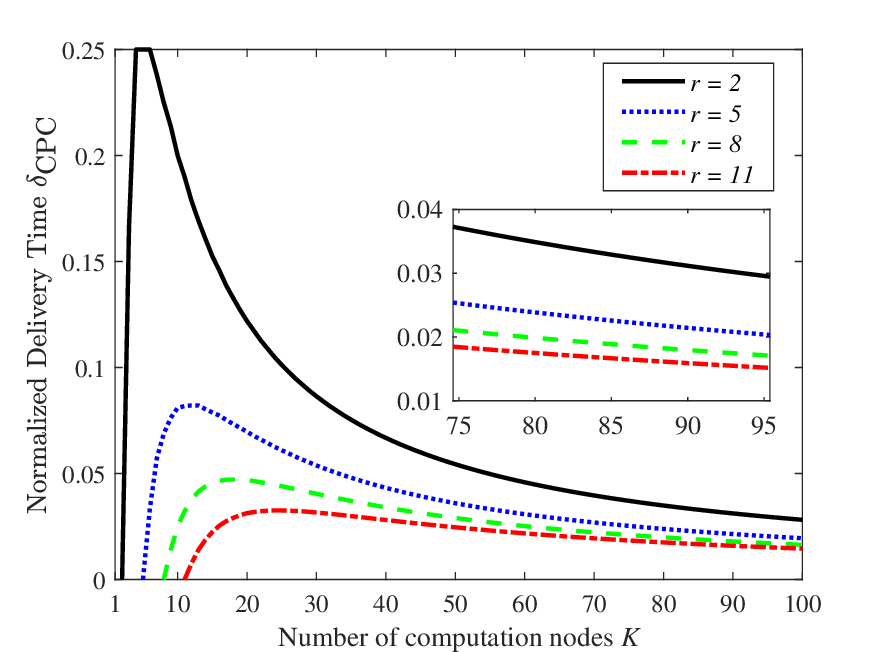}
    \caption{The NDT of CPC scheme versus $ K $ with different $ r $.}
    \label{differentK}
\end{figure}
    %\end{minipage}%
    %\hspace{0.02\linewidth}
    %\begin{minipage}[t]{0.5\linewidth}
\begin{figure}[!t]
    \centering
    \includegraphics[scale=0.5]{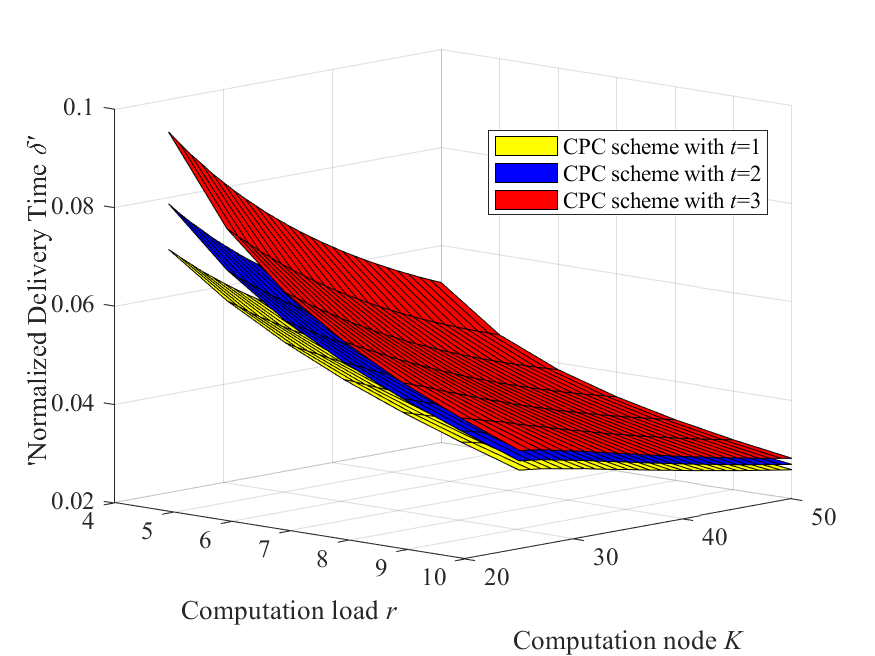}
    \caption{The NDT of CPC scheme versus $ r $ and $ K $ \\ with different $ t $.}
    \label{differentKr}
    %\end{minipage}
\end{figure}

\section{Conclusions} \label{conclusion}
This paper proposed a coded parallel transmission scheme for the half-duplex wireless MapReduce computing systems. Our scheme allows multiple nodes to simultaneously broadcast encoded symbols to other nodes over a cooperative X-multicast channel, followed by leveraging the interference alignment approach to align interference occurred by the parallel transmission. Theoretical analysis and numerical results demonstrated that our scheme can better exploit the computing resource and significantly reduce the communication delay compared with the previously known schemes. Different from all previous schemes which achieve constant communication latency when the number of nodes tends to be infinite, the proposed scheme can approach zero communication latency.

\begin{appendices}

\section{Proof of Lemma \ref{corollary1}}\label{IAproof}
Since the delivery of each message only involves one of the $ \tbinom{K}{K_t} $ partitions, and all the partitions have the same communication structure,  we only describe the transmission of one partition $\pi_{p}$. Suppose that the set of assigned transmitters $\mathcal{T}_p$ are $B_1,B_2,\cdots,B_{K_t}$, the set of receivers $\mathcal{R}_p$ are $D_1,D_2,\cdots,D_{K_r}$. Consider the $\binom{K_t}{t} \times \binom{K_r}{s}$ cooperative X-multicast channel in $\pi_{p}$. A total of $\binom{K_t}{t} \binom{K_r}{s}$ messages need to be sent in $\pi_{p}$. Then we have three different cases {of} Lemma \ref{corollary1} for analysis.

\subsection{Case: $s+t\ge K_\textnormal{r}+1$}
First, we divide the $\binom{K_t}{t} \binom{K_r}{s}$ messages into $\binom{K_t}{t}$ message groups according to the different transmitter groups, i.e., the messages received in each group have the same transmitter nodes. We just consider one of these groups $\mathcal{B}=\{B_1,B_2,\cdots,B_t\}$ as an example to illustrate the achievable transmission scheme. Other message groups can use the same scheme to transmit.

Denote $\mathbf{V}_{\mathcal{D},\mathcal{B}}$ as the transmitted symbol encoded from message $V_{\mathcal{D},\mathcal{B}}$ and set $\{B_1,B_2,\cdots,B_{K_r-s+1}\} \subseteq \mathcal{B}$ as transmitters to send and $\mathcal{D} = \{D_1,D_2,\cdots,D_s\}$ as receivers to receive symbol $\mathbf{V}_{\mathcal{D},\mathcal{B}}$. Using $\Gamma = \binom{K_r-1}{s-1}$-symbol extension to transmit the $\binom{K_r}{s}$ messages, for time slot $d \in [\Gamma]$, the received signal $Y_{j}(d)$, neglecting the noise, at arbitrary receiver $j \in [\mathcal{R}_p]$ is given by
\begin{IEEEeqnarray}{rcl}\label{receivedsignal}
	Y_{j}(d)&=&\sum_{m=B_1}^{B_{K_r-s+1}} h_{j,m}(d) \sum_{\mathcal{D}:|\mathcal{D}|=s} w_{\mathcal{D}, \mathcal{B}, m}(d) \mathbf{V}_{\mathcal{D},\mathcal{B}} \nonumber\\
	&=&\!\!\!\!\!\sum_{\mathcal{D}:|\mathcal{D}|=s, j \in \mathcal{D}} \left[\sum_{m=B_1}^{B_{K_r-s+1}} h_{j,m}(d) w_{\mathcal{D}, \mathcal{B}, m}(d) \right] \mathbf{V}_{\mathcal{D},\mathcal{B}}  \nonumber\\
	&&+\!\!\!\!\!\!\!\sum_{\mathcal{D}:|\mathcal{D}|=s, j \notin \mathcal{D}}\left[\sum_{m=B_1}^{B_{K_r-s+1}} h_{j,m}(d)w_{\mathcal{D}, \mathcal{B}, m}(d) \right] \mathbf{V}_{\mathcal{D},\mathcal{B}},
\end{IEEEeqnarray}
where $h_{j,m}(d)$ is the channel gain from transmitter $m$ to receiver $j$ at time $d \in [\Gamma]$ and $w_{\mathcal{D}, \mathcal{B}, m}(d)$ is the precoder of symbol $\mathbf{V}_{\mathcal{D},\mathcal{B}}$, which equals to the cofactor of $b_m$ in the following $(K_r-s+1) \times (K_r-s+1)$ matrix $\mathbf{H}_{\bar{\mathcal{D}}}(d)$:
\begin{IEEEeqnarray}{rCl} \label{matrix}
	\hspace{-15pt}\begin{aligned}
		\setlength{\arraycolsep}{0.7pt}
		\left(\begin{array}{cccc}
			h_{D_{s+1}, B_1}(d) & h_{D_{s+1}, B_2}(d) & \cdots & h_{D_{s+1}, B_{K_r-s+1}}(d) \\
			h_{D_{s+2}, B_1}(d) & h_{D_{s+2}, B_2}(d) & \cdots & h_{D_{s+2}, B_{K_r-s+1}}(d) \\
			\cdots & \cdots & \cdots & \cdots \\
			h_{D_{K_r}, B_1}(d) & h_{D_{K_r}, B_2}(d) & \cdots & h_{D_{K_r},B_{K_r-s+1}}(d) \\
			b_{1} & b_{2} & \cdots & b_{K_r-s+1}
		\end{array}\right) \!\triangleq\! \mathbf{H}_{\bar{\mathcal{D}}}(d),
	\end{aligned}
\end{IEEEeqnarray}
for any $\{b_{1}, b_{2}, \cdots, b_{K_r-s+1}\}$.

Then at any undesired receiver $D_{\psi}, \psi \in \{s+1,\cdots,K_r\}$, $\sum_{m=B_1}^{B_{K_r-s+1}} h_{D_{\psi}, m}(d) w_{\mathcal{D}, \mathcal{B}, m}(d) =0.$ The received signal in \eqref{receivedsignal} can be written as
\begin{IEEEeqnarray}{rCl}
	Y_{j}(d)&=& \sum_{\mathcal{D}:|\mathcal{D}|=s, j \in \mathcal{D}} \tilde{h}_{j, \mathcal{B}}^{\bar{\mathcal{D}}}(d) \mathbf{V}_{\mathcal{D},\mathcal{B}} \nonumber\\
	&=& \sum_{\mathcal{D}:|\mathcal{D}|=s, j \in \mathcal{D}} \sum_{m=B_1}^{B_{K_r-s+1}} h_{j, m}(d) w_{\mathcal{D}, \mathcal{B}, m}(d) \mathbf{V}_{\mathcal{D},\mathcal{B}}.
\end{IEEEeqnarray}

Since each $\{\tilde{h}_{j, \mathcal{B}}^{\bar{\mathcal{D}}}\}$ has a unique undesired receiver set $\bar{\mathcal{D}}$, it is easy to attain that polynomials $\{\tilde{h}_{j, \mathcal{B}}^{\bar{\mathcal{D}}_1}, \tilde{h}_{j, \mathcal{B}}^{\bar{\mathcal{D}}_2}, \cdots , \tilde{h}_{j, \mathcal{B}}^{\bar{\mathcal{D}}_\Gamma}\}$ are linearly independent, where $\{\bar{\mathcal{D}}_1,\bar{\mathcal{D}}_2,\cdots,\bar{\mathcal{D}}_{\Gamma}\}$ represent the receiver sets that do not need the $\binom{K_r-1}{s-1}$ messages intended to receiver $j$. With the same construction method of $\{w_{\mathcal{D}, \mathcal{B}, m}\}$ and same formation of $\{\tilde{h}_{j, \mathcal{B}}^{\bar{\mathcal{D}}}(d)\}$ at each slot, using \cite[Lemma 3]{Annapureddy2012}, we can receive the following $\Gamma \times \Gamma$ {received signal matrix, which is full-rank with probability 1:}
\begin{IEEEeqnarray}{rCl} \label{eq::matrix}
	\begin{aligned}
		\left(\begin{array}{cccc}
			\tilde{h}_{j, \mathcal{B}}^{\bar{\mathcal{D}}_1}(1) & \tilde{h}_{j, \mathcal{B}}^{\bar{\mathcal{D}}_2}(1) & \cdots & \tilde{h}_{j, \mathcal{B}}^{\bar{\mathcal{D}}_\Gamma}(1) \\
			\tilde{h}_{j, \mathcal{B}}^{\bar{\mathcal{D}}_1}(2) & \tilde{h}_{j, \mathcal{B}}^{\bar{\mathcal{D}}_2}(2) & \cdots & \tilde{h}_{j, \mathcal{B}}^{\bar{\mathcal{D}}_\Gamma}(2) \\
			\cdots & \cdots & \cdots & \cdots \\
			\tilde{h}_{j, \mathcal{B}}^{\bar{\mathcal{D}}_1}(\Gamma) & \tilde{h}_{j, \mathcal{B}}^{\bar{\mathcal{D}}_2}(\Gamma) & \cdots & \tilde{h}_{j, \mathcal{B}}^{\bar{\mathcal{D}}_\Gamma}(\Gamma)
		\end{array}\right).
	\end{aligned}
\end{IEEEeqnarray}
Receiver $k$ can successfully decode its $\binom{K_r-1}{s-1}$ desired messages in $\binom{K_r-1}{s-1}$ time slots. Similar arguments can be applied to other receivers. Therefore, the per-receiver DoF of 1 is achieved.

\subsection{Case:$s+t = K_r$}
\begin{sloppypar}
	First, encode each message $V_{\mathcal{D},\mathcal{B}}$ into a $tn^{\binom{K_r-1}{s}\binom{K_t}{t}}$ symbol vector $\textbf{V}_{\mathcal{D},\mathcal{B}} = \left((\mathbf{V}^1_{\mathcal{D},\mathcal{B}})^T,(\mathbf{V}^2_{\mathcal{D},\mathcal{B}})^T,\cdots,(\mathbf{V}^t_{\mathcal{D},\mathcal{B}})^T\right)^T$, where $n \in \mathbb{N}^+$ and $\mathbf{V}^i_{\mathcal{D},\mathcal{B}}(i \in [t])$ is an $n^{\binom{K_r-1}{s}\binom{K_t}{t}} \times 1$ vector. Using $\Gamma = \binom{K_r-1}{s-1}\binom{K_t}{t}tn^{\binom{K_r-1}{s}\binom{K_t}{t}}+(n+1)^{\binom{K_r-1}{s}\binom{K_t}{t}}$-symbol extension to transmit messages, in each time slot $d \in [\Gamma]$, the received signal $Y_{j}(d)$, neglecting the noise, at arbitrary receiver $j \in [\mathcal{R}_p]$ is given by
	\begin{IEEEeqnarray}{rCl} \label{receivedsignalB}
		Y_{j}(d)&=& \sum_{\mathcal{D}:|\mathcal{D}|=s} \sum_{\mathcal{B}:|\mathcal{B}|=t} \sum_{i=1}^{t} \nonumber\\
		&&\hspace{0.5cm}\left[\sum_{m \in \mathcal{B}}h_{j,m}(d)\left(\mathbf{w}_{\mathcal{D},\mathcal{B},m}^i(d)\right)^T\right] \mathbf{V}^i_{\mathcal{D},\mathcal{B}},
	\end{IEEEeqnarray}
	where $h_{j,m}(d)$ is the channel realization and $\mathbf{w}_{\mathcal{D},\mathcal{B},m}^i(d)$ is $n^{\binom{K_r-1}{s}\binom{K_t}{t}} \times 1$ precoding vector of symbol vector $\mathbf{V}^i_{\mathcal{D},\mathcal{B}}$ at transmitter $m$.
\end{sloppypar}
For an arbitrary symbol vector $\mathbf{V}_{\mathcal{D},\mathcal{B}}^i$ desired by receiver group $\mathcal{D}=\{D_1,D_2,\cdots,D_s\}$ and transmitted by transmitter group $\mathcal{B}=\{B_1,B_2,\cdots,B_t\}$, assuming that the undesired receiver set is $\bar{\mathcal{D}}=\mathcal{R}_p \backslash \mathcal{D}=\{\bar{D}_1,\bar{D}_2,\cdots,\bar{D}_t\}$, it can be neutralized at $t-1$ receivers $\bar{\mathcal{D}}_i = \bar{\mathcal{D}} \backslash \{\bar{D}_i\}$ by interference neutralization same used in Case A. Using $\{h_{j,m}(d), j \in \bar{\mathcal{D}}_i, m \in \mathcal{B}\}$ and any $\{b_{1}, b_{2},\cdots, b_{t}\}$, generate matrix with the same structure in \eqref{matrix}. Let $c_m(d)$ be the cofactor of $b_m$ of this matrix.

For the \textit{p}-th element $w_{\mathcal{D},\mathcal{B},m,p}^i(d)$ of $\mathbf{w}_{\mathcal{D},\mathcal{B},m}^i(d)$, design it as
\begin{IEEEeqnarray}{rCl} \label{designprocode}
	\begin{aligned}
		w_{\mathcal{D},\mathcal{B},m,p}^i(d) = \alpha_{\mathcal{D},\mathcal{B}}^{\bar{\mathcal{D}}_i}(d)c_m(d)z_{\mathcal{D},\mathcal{B},p}^{\bar{\mathcal{D}}_i}(d),
	\end{aligned}
\end{IEEEeqnarray}
where $\alpha_{\mathcal{D},\mathcal{B}}^{\bar{\mathcal{D}}_i}(d)$ is chosen i.i.d. from a continuous distribution for all $\{\mathcal{D},\mathcal{B},\bar{\mathcal{D}}_i,t\}$, $c_m(d)$ is the cofactor of $b_m$ and $z_{\mathcal{D},\mathcal{B},p}^{\bar{\mathcal{D}}_i}(d)$ will be used for interference alignment and determined later. Due to $\sum_{m \in [\mathcal{B}]} h_{\bar{D}_j, m}(d)c_m(d)=0, \bar{D}_j \in \bar{\mathcal{D}}_i$, the received signal in \eqref{receivedsignalB} at receiver $j \in [K_r]$ can be written as
\begin{IEEEeqnarray}{rCl}\label{receivedsignalB2}
	Y_{j}(d)&= & \sum_{\mathcal{D}:|\mathcal{D}|=s,j \in \mathcal{D}} \sum_{\mathcal{B}:|\mathcal{B}|=t} \sum_{i=1}^{t} \nonumber\\
	&&\hspace{1cm} \left[\sum_{m \in \mathcal{B}}h_{j,m}(d)\left(\mathbf{w}_{\mathcal{D},\mathcal{B},m}^i(d)\right)^T\right] \mathbf{V}^i_{\mathcal{D},\mathcal{B}} \nonumber\\
	&& + \sum_{\substack{\bar{\mathcal{D}}_i,\mathcal{D}:|\mathcal{D}|=s, \\ j \notin {\bar{\mathcal{D}}_i \cup \mathcal{D}}}} \sum_{\mathcal{B}:|\mathcal{B}|=t} \nonumber\\
	&&\hspace{1cm} \left[\sum_{m \in \mathcal{B}}h_{j,m}(d)\left(\mathbf{w}_{\mathcal{D},\mathcal{B},m}^i(d)\right)^T\right] \mathbf{V}^i_{\mathcal{D},\mathcal{B}}.
\end{IEEEeqnarray}

Consider the following monomial set:
\begin{IEEEeqnarray}{rCl} \label{monomialset}
	\begin{aligned}
		\mathcal{M}_j[n] = \left\{\prod_{\substack{\mathcal{D},\bar{\mathcal{D}}_i,\mathcal{B}: \\ j \notin \mathcal{D} \cup \bar{\mathcal{D}}_i}} \left[\alpha_{\mathcal{D},\mathcal{B}}^{\bar{\mathcal{D}}_i}\tilde{h}_{j,\mathcal{B}}^{\bar{\mathcal{D}}_i}\right]^{\gamma_{\mathcal{D},\mathcal{B}}^i} \!:\! 1 \leqslant \gamma_{\mathcal{D},\mathcal{B}}^i \leqslant n \right\},
	\end{aligned}
\end{IEEEeqnarray}
where $\tilde{h}_{j,\mathcal{B}}^{\bar{\mathcal{D}}_i}$ is
\begin{IEEEeqnarray}{rCl} \label{h_{k,b}}
	\tilde{h}_{j,\mathcal{B}}^{\bar{\mathcal{D}}_i} &\triangleq& \sum_{m=\mathcal{B}_1}^{\mathcal{B}_t} h_{j,m} c_{m} \nonumber\\
	&=& \setlength{\baselineskip}{0.6pt} \left|\begin{array}{cccc}
		h_{\bar{D}_1, B_1} & h_{\bar{D}_1, B_2} & \cdots & h_{\bar{D}_1, B_t} \\
		\cdots & \cdots & \cdots & \cdots \\
		h_{\bar{D}_{i-1}, B_1} & h_{\bar{D}_{i-1}, B_2} & \cdots & h_{\bar{D}_{i-1}, B_t} \\
		h_{\bar{D}_{i+1}, B_1} & h_{\bar{D}_{i+1}, B_2} & \cdots & h_{\bar{D}_{i+1}, B_t} \\
		\cdots & \cdots & \cdots & \cdots \\
		h_{\bar{D}_t, B_1} & h_{\bar{D}_t, B_2} & \cdots & h_{\bar{D}_t, B_t} \\
		h_{j, B_1} & h_{j, B_2} & \cdots & h_{j, B_t}
	\end{array}\right|.
\end{IEEEeqnarray}
For each element $w_{\mathcal{D},\mathcal{B},m,p}^i(d)$ in $\mathbf{w}_{\mathcal{D},\mathcal{B},m}^i(d)$ satisfying $j \notin \mathcal{D} \cup \bar{\mathcal{D}}_i$, the element $z_{\mathcal{D},\mathcal{B},p}^{\bar{\mathcal{D}}_i}(d)$ in \eqref{designprocode} is given by a unique monomial $M_{\mathcal{D},\mathcal{B},p}^{\bar{\mathcal{D}}_i}(d)$ in $\mathcal{M}_k[n]$ by taking $\{h_{j,m}(d)\} $ and $\{\alpha_{\mathcal{D},\mathcal{B}}^{\bar{\mathcal{D}}_i}(d)\}$ into \eqref{monomialset}. Then the summation $\sum_{m \in \mathcal{B}} h_{j,m}(d) w_{\mathcal{D},\mathcal{B},m,p}^i(d), j \notin \mathcal{D} \cup \bar{\mathcal{D}}_i$ is $\alpha_{\mathcal{D},\mathcal{B}}^{\bar{\mathcal{D}}_i}(d)\tilde{h}_{j,\mathcal{B}}^{\bar{\mathcal{D}}_i}(d)M_{\mathcal{D},\mathcal{B},p}^{\bar{\mathcal{D}}_i}(d)$ and satisfies
%\begin{IEEEeqnarray}{rCl}
$\alpha_{\mathcal{D},\mathcal{B}}^{\bar{\mathcal{D}}_i}(d)\tilde{h}_{j,\mathcal{B}}^{\bar{\mathcal{D}}_i}(d)M_{\mathcal{D},\mathcal{B},p}^{\bar{\mathcal{D}}_i}(d) \in \mathcal{M}_j[n+1](d).$ 
%\end{IEEEeqnarray}
Therefore, the interferences at receiver $j$ are aligned together.

By proving only existing zero factors $s_{\mathcal{B}_0}^{\bar{\mathcal{D}}_0}$ such that $\sum_{\substack{\mathcal{D}_0,\bar{\mathcal{D}}_0,\mathcal{B}_0: \\j \notin \bar{\mathcal{D}}_0, j' \notin \mathcal{D}_0 \cup \bar{\mathcal{D}}_0}} s_{\mathcal{B}_0}^{\bar{\mathcal{D}}_0} \frac{\tilde{h}_{j,\mathcal{B}_{0}}^{\bar{\mathcal{D}}_{0}}}{\tilde{h}_{j^{\prime}, \mathcal{B}_{0}}^{ \bar{\mathcal{D}}_{0}}} = 0$ in \cite[(58)]{taoWireless}, the polynomials in $\{\frac{\tilde{h}_{j,\mathcal{B}_{0}}^{\bar{\mathcal{D}}_{0}}}{\tilde{h}_{j^{\prime}, \mathcal{B}_{0}}^{ \bar{\mathcal{D}}_{0}}} :j \notin \bar{\mathcal{D}}_{0}, j' \notin \mathcal{D}_0 \cup \bar{\mathcal{D}}_0, \forall \mathcal{B} \subseteq \mathcal{T}\}$ are linearly independent. Then the arbitrary subset with the following form is linearly independent:
\begin{IEEEeqnarray}{rCl} \label{eachsubset}
	\begin{aligned}
		&\Big\{\frac{\alpha_{\mathcal{D}_{0}, \mathcal{B}_{0}}^{\bar{\mathcal{D}}_{0}} \tilde{h}_{j,\mathcal{B}_{0}}^{\bar{\mathcal{D}}_{0}}}{\alpha_{ \mathcal{D}_{0}, \mathcal{B}_{0}}^{ \bar{\mathcal{D}}_{0}} \tilde{h}_{j^{\prime}, \mathcal{B}_{0}}^{ \bar{\mathcal{D}}_{0}}} \prod_{ \substack{\mathcal{D}, \bar{\mathcal{D}}_{i}, \mathcal{B} :\\ j' \notin \mathcal{D} \cup \bar{\mathcal{D}}_i} }{ \left[ \alpha_{\mathcal{D},\mathcal{B}}^{\bar{\mathcal{D}}_i} \tilde{h}_{j,\mathcal{B}}^{\bar{\mathcal{D}}_i} \right]^{\gamma_{\mathcal{D},\mathcal{B}}^i} }: \nonumber\\ 
		&\hspace{1cm} j \in \mathcal{D}_{0}, j' \notin \mathcal{D}_0 \cup \bar{\mathcal{D}}_0, \forall \mathcal{B} \subseteq \mathcal{T}\Big\},
	\end{aligned}   
\end{IEEEeqnarray}
where the power of $\alpha_{\mathcal{D},\mathcal{B}}^{\bar{\mathcal{D}}_i}$ is $\gamma_{\mathcal{D},\mathcal{B}}^i$.

Because of the different powers of factors $\{\alpha_{\mathcal{D},\mathcal{B}}^{\bar{\mathcal{D}}_i}\}$, the polynomials in different subsets are linearly independent, combining with \eqref{eachsubset}, the polynomials of desired symbols of receiver $j$ corresponding to the same $j'$ are linearly independent. Moreover, $\{\alpha_{\mathcal{D},\mathcal{B}}^{\bar{\mathcal{D}}_i}:j' \notin \mathcal{D} \cup \bar{\mathcal{D}}_i, \forall \mathcal{B} \subseteq \mathcal{T}\}$ only exist in the polynomials whose transmitted symbols are the interference of receiver $j'$. Thus, polynomials with different $j'$ are linearly independent. Combining the linear independence corresponding to the same $j'$ and different $j'$, the following polynomials
\begin{IEEEeqnarray}{rCl}\label{matr1}
	&\left\{ \alpha_{\mathcal{D},\mathcal{B}}^{\bar{\mathcal{D}}_i} \tilde{h}_{j,\mathcal{B}}^{\bar{\mathcal{D}}_i} M_{\mathcal{D},\mathcal{B},p}^{\bar{\mathcal{D}}_i} : |\mathcal{D}|=s, k \in \mathcal{D}, |\mathcal{B}|=t,\right. \nonumber\\
	& \hspace{1cm} \left.  i \in [t], p \in \left[ n^{ \binom{K_r-1}{s} \binom{K_t}{t} } \right] \right\}  \nonumber\\
	&\cup \left\{ M: M \in \mathcal{M}_k[n+1] \right\}
\end{IEEEeqnarray}
are linear independent. Since the construction method of $\{\mathbf{w}_{\mathcal{D},\mathcal{B},m}^{i}(d)\}$ and the formation of $\{\alpha_{\mathcal{D},\mathcal{B}}^{\bar{\mathcal{D}}_i}(d)\tilde{h}_{j,\mathcal{B}}^{\bar{\mathcal{D}}_i}(d)M_{\mathcal{D},\mathcal{B},p}^{\bar{\mathcal{D}}_i}(d)\}$ and $\{M(d)\}$ are the same at each time slot $d$, based on \cite[Lemma 3]{Annapureddy2012}, putting specific $\{\alpha_{\mathcal{D},\mathcal{B}}^{\bar{\mathcal{D}}_i}(d)\tilde{h}_{j,\mathcal{B}}^{\bar{\mathcal{D}}_i}(d)M_{\mathcal{D},\mathcal{B},p}^{\bar{\mathcal{D}}_i}(d)\}$ and $\{M(d)\}$ into \eqref{matr1}, the $\Gamma \times \Gamma$ received signal matrix is full-rank with probability 1.

Similarly, the polynomials of received symbols at other receivers are also linearly independent. Since each receiver can decode $\binom{K_r-1}{s-1}\binom{K_t}{t}tn^{\binom{K_r-1}{s-1}\binom{K_t}{t}}$ symbols in $\Gamma$ time slots, we can get a per-receiver DoF of
$d = \frac{\binom{K_r-1}{s-1}\binom{K_t}{t}tn^{\binom{K_r-1}{s-1}\binom{K_t}{t}}}{\binom{K_r-1}{s-1}\binom{K_t}{t}tn^{\binom{K_r-1}{s}\binom{K_t}{t}}+(n+1)^{\binom{K_r-1}{s}\binom{K_t}{t}}}$.
When $n \rightarrow \infty$, the per-receiver DoF of $\frac{\binom{K_r-1}{s-1}\binom{K_t}{t}t}{\binom{K_r-1}{s-1}\binom{K_t}{t}t+1}$ is achievable.

\subsection{Case: $s+t \leq K_\textnormal{r}-1$}
In this case, we use two methods to deliver the messages. The first method is similar to the one used in Case A. First, we split each message $V_{\mathcal{D},\mathcal{B}}$ into $\binom{K_r-s}{t-1}$ sub-messages, each associated with a unique receiver set $\mathcal{D} \cup \{D_{s+1},D_{s+2},\cdots,D_{s+t-1}\}$, where $\mathcal{D}$ is the desired receiver set and $\{D_{s+1},D_{s+2},\cdots,D_{s+t-1}\}$ is a set of arbitrary $t-1$ receivers from the rest $K_r-s$ undesired receivers. Then there are $\binom{K_r}{s+t-1}$ different receiver sets in total, and each having $\binom{s+t-1}{s}\binom{K_t}{t}$ sub-messages. In the delivery phase, sub-messages with different receiver sets are delivered individually in a time-division manner, and sub-messages with the same receiver set are delivered together. Through this approach, we observe that the transmitters send $\binom{s+t-1}{s}\binom{K_t}{t}$ sub-messages each slot and each receiver in the corresponding receiver set off these sub-messages desires $\binom{s+t-2}{s-1}\binom{K_t}{t}$ sub-messages, while the rest $K_r-s$ receivers do not desire any sub-message of them. Therefore, we can regard the network each time as a $\binom{K_t}{t} \times \binom{s+t-1}{s}$ cooperative X-multicast network whose per-receiver achievable DoF is 1 in Case A. Note that each receiver only exists in $\binom{K_r-1}{s+t-2}$ of the $\binom{K_r}{s}$ receiver sets in total. Thus, in this method, the per-receiver DoF is $\binom{K_r-1}{s+t-2} \backslash \binom{K_r}{s} = \frac{r}{K_r} $.

The second method is firstly using interference neutralization to neutralize each message at undesired receivers, and then align the rest interference together by interference alignment, which is also used in Case B.

\begin{sloppypar}
	We first consider the case when $t<K_t$ and encode each message $V_{\mathcal{D},\mathcal{B}}$ into a $\binom{K_r-s}{t-1}tn^{(K_r-s-t+1)(K_t-t+1)} \times 1$ symbol vector $\mathbf{V}_{\mathcal{D},\mathcal{B}} = ((\mathbf{V}_{\mathcal{D},\mathcal{B}}^1)^T,(\mathbf{V}_{\mathcal{D},\mathcal{B}}^2)^T,\cdots,(\mathbf{V}_{\mathcal{D},\mathcal{B}}^\sigma)^T)^T,$ where $\sigma \triangleq \binom{K_r-s}{t-1}, n \in \mathbb{N}^+$, and $\mathbf{V}_{\mathcal{D},\mathcal{B}}^i(i \in [\sigma])$ is a $tn^{(K_r-s-t+1)(K_t-t+1)} \times 1$ vector. Using $\Gamma \triangleq \binom{K_r-1}{s-1}\binom{K_t}{t}\binom{K_r-s}{t-1}tn^{(K_r-s-t+1)(K_t-t+1)} + \binom{K_r-1}{s}\binom{K_t}{t-1}\binom{K_r-s-1}{t-1}(n+1)^{(K_r-s-t+1)(K_t-t+1)}$-symbol extension to transmit messages, in each time slot $d$, the received signal, neglecting the Gaussian noise, at an arbitrary receiver $j \in [K_r]$, is
	\begin{IEEEeqnarray}{rCl} \label{receivedsignalC}
		Y_{j}(d)&=& \sum_{\mathcal{D}:|\mathcal{D}|=s} \sum_{\mathcal{B}:|\mathcal{B}|=t} \sum_{i=1}^{\sigma} \nonumber\\
		&&\hspace{1cm}\left[\sum_{m \in \mathcal{B}}h_{j,m}(d)\left(\mathbf{w}_{\mathcal{D},\mathcal{B},m}^i(d)\right)^T\right] \mathbf{V}^i_{\mathcal{D},\mathcal{B}},
	\end{IEEEeqnarray}
	where $h_{j,m}(d)$ is the channel realization and $\mathbf{w}_{\mathcal{D},\mathcal{B},m}^i(d)$ is $n^{(K_r-s-t+1)(K_t-t+1)} \times 1$ precoding vector of symbol vector $\mathbf{V}^i_{\mathcal{D},\mathcal{B}}$ at transmitter $m$.
\end{sloppypar}

To apply interference neutralization, consider an arbitrary symbol vector $\mathbf{V}^i_{\mathcal{D},\mathcal{B}}$ desired by receiver multicast group $\mathcal{D}=\{D_1,D_2,\cdots,D_s\}$, transmitted by transmitter cooperation group $\mathcal{B} = \{B_1,B_2,\cdots,B_t\}$, and whose undesired receiver set is $\bar{\mathcal{D}}=[K_r] \backslash \mathcal{D} = \{\bar{D}_1,\bar{D}_2,\cdots,\bar{D}_{K_r-s}\}$. Assume each $\mathbf{V}^i_{\mathcal{D},\mathcal{B}}$ will be neutralized at a distinct receiver set $\bar{\mathcal{D}}_i \subset \bar{\mathcal{D}}$ with $|\bar{\mathcal{D}}_i|=t-1$. Consider an arbitrary $\mathbf{V}^i_{\mathcal{D},\mathcal{B}}$ with $\bar{\mathcal{D}}_i=\{\bar{D}_{i,1},\bar{D}_{i,2},\cdots,\bar{D}_{i,t-1}\}$. Using the same method in Case A and B, construct the matrix from $\{h_{j,m}(d), j \in \bar{\mathcal{D}}_i, m \in \mathcal{B}\}$ and any $\{b_{1}, b_{2},\cdots, b_{t}\}$ like \eqref{matrix} and let $c_m(d)$ be the cofactor of $b_m$ of matrix. 

For the precoding vector $\mathbf{w}_{\mathcal{D},\mathcal{B},m}^i(d)$, let $w_{\mathcal{D},\mathcal{B},m,p}^i(d)$ be the $p$-th element and designed as
\begin{IEEEeqnarray}{rCl} \label{designprocodeC}
	\begin{aligned}
		w_{\mathcal{D},\mathcal{B},m,p}^i(d) = \alpha_{\mathcal{D},\mathcal{B}}^{\bar{\mathcal{D}}_i}(d)m_j(d)z_{\mathcal{D},\mathcal{B},p}^{\bar{\mathcal{D}}_i}(d),
	\end{aligned}
\end{IEEEeqnarray}
where $\alpha_{\mathcal{D},\mathcal{B}}^{\bar{\mathcal{D}}_i}(d)$ is chosen i.i.d. from a continuous distribution for all $\{\mathcal{D},\mathcal{B},\bar{\mathcal{D}}_i,t\}$, $c_m(d)$ is the cofactor of $b_m$ and $z_{\mathcal{D},\mathcal{B},p}^{\bar{\mathcal{D}}_i}(d)$ will be used for interference alignment and determined later. By such construction, at any undesired receiver $\bar{D}_{i,j} \in \bar{\mathcal{D}}_i$, can get
\begin{IEEEeqnarray}{rCl}
	\begin{aligned}
		\sum_{m \in [\mathcal{B}]} h_{\bar{D}_{i,j}, m}(d) \alpha_{\mathcal{D},\mathcal{B}}^{\bar{\mathcal{D}}_i}(d)c_m(d)z_{\mathcal{D},\mathcal{B},p}^{\bar{\mathcal{D}}_i}(d) =0.
	\end{aligned}
\end{IEEEeqnarray}

Then by interference neutralization, received signal in \eqref{receivedsignalC} at receiver $j \in [K_r]$ can be written as
\begin{IEEEeqnarray}{rCl}\label{receivedsignalC2}
	Y_{j}(d)&= & \sum_{\mathcal{D}:|\mathcal{D}|=s,j \in \mathcal{D}} \sum_{\mathcal{B}:|\mathcal{B}|=t} \sum_{i=1}^{\sigma} \nonumber\\
	&&\hspace{1cm} \left[\sum_{m \in \mathcal{B}}h_{j,m}(d)\left(\mathbf{w}_{\mathcal{D},\mathcal{B},m}^i(d)\right)^T\right] \mathbf{V}^i_{\mathcal{D},\mathcal{B}} \\
	&& + \sum_{\substack{\bar{\mathcal{D}}_i,\mathcal{D}:|\mathcal{D}|=s, \\j \notin {\bar{\mathcal{D}}_i \cup \mathcal{D}}}} \sum_{\mathcal{B}:|\mathcal{B}|=t} \nonumber\\
	&&\hspace{1cm} \left[\sum_{m \in \mathcal{B}}h_{j,m}(d)\left(\mathbf{w}_{\mathcal{D},\mathcal{B},m}^i(d)\right)^T\right] \mathbf{V}^i_{\mathcal{D},\mathcal{B}}.
\end{IEEEeqnarray}

Consider the following monomial set:
\begin{IEEEeqnarray}{rCl} \label{monomialsetC}
	\begin{aligned} 
		\mathcal{M}_{\mathcal{D},\mathcal{B}^c}^{\bar{\mathcal{D}}_i}[n] &= \Big\{ \prod_{\substack{m,j:m \notin \mathcal{B}^c \\ j \notin \mathcal{D} \cup \bar{\mathcal{D}}_i}} \left[\alpha_{\mathcal{D},\{m\}\cup \mathcal{B}^c}^{\bar{\mathcal{D}}_i}\tilde{h}_{j,\{m\}\cup \mathcal{B}^c}^{\bar{\mathcal{D}}_i}\right]^{\gamma_{\mathcal{D},\{m\}\cup \mathcal{B}^c}^{i,j}}:\nonumber\\
		&\hspace{1cm}  1 \leqslant \gamma_{\mathcal{D},\{m\}\cup \mathcal{B}^c}^{i,j} \leqslant n \Big\}, 
	\end{aligned}
\end{IEEEeqnarray}
where $\mathcal{B}^c \subset \mathcal{B}, |\mathcal{B}^c|=t-1$ and
\begin{IEEEeqnarray}{rCl} \label{h_{k,b}C}
	\begin{aligned}
		\tilde{h}_{j,\mathcal{B}}^{\bar{\mathcal{D}}_i} \!\triangleq\!\!\! \sum_{m=\mathcal{B}_1}^{\mathcal{B}_t}\!\! h_{j,m} c_{m} \! =\!\!  \setlength{\arraycolsep}{0.9pt}
		\left|\begin{array}{cccc}
			h_{\bar{D}_{i,1}, B_1} & h_{\bar{D}_{i,1}, B_2} & \cdots & h_{\bar{D}_{i,1}, B_t} \\
			h_{\bar{D}_{i,2}, B_1} & h_{\bar{D}_{i,2}, B_2} & \cdots & h_{\bar{D}_{i,2}, B_t} \\
			\cdots & \cdots & \cdots & \cdots \\
			h_{\bar{D}_{i,t-1}, B_1} & h_{\bar{D}_{i,t-1}, B_2} & \cdots & h_{\bar{D}_{i,t-1}, B_t} \\
			h_{j, B_1} & h_{j, B_2} & \cdots & h_{j, B_t}
		\end{array}\right|.
	\end{aligned}
\end{IEEEeqnarray}

Referring to Case B, the element $z_{\mathcal{D},\mathcal{B},p}^{\bar{\mathcal{D}}_i}(d)$ in \eqref{designprocodeC} is given by a unique monomial $M_{\mathcal{D},\mathcal{B},p}^{\bar{\mathcal{D}}_i}(d)$ in $\mathcal{M}_{\mathcal{D},\mathcal{B}}^{\bar{\mathcal{D}}_i}[n]$ and the summation $\sum_{m \in \mathcal{B}, m \notin \mathcal{D} \cup \bar{\mathcal{D}}_i} h_{j,m}(d) w_{\mathcal{D},\mathcal{B},m,p}^i(d)$ is $\alpha_{\mathcal{D},\mathcal{B}}^{\bar{\mathcal{D}}_i}(d)\tilde{h}_{j,\mathcal{B}}^{\bar{\mathcal{D}}_i}(d)m_{\mathcal{D},\mathcal{B},p}^{\bar{\mathcal{D}}_i}(d)$ and satisfies
%\begin{IEEEeqnarray}{rCl}
%\begin{aligned}
$\alpha_{\mathcal{D},\mathcal{B}}^{\bar{\mathcal{D}}_i}(d)\tilde{h}_{j,\mathcal{B}}^{\bar{\mathcal{D}}_i}(d)M_{\mathcal{D},\mathcal{B},p}^{\bar{\mathcal{D}}_i}(d) \in \mathcal{M}_{\mathcal{D},\mathcal{B}}^{\bar{\mathcal{D}}_i}[n+1](d).$ 
%\end{aligned}
%\end{IEEEeqnarray}
{More specifically}, denote $m_{\mathcal{D},\mathcal{B},p}^{\bar{\mathcal{D}}_i}(d)$ as the monomial selected by $w_{\mathcal{D},\mathcal{B},m,p}^i(d)$ in an arbitrary $\mathcal{M}_{\mathcal{D},\mathcal{B}^c}^{\bar{\mathcal{D}}_i}[n]$, where $\mathcal{B}^{c} \subset \mathcal{B}$. Then
\begin{IEEEeqnarray}{rCl}
	\begin{aligned}
		\alpha_{\mathcal{D},\mathcal{B}}^{\bar{\mathcal{D}}_i}(d)\tilde{h}_{j,\mathcal{B}}^{\bar{\mathcal{D}}_i}(d)M_{\mathcal{D},\mathcal{B}^c,p}^{\bar{\mathcal{D}}_i}(d) \in \mathcal{M}_{\mathcal{D},\mathcal{B}^c}^{\bar{\mathcal{D}}_i}[n+1](d),
	\end{aligned}
\end{IEEEeqnarray}
which means that the symbols intended for the same receiver multicast group $\mathcal{D}$, neutralized at the same receiver set $\bar{\mathcal{D}}_i$, which its precoder $z_{\mathcal{D},\mathcal{B},p}^{\bar{\mathcal{D}}_i}$ constructed from the same monomial set $\mathcal{M}_{\mathcal{D},\mathcal{B}^c}^{\bar{\mathcal{D}}_i}[n]$ are aligned in the same subspace with dimension $(n+1)^{(K_r-s-t-1)(K_t-t+1)}$.

Due to each $\alpha_{\mathcal{D},\{m\}\cup \mathcal{B}^c}^{\bar{\mathcal{D}}_i}\tilde{h}_{j,\{m\}\cup \mathcal{B}^c}^{\bar{\mathcal{D}}_i}$ in \eqref{monomialsetC} has a unique channel coefficient $h_{j,m}$, we can guarantee the Jacobian matrix of polynomials $\{\alpha_{\mathcal{D},\{m\}\cup \mathcal{B}^c}^{\bar{\mathcal{D}}_i}\tilde{h}_{j,\{m\}\cup \mathcal{B}^c}^{\bar{\mathcal{D}}_i}:j \notin \mathcal{D}\cup \bar{\mathcal{D}}_i, m \notin \mathcal{B}^c\}$ is full row rank. Using \cite[Page 135, Theorem 3]{hodge1947methods} and \cite[Lemma 1]{Annapureddy2012}, it can be seen that these polynomials are algebraically independent. Then the polynomials in $\mathcal{M}_{\mathcal{D},\mathcal{B}^c}^{\bar{\mathcal{D}}_i}$ are linearly independent, which is equivalent to linear independence of polynomials for the same $\mathcal{D}, \bar{\mathcal{D}}_i, \mathcal{B}^c$ and $j (j \in \mathcal{D})$. Note that $h_{j,m}$ only exists in polynomials whose symbols are transmitted by $\mathcal{B} \cup \{m\}$, the polynomials $\{\alpha_{\mathcal{D},\mathcal{B}^c \cup \{m\}}^{\bar{\mathcal{D}}_i}\tilde{h}_{k,\mathcal{B}^c \cup \{m\}}^{\bar{\mathcal{D}}_i} M_{\mathcal{D},\mathcal{B},p}^{\bar{\mathcal{D}}_i}\}$ are linearly independent for different $m$. Therefore, the polynomials with the same $\mathcal{D}, \bar{\mathcal{D}}_i$ and $\mathcal{B}^c(j \in \mathcal{D})$:
%\begin{IEEEeqnarray}{rCl}
%\begin{aligned}
$\left\{ \alpha_{\mathcal{D},\mathcal{B}^c \cup \{m\}}^{\bar{\mathcal{D}}_i} \tilde{h}_{k,\mathcal{B}^c \cup \{m\}}^{\bar{\mathcal{D}}_i} M_{\mathcal{D},\mathcal{B}^c,p}^{\bar{\mathcal{D}}_i}: m \notin \mathcal{B}^c, M_{\mathcal{D},\mathcal{B}^c,p}^{\bar{\mathcal{D}}_i} \in \mathcal{M}_{\mathcal{D},\mathcal{B}^c}^{\bar{\mathcal{D}}_i}[n] \right\}$
%\end{aligned}
%\end{IEEEeqnarray}
are linearly independent.

Because of the polynomials for each $\mathcal{B}^c$ {having} a unique factor set $\{\alpha_{\mathcal{D},\mathcal{B}^c \cup \{m\}}^{\bar{\mathcal{D}}_o}:j \notin \mathcal{B}^c\}$, polynomials $\{\alpha_{\mathcal{D},\mathcal{B}^c \cup \{m\}}^{\bar{\mathcal{D}}_i}\tilde{h}_{j,\mathcal{B}^c \cup \{m\}}^{\bar{\mathcal{D}}_i} M_{\mathcal{D},\mathcal{B}^c,p}^{\bar{\mathcal{D}}_i}\}$ for different $\mathcal{B}^c$ are {linearly independent}. Meanwhile, the polynomials corresponding to different $\mathcal{D}$ and $\bar{\mathcal{D}}_i$ have different factors $\{\alpha_{\mathcal{D},\mathcal{B}}^{\bar{\mathcal{D}}_i}\}$, they are linearly independent. Combining these conditions can get the linear independence of the polynomial functions, which is given by
\begin{IEEEeqnarray}{rCl}
	\begin{aligned} \label{matr2}
		&\bigg\{ \alpha_{\mathcal{D},\mathcal{B}}^{\bar{\mathcal{D}}_i}\tilde{h}_{j,\mathcal{B}}^{\bar{\mathcal{D}}_i}M_{\mathcal{D},\mathcal{B},p}^{\bar{\mathcal{D}}_i}: |\mathcal{D}|=s, j \in \mathcal{D}, |\mathcal{B}|=t, \nonumber\\
		&\hspace{1.5cm} i \in [\sigma], p \in \left[n^{(K_r-s-t-1)(K_t-t+1)}\right] \bigg\} \\
		&\hspace{0.5cm} cup \bigg\{ M_{\mathcal{D},\mathcal{B}^c,p}^{\bar{\mathcal{D}}_i}: M_{\mathcal{D},\mathcal{B}^c,p}^{\bar{\mathcal{D}}_i} \in \mathcal{M}_{\mathcal{D},\mathcal{B}^c}^{\bar{\mathcal{D}}_i}[n+1], \nonumber\\
		&\hspace{2cm} j \notin \mathcal{D} \cup \bar{\mathcal{D}}_i, |\mathcal{B}^c| =t-1 \bigg\}.
	\end{aligned}
\end{IEEEeqnarray}

Then putting specific $\{\alpha_{\mathcal{D},\mathcal{B}}^{\bar{\mathcal{D}}_i}(d)\tilde{h}_{j,\mathcal{B}}^{\bar{\mathcal{D}}_i}(d)M_{\mathcal{D},\mathcal{B},p}^{\bar{\mathcal{D}}_i}(d)\}$ and $\{M(d)\}$ into \eqref{matr2}, Similarly to \eqref{matr1}, the $\Gamma \times \Gamma$ received signal matrix is full-rank with probability 1 and successfully decode the $\binom{K_r-1}{s-1}\binom{K_t}{t}$ desired messages of receiver $k$.

Similarly, the polynomials of received symbols at other receivers are linearly independent. Therefore, using \cite[Lemma 3]{Annapureddy2012}, the received signal matrix at each receiver is full rank with probability 1, and each receiver can decode its desired messages successfully. Then a per-receiver DoF of $d = \frac{\binom{K_r-1}{s-1}\binom{K_t}{t}\binom{K_r-s}{t-1}tn^{(K_r-s-t+1)(K_t-t+1)}}{\Gamma}$ is achieved. When $n \rightarrow \infty$, the achievable per-receiver DoF is
\begin{IEEEeqnarray}{rCl} \label{dC1}
	\begin{aligned}
		d = \frac{\binom{K_r-1}{s-1}\binom{K_t}{t}\binom{K_r-s}{t-1}t}{ \binom{K_r-1}{s-1}\binom{K_t}{t}\binom{K_r-s}{t-1}t+ \binom{K_r-1}{s}\binom{K_t}{t-1}\binom{K_r-s-1}{t-1}}.
	\end{aligned}
\end{IEEEeqnarray}

Next, we consider the case when $t= K_t$. Each message $V_{\mathcal{D},[K_t]}$ is encoded into a $\binom{K_r-s}{K_t-1} \times 1$ symbols vector $\mathbf{V}_{\mathcal{D},\mathcal{B}} = \left(v_{\mathcal{D},\mathcal{B}}^1,v_{\mathcal{D},\mathcal{B}}^2,\cdots,v_{\mathcal{D},\mathcal{B}}^\sigma \right)^T,$ where $\sigma \triangleq \binom{K_r-s}{K_t-1}$. Using $\Gamma = \binom{K_r-1}{s-1} \binom{K_r-s}{K_t-1} + \binom{K_r-1}{s}\binom{K_r-s-1}{K_t-1}$ to transmit each message, in each time slot $d$, the received signal $Y_j(d)$ at an arbitrary receiver $j$, neglecting noise, is given by
\begin{IEEEeqnarray}{rCl} \label{receivedsignalC22}
	\begin{aligned}
		Y_{j}(d) = \sum_{\mathcal{D}:|\mathcal{D}|=s} \sum_{i=1}^{\sigma} \left[\sum_{m \in \mathcal{B}} h_{j,m}(d) w_{\mathcal{D},\mathcal{B},m}^i(d) \right] \! v^i_{\mathcal{D},\mathcal{B}},
	\end{aligned}
\end{IEEEeqnarray}
where $h_{j,m}(d)$ is the channel realization and $w_{\mathcal{D},[K_t],m}^i(d)$ is the precoder of symbol $v^i_{\mathcal{D},\mathcal{B}}$ at transmitter $m$.

Design $w_{\mathcal{D},\mathcal{B},m}^i(d) = \alpha_{\mathcal{D},\mathcal{B}}^{\bar{\mathcal{D}}_i}(d)c_m(d)$, where $\alpha_{\mathcal{D},\mathcal{B}}^{\bar{\mathcal{D}}_i}(d)$ is chosen i.i.d. from a continuous distribution for all $\{\mathcal{D},\bar{\mathcal{D}}_i,t\}$, $c_m(d)$ is the cofactor by $b(m)$ in the matrix from $\{h_{j,m}(d), j \in \bar{\mathcal{D}}_i, m \in \mathcal{B}\}$ and any $\{b_{1}, b_{2},\cdots, b_{t}\}$ like \eqref{matrix}.

Then the symbol vectors $w_{\mathcal{D},\mathcal{B}}^i$ undesired by receiver $j \in \bar{\mathcal{D}}_i$ are all neutralized like Case B and received signal in \eqref{receivedsignalC22} at receiver $j \in \mathcal{D}$ can be written as
\begin{IEEEeqnarray}{rCl} \label{receivedsignalC23}
	\begin{aligned}
		Y_{j}(d)&=&  \sum_{\mathcal{D}:|\mathcal{D}|=s,j \in \mathcal{D}} \sum_{i=1}^{\sigma} \left[\sum_{m \in \mathcal{B}}h_{j,m}(d)w_{\mathcal{D},\mathcal{B},m}^i(d)\right] v^i_{\mathcal{D},\mathcal{B}} \nonumber\\
		&&+ \sum_{\mathcal{D},\bar{\mathcal{D}}_i:j \notin \mathcal{D} \cup \bar{\mathcal{D}}_i} \left[\sum_{m \in \mathcal{B}}h_{j,m}(d)w_{\mathcal{D},\mathcal{B},m}^i(d)\right] v^i_{\mathcal{D},\mathcal{B}}\nonumber.
	\end{aligned}
\end{IEEEeqnarray}

Considering the polynomial functions as follows:
\begin{IEEEeqnarray}{rCl}
	\begin{aligned}
		&\left\{ \alpha_{\mathcal{D},\mathcal{B}}^{\bar{\mathcal{D}}_i} \sum_{m \in \mathcal{B}}h_{j,m}c_m \right\}_{i,\mathcal{D}:j\in \mathcal{D},|\mathcal{D}|=s, i \in [\sigma]} \nonumber\\
		&\hspace{2cm}\bigcup \left\{ \alpha_{\mathcal{D},\mathcal{B}}^{\bar{\mathcal{D}}_i} \sum_{m \in \mathcal{B}}h_{j,m}c_m \right\}_{\mathcal{D},\bar{\mathcal{D}}_i: j \notin \mathcal{D} \cup \bar{\mathcal{D}}_i}.
	\end{aligned}
\end{IEEEeqnarray}
Since each polynomials has a unique factor $\alpha_{\mathcal{D},\mathcal{B}}^{\bar{\mathcal{D}}_i}$, these polynomials are linearly independent. Due to the construction method of $\{w_{\mathcal{D},\mathcal{B},m}^i(d)\}$ and the formation of $\{\sum_{m \in \mathcal{B}}h_{j,m}(d)w_{\mathcal{D},\mathcal{B},m}^i(d)\}$ are the same at each time slot $d$, based on \cite[Lemma 3]{Annapureddy2012}, we can ensure the following $\Gamma \times \Gamma$ received signal matrix is full-rank with probability 1. Similarly, the polynomials of received symbols at other receivers are also linearly independent. Therefore, the received signal matrix at each receiver is full rank with probability 1. Since each receiver can decode $\binom{K_r-1}{s-1}\binom{K_r-s}{K_t-1}$ symbols in $\Gamma = \binom{K_r-1}{s-1} \binom{K_r-s}{K_t-1} + \binom{K_r-1}{s}\binom{K_r-s-1}{K_t-1}$-symbol extension, the per-receiver DoF is achievable of
\begin{IEEEeqnarray}{rCl} \label{dC2}
	\begin{aligned}
		d = \frac{\binom{K_r-1}{s-1}\binom{K_r-s}{K_t-1}}{\binom{K_r-1}{s-1} \binom{K_r-s}{K_t-1} + \binom{K_r-1}{s}\binom{K_r-s-1}{K_t-1}}.
	\end{aligned}
\end{IEEEeqnarray}

Combining \eqref{dC1} and \eqref{dC2}, we can get the per-receiver DoF of
\begin{IEEEeqnarray}{rCl}
	\begin{aligned}
		d = \frac{\binom{K_r-1}{s-1}\binom{K_t}{t}\binom{K_r-s}{t-1}t}{ \binom{K_r-1}{s-1}\binom{K_t}{t}\binom{K_r-s}{t-1}t+ \binom{K_r-1}{s}\binom{K_t}{t-1}\binom{K_r-s-1}{t-1}}.
	\end{aligned}
\end{IEEEeqnarray}

For an arbitrary $\binom{K_t}{t} \times \binom{K_r}{s}$ cooperative X-multicast channel. Spilt message into $\binom{t}{t'}(t'<t)$ sub-messages and only transmit by transmitter set $\mathcal{B}' \subset \mathcal{B}$ with $|\mathcal{B}'| =t' $. Then the network topology can be transformed into $\binom{K_t}{t'} \times \binom{K_r}{s}$ cooperative X-multicast channel. Therefore, any achievable DoF in $\binom{K_t}{t'} \times \binom{K_r}{s}$ cooperative X-multicast channel can be achievable in original $\binom{K_t}{t} \times \binom{K_r}{s}$ cooperative X-multicast channel. The optimal DoF in this scheme using the second method is
\begin{align}\label{d'be}
	d'_{s,t} \!\triangleq \! \max \limits_{1\le t'\le t}\left\{\frac{\binom{K_\textnormal{r}-1}{s-1}\binom{K_\textnormal{t}}{t'}\binom{K_\textnormal{r}-s}{t'-1}t'}
	{\binom{K_\textnormal{r}-1}{s-1}\binom{K_\textnormal{t}}{t'}\binom{K_\textnormal{r}-s}{t'-1}t'+\binom{K_\textnormal{r}-1}{s}\binom{K_\textnormal{r}-s-1}{t'-1}\binom{K_\textnormal{t}}{t'-1}}\right\}.
\end{align}

Combining the achievable DoF of $ \frac{r}{K_r}$ using first method and \eqref{d'be} in second method, the achievable per-receiver DoF when $s+t \leq K_\textnormal{r}-1$ is
\begin{align}
	d = \max\left\{d'_{s,t},\frac{s+t}{K_\textnormal{r}}\right\}.
\end{align}

%%%%%%%%%%%%%%%%%%%%%%%%%%%%%%%%%%%%%%%%%%%%%%%%%%%%%%%%%%%%%%%%%%%%%%%%%%%%%%%%
%appendices BC 
%%%%%%%%%%%%%%%%%%%%%%%%%%%%%%%%%%%%%%%%%%%%%%%%%%%%%%%%%%%%%%%%%%%%%%%%%%%%%%%%%%%%%%%%%%%%%	

	\section{Proof of theorem \ref{minmul_throrem2}}\label{proof_th2}
Given $r$ and $K$, the NDT $\delta(r,K)$ of the CPC scheme {will not be} less than $\delta_{\text{CDC}}=\frac{1}{r}(1 - \frac{r}{K})$ in \cite{CDC} when $r\geq K_\textnormal{r}$. We can also see that $\delta_{\text{CPC}}(r,1,K,K_\textnormal{r}^{*}) \leq \frac{1}{r}(1 - \frac{r}{K})$ when $r<K_\textnormal{r}$. Therefore we {only consider} finding $\delta_{min}$ for the case $r\leq K_\textnormal{r}$.

For the case $r = K_\textnormal{r}-1$, if $r$, $K$ are fixed and $t \in\mathbb{N}^+ $, $\delta_{\text{CPC}}(r,t,K,K_\textnormal{r})$ depends on the factor $\frac{1}{\binom{r}{t} \binom{K-K_\textnormal{r}}{t}t}$. Let $\Phi(t) = \frac{1}{\binom{r}{t} \binom{K-K_\textnormal{r}}{t}\cdot t}$, $t \in\mathbb{N}^+ $, it can be verified that for any $t \leq \lfloor 1 + \frac{-r^2 + rK -r}{K} \rfloor$, $\Phi(t-1) \geq \Phi(t)$ and the opposite for $t \geq \lfloor 1 + \frac{-r^2 + rK -r}{K} \rfloor$, i.e., the minimum of $\Phi(t)$ is achieved at $t^{*} = \lfloor 1 + \frac{-r^2 + rK -r}{K} \rfloor$. Therefore, the minimum NDT in the case $r=K_\textnormal{r}-1$ is $\frac{1}{r+1}(1-\frac{r}{K})(1+\frac{1}{\binom{r}{t^*}\binom{K-r-1}{t^*}t^*})$.

For the case $r \leq K_\textnormal{r}-2$, it can be seen that $\frac{1}{\tau_{r,t}}=1+\frac{K_\textnormal{r}-r-j}{(r+1-t)(K_\textnormal{t}-j+1)}$ is an increasing function of $j$ when $K_t \leq K_r - (r+1-t)$, and is a decreasing function of $j$ when $K_t \geq K_r - (r+1-t)$. The minimum of $\delta_{\text{CPC}}(r,t,K,K_\textnormal{r})$ in this case can be written as
\begin{align}\label{ndt2min}
	%\textnormal{NDT}_2(r,t,K) = 
	\frac{1}{r} (1\!-\!\frac{r}{K}) \frac{r}{K_\textnormal{r}}\cdot \min & \bigg\{ 1+ \frac{t(K_\textnormal{r}\!-\! r) }{ (r+1-t)(K_\textnormal{t}\!-\! t \!+\! 1) },\nonumber\\
    &\hspace{1.5cm} 1\!+\!\frac{K_\textnormal{r}\!-\!(r \! +\! 1 \!-\! t)}{(r \!+\! 1 \!-\! t)K_\textnormal{t}}, \frac{K_r}{r} \bigg\}.
\end{align}
It can be seen that the optimal $t$ for \eqref{ndt2min} is $t=1$, and the minimum of $\delta_{\text{CPC}}(r,t,K,K_\textnormal{r})$ in this case is $\frac{1}{r} (1 - \frac{r}{K}) \frac{K_\textnormal{t} r + K_\textnormal{r} - r}{K_\textnormal{t} K_\textnormal{r}}$.

Then we want to find the optimal $K_\textnormal{r}$ considering the $r$ and $K$ is fixed and $r\leq K$. Note that the value of $\delta_{\text{CPC}}(r,1,K,K_\textnormal{r})$ depends on $\frac{ (K-K_\textnormal{r})r + K_\textnormal{r} - r }{ (K-K_\textnormal{r})K_\textnormal{r} }$. Let the factor {be}
\begin{align}
	h(K_\textnormal{r})  &= \frac{ (K-K_\textnormal{r})\cdot r + K_\textnormal{r} - r }{ (K-K_\textnormal{r})\cdot K_\textnormal{r} }, \nonumber\\
	h(K_\textnormal{r}-1)  &= \frac{ (K-K_\textnormal{r})\cdot r + K_\textnormal{r} - 1 }{ (K-K_\textnormal{r}+1)(K_\textnormal{r}-1) }.
\end{align}
When \begin{align}
	K_\textnormal{r} \leq K_\textnormal{r}^{*} = \begin{cases}
		\lfloor \frac{K+1}{2} \rfloor , & r=1, \\
		\lfloor \frac{rK-r+\frac{r-1}{2} - \sqrt{r(K-1)(K-r) +\frac{(r-1)^2}{4}}}{r-1} \rfloor , & r > 1,
	\end{cases}
\end{align} it can be verified	that
\begin{align}
	\frac{ (K-K_\textnormal{r})r + K_\textnormal{r} - r }{ (K-K_\textnormal{r})K_\textnormal{r} } \frac{ (K-K_\textnormal{r}+1)(K_\textnormal{r}-1) }{ (K-K_\textnormal{r})r + K_\textnormal{r} - 1 } >1,
\end{align}
which is equivalent to $\frac{ h(K_\textnormal{r}) }{ h(K_\textnormal{r}-1)} \geq 1$. The inequation $\frac{ h(K_\textnormal{r}) }{ h(K_\textnormal{r}-1)} < 1$ {also holds},
when \begin{align}
	K_\textnormal{r} > K_\textnormal{r}^{*} = \begin{cases}
		\lfloor \frac{K+1}{2} \rfloor , & r=1, \\
		\lfloor \frac{rK-r+\frac{r-1}{2} - \sqrt{r(K-1)(K-r) +\frac{(r-1)^2}{4}}}{r-1} \rfloor , & r > 1.
	\end{cases}
\end{align}
Therefore, the optimal $K_\textnormal{r}$ is $K_\textnormal{r}^{*}$ and $K_\textnormal{t}^{*} = K - K_\textnormal{r}^{*}$ . This completes the proof of Theorem \ref{minmul_throrem2}.

\section{Proof of theorem \ref{teq1_theorem3}}\label{proof_th3}
When $K \leq 5$, we have $t^{*} = \lfloor 1 + \frac{-r^2 + rK -r}{K} \rfloor = 1$. Combining Appendix \ref{proof_th2}, we know $\delta_{min}$ is achieved by setting $t = 1$ in the case $r \leq K_\textnormal{r}-1$ and $\delta_{min} = \delta_{\text{CPC}}(r,1,K,K_\textnormal{r}) = \textnormal{NDT}_2$.

When $K \geq \max\{r+4+\frac{4}{r-1}, \frac{r+4+\sqrt{r^2+16r}}{2}\}$, we want to prove $\delta_{min}$ is also attained when $t=1$. Recalling that
\begin{align}\label{ndt2min2}
	\textnormal{NDT}_2 = \frac{1}{r} (1 - \frac{r}{K}) \frac{K_\textnormal{t}^{*} r + K_\textnormal{r}^{*} - r}{K_\textnormal{t}^{*} K_\textnormal{r}^{*}},
\end{align}
ignoring the integer constraints, we set $K_\textnormal{t}^{*}=\frac{r - K + \sqrt{r(K-1)(K-r)}}{r-1}$ and $K_\textnormal{r}^{*}=K-K_\textnormal{t}^{*} = \frac{r K - r -\sqrt{r(K-1)(K-r)}}{r-1}$.
When $K \geq \frac{r+4+\sqrt{r^2+16r}}{2}$, it can be verified that
\begin{align}\label{Kt & 2}
	K_\textnormal{t}^{*} \geq 2 .
\end{align}
Meanwhile, when $K\geq r+4+\frac{4}{r-1}$, it can be verified that $K_\textnormal{r}^{*} \geq \frac{(r+1)r}{r-1}$, which is equivalent to
\begin{align}\label{Kr & r2}
	\frac{1}{r+1} - \frac{1}{K_\textnormal{r}^{*}} \geq \frac{1}{2} (\frac{1}{r} + \frac{1}{K_\textnormal{r}^{*}}).
\end{align}
Combining \eqref{Kt & 2} and \eqref{Kr & r2} we get
	$\frac{1}{r+1} - \frac{1}{K_\textnormal{r}^{*}} \geq \frac{1}{K_\textnormal{t}^{*}} (\frac{1}{r} + \frac{1}{K_\textnormal{r}^{*}})$,
which is equivalent to
\begin{align}\label{ineqrelex}
	\frac{r}{r+1} \geq \frac{K_\textnormal{t}^{*} r + K_\textnormal{r}^{*} - r}{K_\textnormal{t}^{*} K_\textnormal{r}^{*}}.
\end{align}
Combining \eqref{ndt1}, \eqref{ndt2} and \eqref{ineqrelex}, we get $
	\textnormal{NDT}_1(r,K) > \textnormal{NDT}_2(r,K).
$ 
Therefore, $\delta_{min}$ is attained by letting $t=1$ when $K \geq \max\{r+4+\frac{4}{r-1}, \frac{r+4+\sqrt{r^2+16r}}{2}\}$ and
\begin{align}
	\delta_{min} = \textnormal{NDT}_2(r,K) = \frac{1}{r} (1 - \frac{r}{K}) \frac{K_\textnormal{t}^{*} r + K_\textnormal{r}^{*} - r}{K_\textnormal{t}^{*} K_\textnormal{r}^{*}} .
\end{align}
This completes the proof of Theorem \ref{teq1_theorem3}.

\section{Proof of Theorem \ref{RemCompa}} \label{prooftheorem3}
%{\color{blue}	When   $r\leq \frac{K}{2}$, we have $\delta^\text{half-duplex}_\text{BW}(r,K)=    (1-\frac{r}{K})\geq {\delta}_\text{CDC}(r,K)=\frac{1}{r} (1-\frac{r}{K}).$ 
%When $r>  \frac{K}{2}$, we can obtain that $\frac{K[r(K-1)+K-r-1]}{r(K-1)^2+r(K-2)}>1$. Thus, we have $\delta^\text{half-duplex}_\text{BW}(r,K)\geq    (1-\frac{r}{K})\geq {\delta}_\text{CDC}(r,K)= \frac{1}{r} (1-\frac{r}{K}).$ Obivously, from \eqref{TDBW}, if $r\ll K$, we have ${\color{blue}\lim_{K\to\infty} \delta^\text{half-duplex}_\text{BW}(r,K)=\lim_{K\to\infty} (1-\frac{r}{K})=1.} $}

It is easy to obtain that when $K\in\{ 1,2\}$, $\bar{\delta}_\text{CPC}(r,K)={\delta}^{\text{HD}}_\text{OSL}(r,K)={\delta}_\text{CDC}(r,K)$. Thus, we only focus on the case with $K> 2$. 
Since $ \bar{\delta}_\text{CPC}=0$ when $ r = K $, to prove the first inequality in \eqref{CPCCDC}, it is sufficient to prove $\frac{ K_\textnormal{t}r +K_\textnormal{r} -r }{K_\textnormal{t}K_\textnormal{r}}\leq 1$ for any $r\in[K-1]$. First, consider the regime $ r\in[K-2]$. Let $ K_\textnormal{t} = K-r-1 $ and $ K_\textnormal{r} = r+1 $, then the multiplicative factor
\begin{IEEEeqnarray*}{rCl}
	\frac{ K_\textnormal{t}r +K_\textnormal{r} -r }{K_\textnormal{t}K_\textnormal{r}} = \frac{(K-r-1)r+1}{(K-r-1)r+K-r-1} < 1.
\end{IEEEeqnarray*}
Next, consider the rest regime $ r = K-1$. Let $ K_\textnormal{t} = K-r $ and $ K_\textnormal{r} = r $, then the multiplicative factor $\frac{ K_\textnormal{t}r +K_\textnormal{r} -r }{K_\textnormal{t}K_\textnormal{r}} = \frac{(K-r)r+r-r}{(K-r)r} = 1$. 
For the second inequality in \eqref{CPCCDC}, it is obviously that $\frac{1}{r}\left(1-\frac{r}{K}\right) \leq \frac{1}{\min\{K/2,r\}}\left(1-\frac{r}{K}\right)$.

Here we prove \eqref{CPCBW}. When $r\geq \frac{K}{2}$, the inequation holds since $\delta^{\text{HD}}_{\text{BW}}(r,K)=\delta^{\text{HD}}_{\text{OSL}}(r,K)$. When $r<\frac{K}{2}$, let $K^{*}_\textnormal{r}=\frac{K}{2}$, $\bar{\delta}_\text{CPC}(r,K)=(1-\frac{r}{K})\frac{2K+\frac{2K}{r}-4}{K^2}$ and $\delta^{\text{HD}}_{\text{BW}}(r,K)=(1-\frac{r}{K})\frac{ 2K+\frac{2(K-1)}{r}-4 }{K^2-K-1}$. It is sufficient to prove $\frac{2K+\frac{2K}{r}-4}{K^2}< \frac{ 2K+\frac{2(K-1)}{r}-4 }{K^2-K-1}$, which is equivalently to prove $K^2+K(\frac{1}{r}-1)-2\geq 0$. It is easy to be verified that $K^2+K(\frac{1}{r}-1)-2\geq 0$ when $K\geq 2$. Hence, the inequation holds when $r<\frac{K}{2}$.

  Observe that if $ K \geq 2(r+1+\sqrt{r^2+1}) $, then we have ${\min\{K,2r\}}=2r$, and by choosing $ K_\textnormal{t} = \lfloor \frac{K}{2} \rfloor$ and $K_\textnormal{r}=K-K_\textnormal{t}$, we obtain that
%\begin{IEEEeqnarray*}{rCl}
	$\min_{K_\textnormal{t}\in[K-r]} \frac{K_\textnormal{t}K_\textnormal{r}r}{ K_\textnormal{t}r +{K_\textnormal{r}}-r } \geq 2r ={\min\{K,2r\}}.$ %\end{IEEEeqnarray*}
% which complete the proof of \eqref{CPCOneshot}.
Hence, $\bar{\delta}_{\text{CPC}}(r,K) \leq {\delta}^{\text{FD}}_\text{OSL}(r,K)$,  if $K \geq 2(r+1+\sqrt{r^2+1})$.

Finally, when $ K \to \infty $ and $r\ll K$, it is easy to obtain that $\lim_{K\to\infty}\delta_\text{CDC}\to 1/r $, $\lim_{K\to\infty} \delta^{\text{FD}}_\text{OSL} \to 1/(2r) $, $\lim_{K\to\infty}\delta^{\text{HD}}_{\text{OSL}}\to\frac{1}{r}$ and $\lim_{K\to\infty}\delta^{\text{HD}}_{\text{BW}}\to 0$. For $\bar{\delta}_\text{CPC}(r,K)$, let $K_\textnormal{t}=K_\textnormal{r}=K/2$, then we obtain that
%\begin{IEEEeqnarray*}{rCl}
	$\lim_{K\to\infty} \frac{1}{r} \big({1}- \frac{r}{K}\big) \frac{ K r/2 + K/2-r }{K^2/4} = 0.$
%\end{IEEEeqnarray*}
This completes the proof of Theorem \ref{RemCompa}.

{
\section{Proof of Lemma \ref{lma:dof_upper_bound}} \label{sec:dof_upper_bound_proof}
	Denote by $\mathcal{H}$ the set of all channel coefficients to all nodes in the system. Define $\mathcal{W}_c  \triangleq  \mathcal{A} \backslash (	\mathcal{W}_r  \cup 	\mathcal{W}_{t1} \cup \mathcal{W}_{t2} )$, and define $\mathcal{F} = [K]\backslash(\cT \cup \cR)$. 
	Since channel coefficients and IVs are independent, we have
	{
	\begin{IEEEeqnarray}{rCl}
		\lefteqn{	H(\mathcal{W}_{t1}, \mathcal{W}_{t2}, \mathcal{W}_r) }\nonumber \\
 		&=& H(\mathcal{W}_{t1}, \mathcal{W}_r|\mathcal{W}_{t2}, \mathcal{W}_c,\mathcal{H}) +  H( \mathcal{W}_{t2}|\mathcal{W}_{t1}, \mathcal{W}_r, \mathcal{W}_c,\mathcal{H})\nonumber\\
		&=& I(\mathcal{W}_{t1}, \mathcal{W}_r; \*Y_{\mathcal{R}}, |\mathcal{W}_{t2}, \mathcal{W}_c,\mathcal{H}) + 
		H(\mathcal{W}_{t1},\mathcal{W}_r| \mathcal{W}_{t2}, \mathcal{W}_c,\*Y_{\mathcal{R}}, \mathcal{H}) \nonumber\\ &&+ I(\mathcal{W}_{t2}; \*Y_{\cT}, |\mathcal{W}_{t1}, \mathcal{W}_r, \mathcal{W}_c,\mathcal{H}) + 
		H(\mathcal{W}_{t2}| \mathcal{W}_{t1}, \mathcal{W}_r, \mathcal{W}_c, \*Y_{\cT}, \mathcal{H}) \IEEEeqnarraynumspace\nonumber\\
		&=& h(\*Y_{\mathcal{R}}| \mathcal{W}_{t2}, \mathcal{W}_c,\mathcal{H}) - h(\*Z_{\mathcal{R}}) + H(\mathcal{W}_r| \mathcal{W}_{t2}, \mathcal{W}_c,\*Y_{\mathcal{R}}, \mathcal{H})  \nonumber \\ 
		&&+ H(\mathcal{W}_{t1}|\mathcal{W}_r,\mathcal{W}_{t2}, \mathcal{W}_c, \*Y_{\cR}, \mathcal{H}) + h(\vect{Y}_{\cT}|\mathcal{W}_{t1}, \mathcal{W}_{r}, \mathcal{W}_c, \mathcal{H}) \nonumber \\
		&& - h(\vect{Z}_{\cT}) + H(\mathcal{W}_{t2}|\mathcal{W}_{t1}, \mathcal{W}_r,\mathcal{W}_c,  \*Y_{\mathcal{T}}, \mathcal{H})\nonumber \\
		&\leq& h(\*Y_{\mathcal{R}}|\mathcal{W}_{t2}, \mathcal{W}_c,\mathcal{H}) + h(\*Y_{\cT}| \mathcal{W}_{t1}, \mathcal{W}_r, \mathcal{W}_c,\mathcal{H}) - h(\*Z_{\cR}) \nonumber \\
		&& - h(\*Z_{\cT}) +  {  T_1} \epsilon_{  T_1}
		+ H(\mathcal{W}_{t1}|\mathcal{W}_r,\mathcal{W}_{t2}, \mathcal{W}_c, \*Y_{\mathcal{R}}, \mathcal{H}) + {  T_2} \epsilon_{  T_2} ,\label{eq:a2}
	\end{IEEEeqnarray}}
	where we defined $\*Y_{\mathcal{A}} \triangleq [\*Y_j]_{j \in \mathcal{A}}$ for a set $\mathcal{A}\subseteq [ K]$ ,  also assume the length of $\*Y_{\cR}$ is $  T_1$, the length of $\*Y_{\cT}$ is $  T_2$ and and $  T =   T_1 +   T_2$, $\epsilon_{  T_1}$ and $\epsilon_{  T_2}$ are vanishing sequences as $  T\to \infty$. Here the inequality holds by Fano's inequality, because $\mathcal{W}_r$ is decoded from  $\*Y_{\mathcal{R}}$ and $\mathcal{W}_c$, and because we impose vanishing probability of error \eqref{eq:error_computing}. 
	
	Again by Fano's inequality and by \eqref{eq:error_computing}, there exists a vanishing sequence $\epsilon_{  T_1}'$ such  that
        {
            \begin{IEEEeqnarray}{rCl} 
		  \lefteqn{H(\mathcal{W}_{t1}|\mathcal{W}_r,\mathcal{W}_{t2}, \mathcal{W}_c, \*Y_{\mathcal{R}},\mathcal{H}) }\hspace{1cm}  \nonumber \\
		  &\leq & I (\mathcal{W}_{t1}; \*Y_{\mathcal{F}}|\mathcal{W}_r,\mathcal{W}_{t2},\mathcal{W}_c, \*Y_{\mathcal{R}}, \mathcal{H}) 
		  +  {T_1} \epsilon_{  T_1}' \\
		  &=& h(\*Y_{\mathcal{F}}| \mathcal{W}_r,\mathcal{W}_{t2},\mathcal{W}_c, \*Y_{\mathcal{R}}, \mathcal{H}) \\
		  &&- h(\*Y_{\mathcal{F}}| \mathcal{W}_r, \mathcal{W}_{t2},\mathcal{W}_{t1},\mathcal{W}_c, \*Y_{\mathcal{R}},  \mathcal{H}) +  {T_1} \epsilon_{  T_1}' \\
		  &\leq& h(\bar{\*Y}_{\mathcal{F}}|\bar{\*Y}_{\mathcal{R}}, \mathcal{H})   - h(\*Z_{\mathcal{F}}) +  {T_1} \epsilon_{  T_1}' , \IEEEeqnarraynumspace \label{eq:a}
	\end{IEEEeqnarray}
        }
	where  $\bar{\*Y}_{\mathcal{A}} \triangleq [\bar{\*Y}_j]_{j \in \mathcal{A}}$ and $\bar{\*Y}_j$ denotes Node $j$'s ``cleaned" signal without the inputs that do not depend on files in $\mathcal{M}$ but   only  on IVs $\mathcal{W}_r \cup  \mathcal{W}_c$: 
	\begin{equation*}
		\bar{\*Y}_j \triangleq \*H_{j,{\mathcal{T}}}\vect{X}_{\mathcal{T}} + \*Z_j, \qquad  j \in \mathcal{F}.
	\end{equation*}
	Here,	$\*H_{\mathcal{A},\mathcal{B}}$ denotes the channel matrix from  set $\mathcal{B}$ to  set $\mathcal{A}$. 
	
	To bound the first term in \eqref{eq:a}, we introduce a random variable $E$ indicating whether the matrix $\mat{H}_{\mathcal{R},\mathcal{T}}$ is invertible $(E=1)$  or not $(E=0)$. If this matrix is invertible and $E=1$, then  the input vector  $\vect{X}_{\mathcal{T}}$ can be computed from  $\bar{\*Y}_{\mathcal{R}}$ up to noise terms.   Based on this observation and defining the residual noise terms
	\begin{equation}
		\bar{\*Z}_j \triangleq \*Z_j - \*H_{j,\mathcal{T}}\*H_{\mathcal{R},\mathcal{T}}^{-1}\*Z_{\mathcal{R}}, \quad \textnormal{ if }E=1,
	\end{equation}
	we obtain: 
	\begin{IEEEeqnarray}{rCl}
		\lefteqn{	h(\bar{\*Y}_{\mathcal{F}}| \bar{\*Y}_{\mathcal{R}}, \mathcal{H}) }\nonumber\\
		&\leq & \mathbb{P}(E=1) \cdot h\left(\bar{\mat{Z}}_{\mathcal{F}} |\bar{\*Y}_{\mathcal{R}},\mathcal{H}, E=1\right)  \nonumber\\
		&&+ \mathbb{P}(E=0) \cdot h\left(\bar{\*Y}_{\mathcal{F} } | \bar{\*Y}_{\mathcal{R}},\mathcal{H}, E=0\right) \\
		& \leq & h\left(\bar{\mat{Z}}_{\mathcal{F}}\right)  + \mathbb{P}(E=0)h\left(\bar{\*Y}_{\mathcal{F}} | \bar{\*Y}_{\mathcal{R}},\mathcal{H}, E=0\right).\IEEEeqnarraynumspace
	\end{IEEEeqnarray}
	Since  the channel coefficients follow continuous distribution, $\mat{H}_{\mathcal{R},\mathcal{T}}$ is invertible almost surely,  implying $\mathbb{P}(E=0)=0$. By the boundedness of the entropy term $h\left(\bar{\*Y}_{\mathcal{F}} | \bar{\*Y}_{\mathcal{R}},\mathcal{H}, E=0\right)$ (since power $  P$ and channel coefficients are bounded), this implies 
        {
            \begin{IEEEeqnarray}{rCl} 
		  \lefteqn{H(\mathcal{W}_{t1}, \mathcal{W}_{t2}, \mathcal{W}_r)} \nonumber\\
		  &\leq&  h(\*Y_{\mathcal{R}}|\mathcal{W}_{t2}, \mathcal{W}_c,\mathcal{H}) + h(\*Y_{\cT}| \mathcal{W}_{t1}, \mathcal{W}_r, \mathcal{W}_c,\mathcal{H}) - h(\*Z_{\cR})  \nonumber\\
            &&  - h(\*Z_{\cT}) + h(\bar{\*Z}_{\mathcal{F}}) - h(\*Z_{\mathcal{F}})  +  {  T_1} (\epsilon_{  T_1} + \epsilon_{  T_1}') + {  T_2}\epsilon_{  T_2} \nonumber\\
		  &\leq& {  T}|\mathcal{R} |\log(  P) + {  T}C_{  T, \mathcal{H}}, \label{eqn:converse_part4_2}
	   \end{IEEEeqnarray}
        }
	where $C_{  T, \mathcal{H}}$ is a function that is uniformly bounded over all realizations of channel matrices and powers $  P$. Noticing  
	\begin{equation}
		H(\mathcal{W}_{t1}, \mathcal{W}_{t2}, \mathcal{W}_r) = {\eta_2}  B( |\mathcal{W}_{t1}|+|\mathcal{W}_{t2}| + | \mathcal{W}_r|),
	\end{equation} dividing  \eqref{eqn:converse_part4_2} by $  T \log (  P)$,  and letting $  P \to \infty$, establishes the lemma because $|\mathcal{R}|=|\mathcal{T}|$ and  ${  T}C_{  T, \mathcal{H}}$ is bounded. 
}

\end{appendices}

\bibliographystyle{IEEEtran}
\bibliography{paper.bib}

\end{document}